\documentclass[a4paper,11pt,DIV=12,abstract=true,numbers=noenddot,titlepage=false,twocolumn=false,draft=false]{scrartcl}
\pdfoutput=1

\usepackage[ttscale=0.9]{libertine}
\usepackage[T1]{fontenc}
\usepackage[utf8]{inputenc}

\makeatletter
\DeclareOldFontCommand{\rm}{\normalfont\rmfamily}{\mathrm}
\DeclareOldFontCommand{\sf}{\normalfont\sffamily}{\mathsf}
\DeclareOldFontCommand{\tt}{\normalfont\ttfamily}{\mathtt}
\DeclareOldFontCommand{\bf}{\normalfont\bfseries}{\mathbf}
\DeclareOldFontCommand{\it}{\normalfont\itshape}{\mathit}
\DeclareOldFontCommand{\sl}{\normalfont\slshape}{\@nomath\sl}

\usepackage[usenames,dvipsnames]{color}
  \definecolor{hgreen}{rgb}{0,.3,0}
  \definecolor{hred}{rgb}{.3,0,0}
  \definecolor{hblue}{rgb}{0,0,.3}
  \definecolor{LightGray}{gray}{0.95}
  \definecolor{gray}{gray}{0.6}
\usepackage[
	    colorlinks=true,
	    linkcolor=hred,
	    citecolor=hgreen,
	    filecolor=hblue,
	    urlcolor=hred
	    ]{hyperref}

\usepackage{scrlayer-scrpage}
\usepackage[titletoc,title]{appendix}
\usepackage[table]{xcolor}
\usepackage[numbers,sort&compress]{natbib}
\usepackage[protrusion=true,expansion,kerning=true,tracking=true,final]{microtype}
\usepackage[affil-it]{authblk}
\usepackage[intlimits]{amsmath}

\usepackage{amssymb}
\usepackage{mathrsfs}
\usepackage{slashed}

\usepackage{float}
\usepackage{soul}
\usepackage[compat=1.1.0]{tikz-feynman}
\usepackage{booktabs}
\usepackage{physics}
\usepackage{comment}
\usepackage{braket}
\usepackage{graphicx}
\usepackage{tensor}
\usepackage{array}
\usepackage{colortbl}
\usepackage{multirow}

\DeclareMathAlphabet{\mathbbold}{U}{bbold}{m}{n}

\setcapindent{1em}
\setkomafont{captionlabel}{\bfseries}

\KOMAoptions{headinclude=false,
             footinclude=false,
             parskip=false,
             draft=false,
             abstract=true,
             numbers=noenddot,
             titlepage=false,
             twocolumn=false,
             twoside=false,
             DIV=12}

\hypersetup{
  colorlinks=true,
  linkcolor=red,
  citecolor=blue,
  urlcolor=blue
}

\numberwithin{equation}{section}

\newcommand{\CC}[2]{\mathbb{C}^{#1}_{#2}}

\newcommand{\Lag}{\mathscr{L}}
\newcommand{\eps}{\epsilon}

\newcommand{\kag}{\kappa_{\V}}
\newcommand{\kap}{\kappa_{P^\prime}}
\newcommand{\kaV}{\kappa_{\mathcal{V}}}
\newcommand{\kaB}{\kappa_{B^\prime}}
\newcommand{\kal}{\kappa_{\ell^\prime}}

\newcommand{\MeV}{\, {\rm MeV}}
\newcommand{\GeV}{\, {\rm GeV}}
\newcommand{\TeV}{\, {\rm TeV}}

\newcommand{\V}{V^\prime}

\newcommand{\rD}{{\rm D}}
\newcommand{\rV}{{\rm V}}
\newcommand{\rA}{{\rm A}}
\newcommand{\rL}{{\rm L}}
\newcommand{\rR}{{\rm R}}
\newcommand{\rT}{{\rm T}}
\newcommand{\rTV}{{\rm TV}}
\newcommand{\rTA}{{\rm TA}}

\allowdisplaybreaks

\begin{document}
\renewcommand\Authands{, }

\titlehead{\hfill P3H-24-013, TTP24-003}

\title{\boldmath
Flavor Phenomenology of Light Dark Vectors}

\date{\today}

\author[a]{Jordi Folch Eguren%
\thanks{\texttt{jordi.eguren@tu-dortmund.de}}}

\author[b]{Sophie Klingel}

\author[a]{Emmanuel Stamou%
\thanks{\texttt{emmanuel.stamou@tu-dortmund.de}}}

\author[a]{\\Mustafa Tabet%
\thanks{\texttt{mustafa.tabet@tu-dortmund.de}}}

\author[b,c]{Robert Ziegler%
\thanks{\texttt{robert.ziegler@kit.edu}}}

\affil[a]{{\large Department of Physics, TU Dortmund, D-44221 Dortmund, Germany}}
\affil[b]{{\large Institute for Theoretical Particle Physics, KIT, 76128 Karlsruhe, Germany }}
\affil[c]{{\large Institute for Theoretical Physics, Heidelberg University, 69120 Heidelberg, Germany }}

\maketitle

\begin{abstract}
Light dark matter with flavor-violating couplings to fermions may be copiously
produced in the laboratory as missing energy from decays of SM particles. Here
we study the effective Lagrangian of a light dark vector with generic dipole or
vector couplings. We calculate the resulting two-body decay rates of mesons,
baryons and leptons as a function of the dark vector mass and show that existing
experimental limits probe UV scales as large as $10^{12} \GeV$. We also derive
the general RGEs in order to constrain the flavor-universal UV scenario, where
all flavor violation arises radiatively proportional to the CKM matrix.
\end{abstract}

\section{Introduction\label{sec:introduction}}
In recent years light new particles interacting very weakly with the Standard
Model (SM) have gained increased interest. The so far negative results on
searches for heavy particles above the electroweak scale at the LHC and 
high-intensity experiments have increased the interest in less explored scenarios, with
additional degrees of freedom beyond the SM with masses at sub-GeV scales. Such
particles can be motivated by dynamics addressing the Strong CP Problem (in case
of the QCD axion) or the origin of neutrino masses (in case of sterile
neutrinos), but probably the main motivation is the possibility that such light
particles could be connected to the origin of particle dark matter
(DM)~\cite{Knapen:2017xzo}.

In this context a popular scenario is the dark photon~\cite{Holdom:1985ag,
Dobrescu:2004wz}, which is either itself DM or is the only mediator (``Vector
Portal'') between the SM and a hidden ``dark sector'', which contains one or
several DM particles~\cite{Hambye:2019dwd, Chun:2010ve}, see
Ref.~\cite{Fabbrichesi:2020wbt} for a review.  The term ``dark photon'' usually
refers to a light vector particle coupled to the SM only via kinetic mixing or
dipole operators and that is often taken as the only new degree of freedom.
Instead, the term $Z^\prime$ is typically reserved for the vast model space
of theories of gauged $U(1)^\prime$ extensions of the SM, where also a
complete Higgs sector for $U(1)^\prime$ breaking is explicitly present,
besides additional matter needed for anomaly cancellation, see, e.g.,
Ref.~\cite{Allanach:2018vjg} for a classification. While the $Z^\prime$
vector boson is often taken to be heavy, with a mass much above the electroweak (EW) scale,
this particle can also be much lighter. The resulting coupling patterns are often
related to the underlying UV symmetries, see, e.g.,
Refs.~\cite{Williams:2011qb, Smolkovic:2019jow,Kahn:2016vjr,Bauer:2018onh}, and can 
leave imprints in low-energy phenomenology/anomalies in current data, e.g., 
in $(g-2)_\mu$~\cite{Greljo:2022dwn} or in low-energy QCD~\cite{Bause:2022jes}.
Beyond perturbative models, light vector particles
can also be in the spectrum of light resonances of low-energy, dark strongly
coupled sectors, see, e.g., Ref.~\cite{Brod:2014loa}.
To encompass all these cases, we employ in the current work the term 
``light dark vector'' (LDV), which is a massive vector boson with 
mass much below the EW scale, and sufficiently suppressed couplings to 
SM particles such that it is stable on collider scales. 
For the purpose of low-energy phenomenology we leave its UV
origin unspecified.

While constraints on light particles have been extensively studied in the
context of colliders, beam-dump experiments, astrophysics, and cosmology, their
phenomenology at precision flavor experiments has so far received less
attention (see Ref.~\cite{Badin:2010uh, Kamenik:2011vy} for early studies).
Even if flavor-violating couplings may be considered more model-dependent than
flavor-diagonal couplings, they can provide for an efficient production of
light invisible particles from decays of SM leptons, mesons or baryons.
Interestingly, direct searches at laboratory experiments for such two-body
decays with missing energy have the potential to probe enormously large scales,
as the relevant Lagrangian interactions can be dimension-five, instead of
dimension-six as in the case of heavy New Physics. For example, in models with
sufficiently light invisible bosons like the QCD axion, precision flavor
experiments are sensitive to scales as large as $10^{12} \GeV$ from $K \to \pi
+ {\rm invis.}$ searches at NA62~\cite{Goudzovski:2022vbt}, $10^{10} \GeV$
from $\mu \to e + {\rm invis.}$ searches at MEG-II~\cite{Calibbi:2020jvd,
Jho:2022snj}, Mu3e~\cite{Knapen:2023zgi}, Mu2e or COMET~\cite{Hill:2023dym},
and $10^{8} \GeV$ for $b \to d/s$ transitions at
Belle~II~\cite{MartinCamalich:2020dfe}.\footnote{For the flavor phenomenology of
the QCD axion and light invisible axion-like particles see
Refs.~\cite{Feng:1997tn, Kamenik:2011vy, Bjorkeroth:2018dzu,
MartinCamalich:2020dfe, Calibbi:2020jvd, Ziegler:2023aoe}.}

The aim of the current work is to systematically study the flavor phenomenology of
light dark vector particles (LDVs), both in the quark and the lepton sectors. We
restrict the discussion to invisible particles, since after all the main (only) motivation for
these particles is the observed DM abundance, and we have in mind scenarios
where either the LDV is itself stable on cosmological scales or promptly decays
to stable DM particles. This analysis includes scenarios where the
LDVs are just sufficiently long-lived to appear as missing energy.  This is
particularly justified for vector particles lighter than the electron, as their
decay into two photons is forbidden by the Landau--Yang
theorem~\cite{Landau:1948kw,Yang:1950rg}.
As we shall discuss, the resulting limits on flavor-violating
interactions can be as strong as in the axion case, which is not unexpected due
to the Goldstone-boson equivalence theorem. In light of past and ongoing
experimental searches, it is thus important to systematically study the 
phenomenological differences between light dark scalars and vectors 
originating from their distinct  helicity and coupling structure.

Earlier works have focused on the case of flavor-violating dipole couplings of
a massless dark photon in $\mu \to e$ and $s \to d$
transitions~\cite{Gabrielli:2016cut, Fabbrichesi:2017vma, Su:2019ipw,
Fabbrichesi:2020wbt, Su:2020xwt, Camalich:2020wac},
or considered general interactions and masses, but using only 
the available experimental limits on three-body decays to neutrinos
to study limits from $s \to d$ and $b \to s$ transitions~\cite{Kamenik:2011vy}.
Here instead we consider the case
of a light vector particle with generic mass and either dipole or minimal
couplings to SM fermions.  We work within the framework of a general
effective-field-theory (EFT) approach and consider all possible quark
flavor-violating transitions except those involving the top quark (where
constraints are very weak), and all possible lepton flavor-violating (LFV)
transitions.  We also discuss the decays of polarized leptons, which play an
important role in separating signal from SM background.  We derive bounds in
the general parameter plane of light-vector mass and the appropriate flavor-changing
coupling by comparing theoretical predictions for the decay rates to the
experimental bounds from various flavor factories, such as
NA62~\cite{NA62:2020pwi, NA62:2021zjw},
BaBar~\cite{BaBar:2004xlo,BaBar:2013npw}, CLEO~\cite{CLEO:2008ffk}, Belle
II~\cite{Belle-II:2022heu, Belle-II:2023esi}, BES~III~\cite{BESIII:2022vrr},
and TWIST~\cite{TWIST:2014ymv}. Whenever not available (as in the case of,
e.g., $B \to K/K^*/\pi + {\rm invis.}$ or $D \to \pi +{\rm invis.}$ decays), we
derive model-independent limits on the two-body decay rate as a function of the
invisible particle mass by recasting experimental data on the three-body decay
with two invisible neutrinos.  Finally, we also discuss the scenario where the light vector has only flavor-universal couplings to SM fermions in the UV, so that all quark flavor-changing effects in the IR are
induced radiatively by the Cabibbo--Kobayashi--Maskawa (CKM) matrix, satisfying the paradigm of Minimal Flavor-Violation (MFV)~\cite{DAmbrosio:2002vsn,
Isidori:2012ts}. 
For this analysis we derive the relevant
renormalization-group equations (RGEs) for both dipole and minimal couplings,
and use our results to convert limits on the flavor-changing interactions into
limits on flavor-diagonal couplings.

This work is organized as follows. In Section~\ref{sec:theoreticalsetup} we
define our basic setup by providing the effective Lagrangian for dipole and
minimal (vector) interactions of the LDV. The resulting phenomenology is
studied in the subsequent sections, separately for the quark
(Section~\ref{sec:quarkflavor}) and lepton (Section~\ref{sec:leptonflavor})
sectors, where we present our main results, the model-independent bounds on
generic flavor-violating LDV couplings as a functions of its mass. In
Section~\ref{sec:MFV} we use these constraints to derive bounds on flavor-universal UV couplings with either dipole or vector interactions from RG-induced flavor violation. We
conclude in Section~\ref{sec:conclusions}. Many technical details are deferred to
appendices: Appendix~\ref{sec:recast} contains the details and results of our
recast of two-body flavor-violating decays with missing energy for generic
masses of the invisible particles (extending the analysis for a massless
invisible particle in Ref.~\cite{MartinCamalich:2020dfe}).
Appendix~\ref{sec:LRbounds} contains the bounds on flavor-violating couplings
in the chiral $\rL/\rR$ basis (as opposed to the V/A basis in Section~\ref{sec:quarkflavor} 
and \ref{sec:leptonflavor}). The complete set of RGEs
relevant for Section~\ref{sec:MFV} is given in Appendix~\ref{sec:rges}, and
Appendix~\ref{sec:2bodydecays} contains the full expressions of two-body decay
rates of mesons, baryons, and polarized leptons, for a generic mass for the
light vector. We have also collected the hadronic matrix elements entering the
numerical analysis in Appendix~\ref{sec:formfactors}. Finally,
Appendix~\ref{sec:UV} contains a discussion of the EFT description of
flavor-violating vector couplings and their possible UV origin.

\section{Setup\label{sec:theoreticalsetup}}

We extend the SM by a new, neutral, massive vector boson $\V_{\mu}$ with a small mass
$m_{\V}$, which arises either by spontaneous symmetry breaking of, e.g., a $U(1)^\prime$ gauge 
symmetry or by the Stueckelberg mechanism~\cite{Ruegg:2003ps, Kors:2004dx,Kors:2005uz}.  Here we
focus on the case where this mass is much below the electroweak scale, and the
light dark vector (LDV) is either stable on collider scales or decays into
stable invisible particles.

The most general interactions of the LDV with the SM fermions can be
parametrized using an EFT approach, by considering the most general operators
that respect the unbroken part of the SM gauge group, $SU(3)_c\times U(1)_{\rm
em}$.  Here we focus on flavor-violating interactions written without loss of
generality in the fermion-mass basis. We can further assume that a possible
kinetic mixing between the photon and the LDV, i.e., $\propto \epsilon A^\mu
\V_\mu$, has been diagonalized such that $\V_\mu$ is also in
the mass-eigenstate basis. This diagonalization can be performed equally well
for a massless $\V_\mu$ (cf.~Ref.~\cite{Fabbrichesi:2017vma}), and the
difference with respect to the massive case is merely that for massless vectors
there remains an unphysical ambiguity in the choice of ``mass-eigenstate''
basis, due to the presence of an unbroken $SO(2)$ symmetry of the free
Lagrangian. Thus our setup applies equally well to the ``massless dark photon''
considered in Ref.~\cite{Fabbrichesi:2017vma} in the limit of $m_{\V} \to 0$.

Below the EW scale the lowest dimensional interactions of the LDV
are described by two classes of operators: dipole and vector interactions.
Firstly, we consider flavor-violating, dimension-five dipole interactions of
the form
\begin{equation}\label{eq:dipole}
\begin{aligned}
  \Lag_{\rD}&=
  -\frac{1}{4} \V_{\mu\nu}V^{\prime \mu\nu}+\frac{m_{\V}^2}{2} \V_\mu V^{\prime \mu}
  +\frac{1}{\Lambda} \V_{\mu \nu} \, \overline{f}_i \sigma^{\mu \nu} \left( \CC{\rD }{ij} + i \CC{\rD5}{ij} \gamma_5 \right) f_j \, ,
\end{aligned}
\end{equation}
where $\V_{\mu \nu} = \partial_\mu \V_{\nu} - \partial_\nu \V_{\mu}$ is
the LDV field strength, $\sigma^{\mu \nu} = \tfrac{i}{2} [\gamma^\nu, \gamma^\nu]$,
and $i \ne j$ denote SM quark or lepton flavors.
$\Lambda$ is the UV-completion scale of the associated
dipole couplings $\CC{\rD }{ij}$ and $\CC{\rD5}{ij}$, which
are hermitian matrices in flavor space,
$\big(\CC{\rD}{ij}\big)^* = \CC{\rD}{ji}$ and $\big(\CC{\rD5}{ij}\big)^* = \CC{\rD5}{ji}$.

Secondly, we consider flavor-violating couplings of the LDV to SM vector and
axial-vector currents.  Naively these are dimension-four interactions below the
EW scale.  However, such flavor-violating couplings violate $U(1)^\prime$ gauge
invariance (flavor-violating currents are not conserved), and thus must be
proportional to some power of the $U(1)^\prime$-breaking order parameter, which
we take as the vacuum expectation value (VEV) in the dark sector.  Therefore,
the flavor-violating vector couplings are actually dimension-five or higher,
depending on the underlying UV model. In perturbative UV completions the lowest
possible scaling is proportional to a single power of the dark VEV, which upon
including the dark gauge coupling becomes the LDV mass $m_{\V}$.
Normalizing by some UV scale $\Lambda$, the flavor-violating vector interactions are
\begin{equation}\label{eq:vector}
  \Lag_{\rV}=
 -\frac{1}{4} \V_{\mu\nu}V^{\prime \mu\nu}+\frac{m_{\V}^2}{2} \V_\mu V^{\prime \mu}
 + \frac{m_{\V}}{\Lambda} \V_\mu \, \overline{f}_i \gamma^\mu \left( \CC{\rV }{ij} + \CC{\rV5}{ij}  \gamma_5 \right)f_j\, ,
\end{equation}
where again $i \ne j$ denote SM quark or lepton flavors and the vector
couplings $\CC{\rV }{ij}$ and $\CC{\rV5}{ij}$ are hermitian matrices in flavor
space, $\big(\CC{\rV}{ij}\big)^* = \CC{\rV}{ji}$ and $\big(\CC{\rV5}{ij}\big)^* =
\CC{\rV5}{ji}$.

By choosing a scaling that is linear in $m_{\V}/\Lambda$, we ensure that the growth of
amplitudes with longitudinally polarized LDVs in initial and/or final states
$\propto E/m_{\V}$ as $m_{\V}\to 0$ is cancelled by the $m_{\V}$ dependence
in the interaction. This leads to finite amplitudes in the $m_{\V} \to 0$ limit
(see Refs.~\cite{Kamenik:2011vy, DiFranzo:2015nli, Zaazoua:2021xls, Baek:2021hnl, Arcadi:2023gox}
for related discussions),
which are just the amplitudes with the corresponding Goldstone bosons as
initial/final states.
An explicit example for a UV model that provides this linear scaling is
provided by Froggatt--Nielsen type models~\cite{Froggatt:1978nt}, discussed in
Appendix~\ref{sec:UV}. However, the linear scaling with $m_{V^{\prime}}$ is
only one possibility. For example, in UV models in which SM fermions do not
carry $U(1)^\prime$ charges the scaling can be quadratic in the dark VEV,  as
the coefficients involve additional powers of the $U(1)^\prime$ breaking scale
$v^\prime$, $\propto m_{\V}v^\prime/\Lambda^2$. An explicit realization of this
scenario is also discussed in Appendix~\ref{sec:UV}.

The interactions in Eq.~\eqref{eq:dipole} and ~\eqref{eq:vector} can also be
written in the chiral basis, which is more suited to match explicit UV models.
In this basis
\begin{equation}\label{eq:LRsetup}
\begin{aligned}
\Lag_{\rD}&=\frac{1}{\Lambda} \V_{\mu \nu} \, \overline{f}_i \sigma^{\mu \nu} \left( \CC{\rD \rL }{ij} P_{\rL} +  \CC{\rD \rR }{ij} P_{\rR}  \right) f_j \, ,\\
\Lag_{\rV}&= \frac{m_{\V}}{\Lambda} \V_\mu \, \bar{f}_i \gamma^\mu \left( \CC{\rV \rL }{ij} P_{\rL} + \CC{\rV \rR }{ij} P_{\rR} \right)f_j\, ,
\end{aligned}
\end{equation}
where $\CC{\rD \rL }{ij} = ( \CC{\rD \rR }{ji})^*$,  
$\CC{\rV \rL }{ij} = ( \CC{\rV \rL }{ji})^*$, $\CC{\rV \rR }{ij} = ( \CC{\rV \rR }{ji})^*$ 
and the relations between the ``$\rV/\rA$'' and the ``$\rL/\rR$'' bases are
\begin{align}\label{eq:LRbasis}
\CC{\rD}{ij}&=\frac{1}{2}\left(\CC{\rD \rL }{ij}+\CC{\rD \rR }{ij}\right) = \frac{1}{2}\left(\big(\CC{\rD \rR }{ji}\big)^*+\CC{\rD \rR }{ij}\right) \,,&
\CC{\rV}{ij}&=\frac{1}{2}\left(\CC{\rV \rL }{ij} +\CC{\rV \rR }{ij}\right) \,,\\
\CC{\rD5}{ij}&=\frac{i}{2}\left(\CC{\rD \rL }{ij}- \CC{\rD \rR }{ij}\right) = \frac{i}{2}\left(\big(\CC{\rD \rR }{ji}\big)^*- \CC{\rD \rR }{ij}\right)\,,&
\CC{\rV5}{ij}&=\frac{1}{2}\left(\CC{\rV \rR }{ij}-\CC{\rV \rL }{ij}\right)  \, .
\end{align}

Above the EW scale the operators must be expressed in a manifestly
$SU(2)_{\rL}\times U(1)_Y$ invariant manner. For $\Lag_{\rV}$ this is directly
the case after embedding the left- and right-handed fermions in the
corresponding $SU(2)_{\rL}$ doublets and singlets, respectively. Instead, the
dipole operators in $\Lag_{\rD}$ require an additional Higgs insertion, 
making them dimension-six operators
\begin{align}
\label{eq:EW}
\Lag_{\rD6}&=\frac{1}{\Lambda_6^2} \V_{\mu \nu} \left(\overline{F}_i H C^{\rD}_{ij} \sigma^{\mu \nu} P_{\rR}  f_j +  \text{h.c.}\right) \, ,
\end{align}
with $F_i$ and $f_j$ denoting here $SU(2)_{\rL}$ doublets and singlets,
respectively, and $H \to \widetilde{H}$, depending on the fermion sector and
the hypercharge conventions.  The matching to $\Lag_{\rD}$ is provided by
identifying $\Lambda_6 = \sqrt{v \Lambda}$, where $v = 174\GeV$ is the Higgs
VEV.

In the following we derive bounds on the flavor-violating couplings in
Eq.~\eqref{eq:dipole} and \eqref{eq:vector} from hadronic and leptonic decays
with missing energy in the final state. This discussion is unaffected by other
possible interactions of the LDV with SM fields, in particular flavor-diagonal
couplings, as long as these couplings are sufficiently small to ensure that the
LDV is invisible on collider scales.

For massive LDVs neither the flavor-violating dipole (Eq.~\eqref{eq:dipole})
nor the vector (Eq.~\eqref{eq:vector}) interactions are UV complete. The UV
completion depends on the origin of the mass for the LDV and the corresponding
(highly model-dependent) radial mode required for the unitarity of the theory.
In turn this implies that unless the complete dark Higgs sector of the
theory is specified, there exist perturbative unitarity constraints on the
couplings of the LDV, similar to the unitarity constraints from $WW\to WW$
scattering in the Higgs-less SM.  We briefly note that, as long as the flavor
bounds are applicable, i.e., LDV masses in the kinematically allowed
region, unitarity of $2\to 2$ scattering poses constraints on the
corresponding couplings that are weaker than those limits by order of
magnitudes. We thus refrain from elaborating upon these constraints in the
current work. For the case of unitarity bounds on massless fermions with
flavor-diagonal couplings coupled to transversely polarized vectors see,
e.g., Ref.~\cite{Barducci:2023lqx}. 
The more general case including massive fermions with
flavor-violating couplings to LDVs will be presented in
Ref.~\cite{ESTZ:2024}.

\section{Quark Phenomenology of Light Dark Vectors\label{sec:quarkflavor}}

\begin{figure}[t]
\centering
\includegraphics{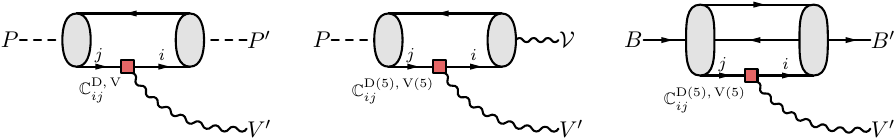}
\caption{Illustrative Feynman diagrams
  with a flavor-violating $q_j\to q_i$ transition in 
  two-body decays of type
  $P\to P'+\V$,  $P\to \mathcal{V}+\V$, and $B\to B'+\V$, in the left, middle, and right panel, respectively.
  \label{fig:twobodydecays}}
\end{figure}

In this section we derive bounds on the flavor-violating couplings 
$\CC{\rD(5)}{ij}$ in Eq.~\eqref{eq:dipole} and $\CC{\rV(5)}{ij}$ in Eq.~\eqref{eq:vector} 
for the quark-flavor transitions: $s \to d$, $b \to s$, $b \to d$, and $c \to u$.
We employ the following three types of two-body decays containing
the LDV as an invisible final state\footnote{Three-body decays
and neutral meson mixing typically give weaker constraints, e.g., for example
LHCb constraints on $B_{(s)} \to \mu \mu a$ cannot compete with Belle~II
limits~\cite{Albrecht:2019zul}.}
\begin{itemize}
  \item$P\to P^{\prime}+\V$: pseudoscalar meson to pseudoscalar meson and LDV,
  \item$P\to \mathcal{V}+\V\,\,$: pseudoscalar meson to vector meson and LDV,
  \item$B\to B^{\prime}+\V$: baryon to baryon and LDV. 
\end{itemize}
Figure~\ref{fig:twobodydecays} shows representative Feynman diagrams for the
three types of decays.

Appendix~\ref{sec:2bodydecays} contains the analytical expressions for the 
corresponding decays rates (including the dependence on $m_\V$); 
the relevant form factors are collected in Appendix~\ref{sec:formfactors}. 
Comparing the decay rates to the experimental
upper limits on the branching ratios, we set upper bounds on the couplings in
the $\rV/\rA$ basis\footnote{
In Appendix~\ref{sec:LRbounds} we show the bounds in the $\rL/\rR$ basis.}
of Eq.~\eqref{eq:dipole} and Eq.~\eqref{eq:vector}, i.e. on
the set $\{\CC{\rD }{ij}$, $\CC{\rD5}{ij}$, $\CC{\rV }{ij}$, $\CC{\rV5}{ij}\}$. The
limits are determined as a function of the LDV mass, with range $0\leq
m_{\V}^2\leq (m_{\rm I}-m_{\rm F})^2\equiv m_{\V\text{max}}^2$ depending on the
masses of the initial, $m_{\rm I}$, and final, $m_{\rm F}$, states of the decay at hand.
Crucially, the form factors depend on the LDV mass and it is, therefore,
essential to consider the full form-factor parametrization for an accurate
analysis.

The available theoretical and experimental information is summarized in
Table~\ref{tab:experimentalinput}, where we collect the references for the form
factors and relevant experimental limits. 
Often the experimental collaborations do not provide
limits on two-body decays with missing energy.
Yet, in some cases there is enough information to extract this
bound from available data.
We indicate this case by a subindex ``$r$'' in the last column of the table, 
and either use existing recasts in the literature or perform our own recast, e.g.,
to find a bound on $B\to \pi / K /K^* +{\rm invis.}$ from
BaBar data on the corresponding three-body decays~\cite{BaBar:2004xlo,BaBar:2013npw}, 
see Appendix~\ref{sec:recast} for details.

Concretely we use our recast for $B \to K^{(*)} +{\rm invis.}$ only for LDV masses above $3\GeV$. 
Note that we can recast only the experimental results of the BaBar collaboration and cannot use
the newer Belle measurements, since the Belle collaboration does not provide
the event count as a (binned) function of the missing-momentum distribution. 
We use existing recasts for $B \to \rho +{\rm invis.}$ decays from
LEP~\cite{ALEPH:2000vvi, Alonso-Alvarez:2023mgc}, $B\to K +{\rm invis.}$ decays from
Belle~II~\cite{Altmannshofer:2023hkn, Belle-II:2023esi} (this recast is limited
to masses below $m_{\V}= 3\GeV$), $B\to K^* +{\rm invis.}$ decays from
BaBar~\cite{Altmannshofer:2023hkn, BaBar:2013npw} (below $m_{\V}= 3 \GeV$).
For invisible baryon decays for which there is no analysis, we derive
limits using the total lifetime from the PDG~\cite{ParticleDataGroup:2022pth} after 
subtracting all observed channels as in Ref.~\cite{MartinCamalich:2020dfe}.

For the bound based on $D\to\pi +{\rm invis.}$ decays we use the result of 
Ref.~\cite{MartinCamalich:2020dfe} for $m_{\V} \approx 0$, obtained 
from recasting CLEO data on $D \to (\tau \to \pi \nu) \nu$~\cite{CLEO:2008ffk}.  
We also perform a recast of these data
for LDV masses up to $m_{\V}\approx 0.5\GeV$ (which is the upper range of the
CLEO data set), assuming the efficiency in all bins to be the same as for
$m_{\V}\approx 0$. Note that recasting BES~III data~\cite{BESIII:2021slf} on 
$D \to \pi \nu \overline{\nu}$ gives weaker
constraints~\cite{MartinCamalich:2020dfe}, although this result does not use
the full experimental information. It would be interesting if BES~III would
provide an explicit two-body recast of their full data set.
The collaboration actually
does this for the case of two-body hyperon decays $\Lambda_c \to p + {\rm
invis.}$, albeit only for ``massless'' invisible particles.  Their signal region
in fact covers invisible masses up to $316\,\MeV$, and leads to limits that are
much stronger than the ones obtained by saturating the total $\Lambda_c$
lifetime~\cite{MartinCamalich:2020dfe}.
As a conservative limit, to be replaced by a dedicated experimental
analysis, we multiply their limit for the massless case by a factor 1/2
(since close to the endpoint of the signal region half of the signal events are lost
due to energy resolution).
We use the resulting bound ${\rm BR} (\Lambda_c \to
p V^\prime)  < 1.6 \times 10^{-4}$ for LDV masses up to $316\,\MeV$, and take
lifetime limits above $316\,\MeV$.
We notice that a search for the decay $D \to \pi +X$ would not suffer
from two-body SM backgrounds in contrast to hyperon decays, 
where $\Lambda_c \to p + \gamma$ contributes to the signal of a massless $X$, 
if the photon is missed.

\begin{table}[t]
\centering
\begin{tabular}{>{\centering\arraybackslash}p{3cm}p{3.4cm}>{\centering\arraybackslash}p{2.8cm}p{4.5cm}}
\toprule
\multicolumn{1}{c}{\bfseries Quark Transition}& {\bfseries Hadronic Process} & {\bfseries Form Factors} & {\bfseries Experimental Limit}\\
\midrule
\multirow{6}{*}{$s \to d$} 
&  $K^{+}\to \pi^{+}+\V$ &    \cite{Baum:2011rm, Lubicz:2009ht} &NA62~\cite{NA62:2021zjw, NA62:2020pwi, Goudzovski:2022vbt}\\
& $\Sigma^{+}\to p+\V$ &   \cite{Camalich:2020wac, Ledwig:2014rfa, Cabibbo:2003cu, Gaillard:1984ny} &BES~III~\cite{BESIII:2023utd}, Lifetime$_r$\cite{ParticleDataGroup:2022pth, MartinCamalich:2020dfe} \\
& $\Xi^{-}\to \Sigma^{-}+\V$ &     \cite{Camalich:2020wac, Ledwig:2014rfa, Cabibbo:2003cu, Gaillard:1984ny}&Lifetime$_r$\cite{ParticleDataGroup:2022pth, MartinCamalich:2020dfe}  \\
&  $\Xi^{0}\to \Sigma^{0}+\V$ &    \cite{Camalich:2020wac, Ledwig:2014rfa, Cabibbo:2003cu, Gaillard:1984ny}&Lifetime$_r$\cite{ParticleDataGroup:2022pth, MartinCamalich:2020dfe}  \\
&  $\Xi^{0}\to \Lambda+\V$ &       \cite{Camalich:2020wac, Ledwig:2014rfa, Cabibbo:2003cu, Gaillard:1984ny}&Lifetime$_r$\cite{ParticleDataGroup:2022pth, MartinCamalich:2020dfe}   \\
&  $\Lambda\to n+\V$ &    \cite{Camalich:2020wac, Ledwig:2014rfa, Cabibbo:2003cu, Gaillard:1984ny} &Lifetime$_r$\cite{ParticleDataGroup:2022pth, MartinCamalich:2020dfe}  \\
\midrule
&  $B^{+}\to K^{+}+\V$& \cite{Gubernari:2018wyi, Gubernari:2018wyi}&BaBar$_r$~\cite{BaBar:2013npw}, Belle~II$_r$~\cite{Altmannshofer:2023hkn, Belle-II:2023esi}  \\
$b \to s$  
&  $B\to K^{\ast}+\V$& \cite{Gubernari:2018wyi, Gubernari:2018wyi}&BaBar$_r$~\cite{BaBar:2013npw, Altmannshofer:2023hkn} \\
&  $\Lambda_b\to \Lambda+\V$& \cite{Detmold:2016pkz, Detmold:2016pkz}&Lifetime$_r$\cite{ParticleDataGroup:2022pth, MartinCamalich:2020dfe}\\
\midrule
&  $B^{+}\to \pi^{+}+\V$& \cite{Leljak:2021vte, Gubernari:2018wyi}& BaBar$_r$~\cite{BaBar:2004xlo}  \\
$b \to d$
&  $B\to \rho+\V$& \cite{Gubernari:2018wyi, Gubernari:2018wyi}& LEP$_r$~\cite{ALEPH:2000vvi, Alonso-Alvarez:2023mgc}  \\
&  $\Lambda_b\to n+\V$& \cite{Detmold:2016pkz, Detmold:2015aaa}&Lifetime$_r$~\cite{ParticleDataGroup:2022pth, MartinCamalich:2020dfe}\\
\midrule
\multirow{2}{*}{$c \to u$} 
&  $D^{+}\to\pi^{+}+\V$& \cite{Lubicz:2018rfs, Lubicz:2017syv}&CLEO$_r$~\cite{CLEO:2008ffk, MartinCamalich:2020dfe}\\ 
&  $\Lambda_{c}\to p+\V$& \cite{Meinel:2017ggx, Meinel:2017ggx}&  BES~III~\cite{BESIII:2022vrr}, Lifetime$_r$~\cite{ParticleDataGroup:2022pth, MartinCamalich:2020dfe} \\
\bottomrule
\end{tabular}
\caption{
Overview of considered hadron decays with invisibles in the final state.
 The first column shows the underlying quark-flavor transition, 
 the second the specific hadronic process. The
 relevant vector and dipole form factors are taken from the references in the
 third column. The last column contains the references for the
 experimental upper limits on the respective branching ratios. 
 A subindex ``$r$'' indicates that a recast of experimental data was needed, see text and
 Appendix~\ref{sec:recast} for details.\label{tab:experimentalinput}
}
\end{table}

To set constraints on the couplings $\{\CC{\rD }{ij}$, $\CC{\rD5}{ij}$,
$\CC{\rV }{ij}$, $\CC{\rV5}{ij}\}$ we consider dipole ($\Lag_{\rD}$) and vector
interactions ($ \Lag_{\rV}$) separately, and turn on a single coupling at a
time. We use the theory predictions in Appendix~\ref{sec:2bodydecays}
together with the form factors in Table~\ref{tab:experimentalinput}
(see also Appendix~\ref{sec:formfactors})
to calculate the decay rates as a function of the couplings and the LDV mass.
The rates are then compared to the experimental limits to obtain the bounds
in the mass--coupling plane. 
We include statistical and systematic uncertainties
as follows. For the theory predictions we only use the systematic uncertainties
associated with hadronic form factors (these are the most relevant ones), while
the treatment of uncertainties of experimental limits depend on their nature:
for decays where the experimental collaborations provide two-body
interpretations (or a theory recast exists), we add the experimental and 
form-factor uncertainties in quadrature. In the case where we performed our own 
two-body recast (as described in Appendix~\ref{sec:recast}) we treat theory
uncertainties as Gaussian uncertainties smearing the expectation values of the
underlying Poisson probability distribution functions.

Our results are summarized in Figures~\ref{fig:quarkdipoleVA} and
\ref{fig:quarkvectorVA} in which we show the lower bounds on the effective inverse
coupling $\Lambda/\CC{}{ij}$ for given LDV mass $m_{\V}$. The plots are
organized according to the underlying flavor transition, i.e, $s \to d$,
$b \to s$, $b \to d$, and $c \to u$ and we separate dipole 
$\{\CC{\rD }{ij}$, $\CC{\rD5}{ij}\}$ (Figure~\ref{fig:quarkdipoleVA}) and 
vector couplings $\{\CC{\rV }{ij}$, $\CC{\rV5}{ij}\}$ (Figure~\ref{fig:quarkvectorVA}). 
Each plot shows the bound on
a single coupling for a given quark-flavor transition, with each line
corresponding to a particular hadronic decay, excluding the region below.
Note that $P\to P^{\prime}+\V$ decays are only sensitive to $\{\CC{\rD }{ij}$ and 
$\CC{\rV }{ij}\}$ couplings, which follows from parity conservation of the strong
interactions and the Lorentz structure of the form factors 
(see Appendix~\ref{sec:formfactors}). Also note that dipole
operators are dimension-six above the electroweak scale, so in fact the actual
UV scale probed is $\Lambda_6 = \sqrt{v \Lambda}$ in all transitions.

\subsection{Dark Dipole Interactions\label{sec:dipolesquark}}

\begin{figure}[H]
  \begin{center}
   \includegraphics[width=0.47\textwidth]{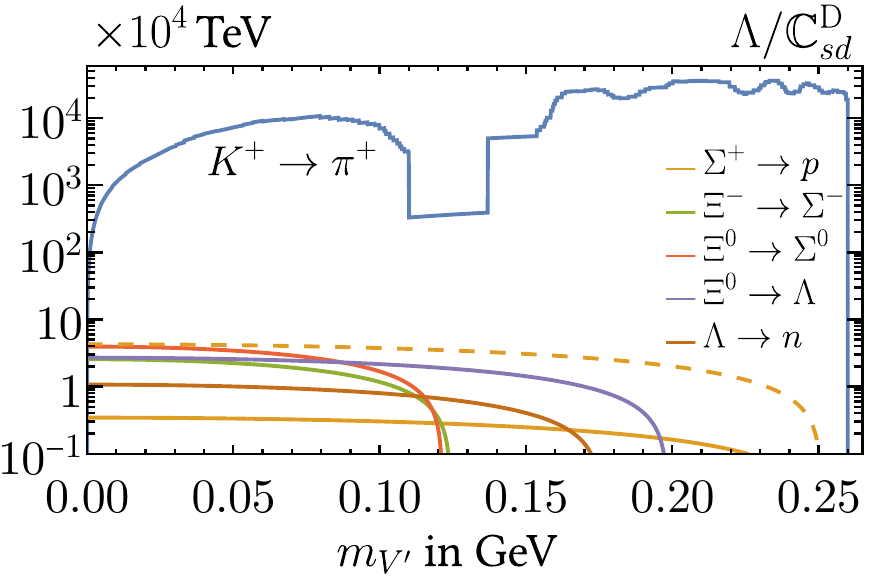}
   \quad
   \includegraphics[width=0.47\textwidth]{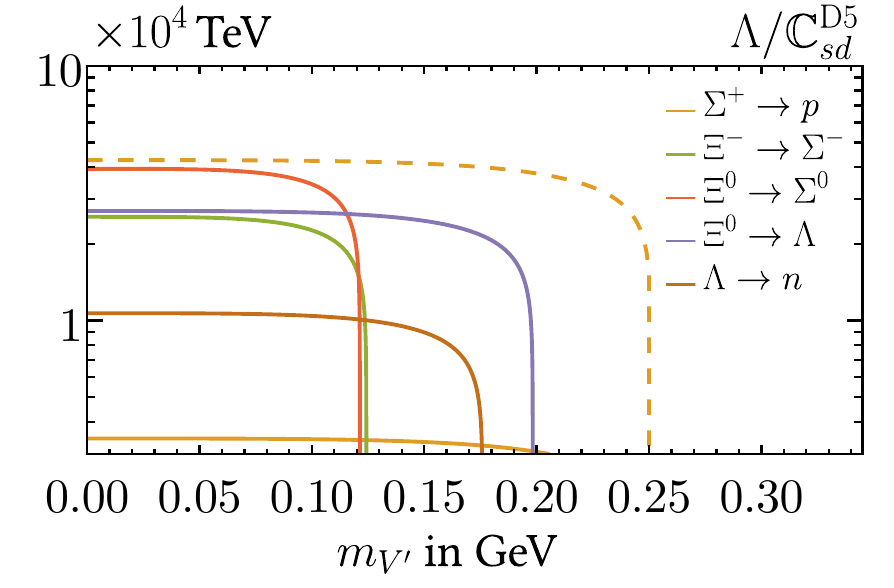}
   \\[0.5em]
   \includegraphics[width=0.47\textwidth]{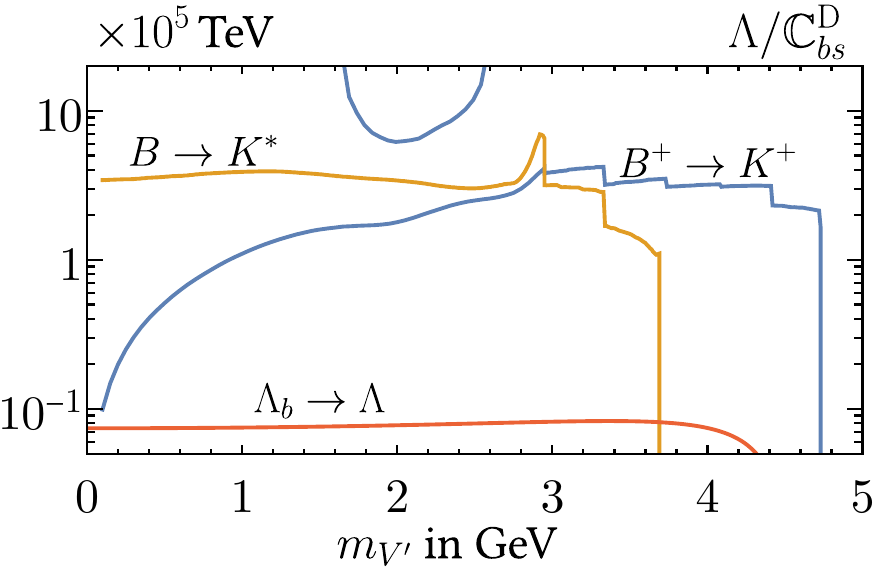}
   \quad
   \includegraphics[width=0.47\textwidth]{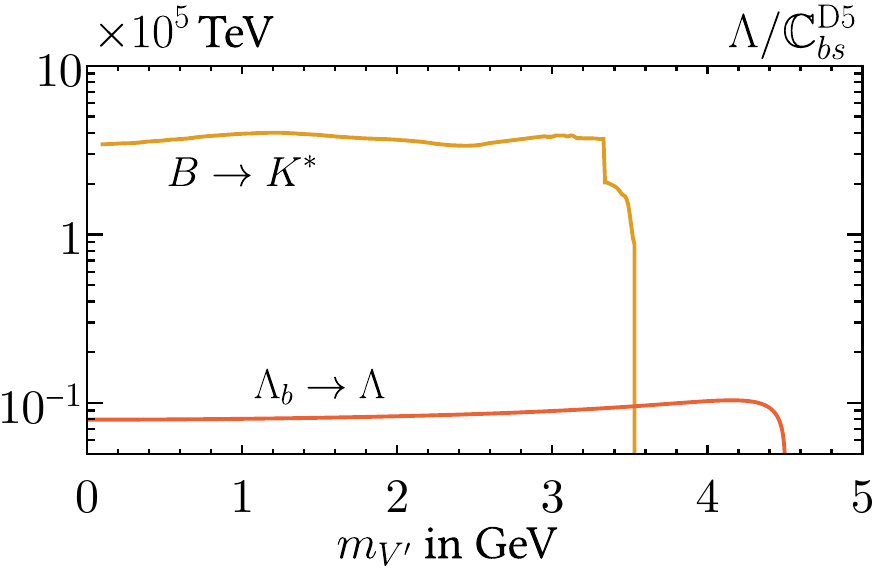}
   \\[0.5em]
   \includegraphics[width=0.47\textwidth]{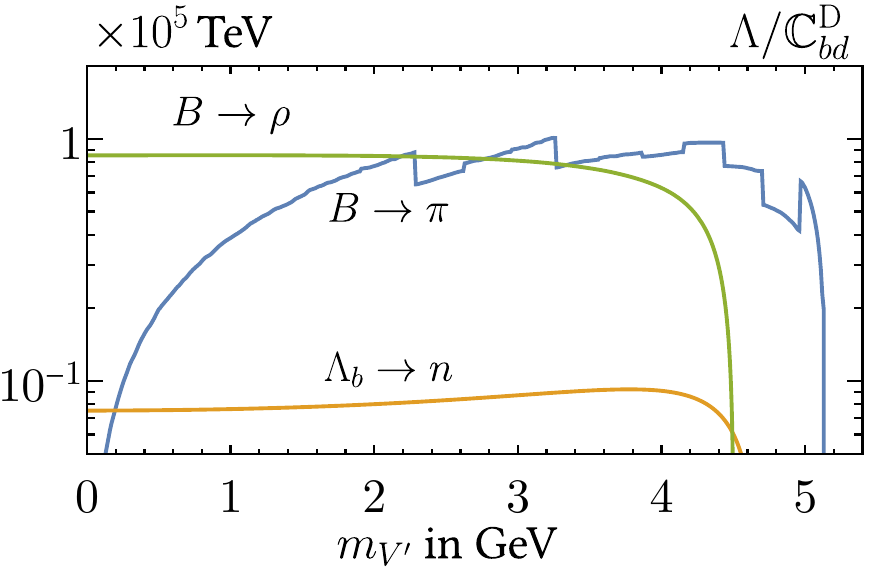}
   \quad
   \includegraphics[width=0.47\textwidth]{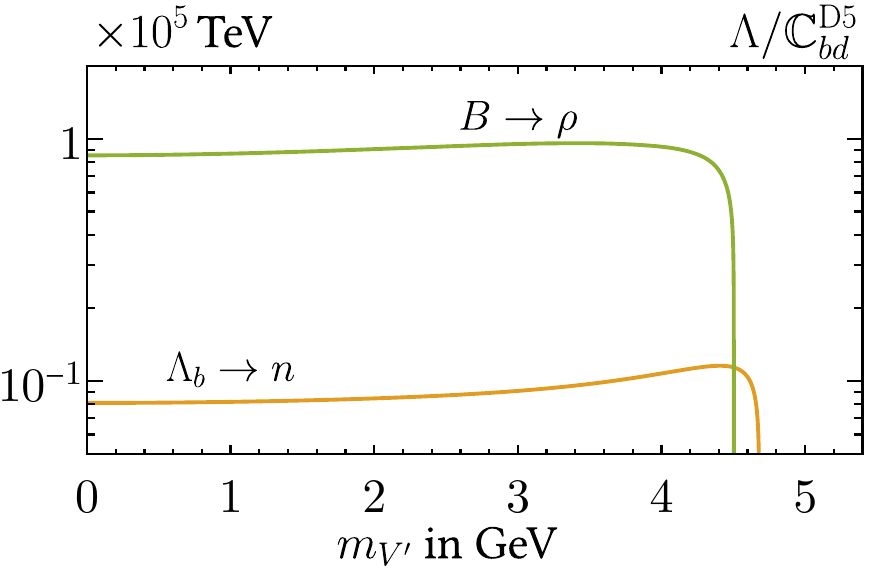}
   \\[0.5em]
   \includegraphics[width=0.47\textwidth]{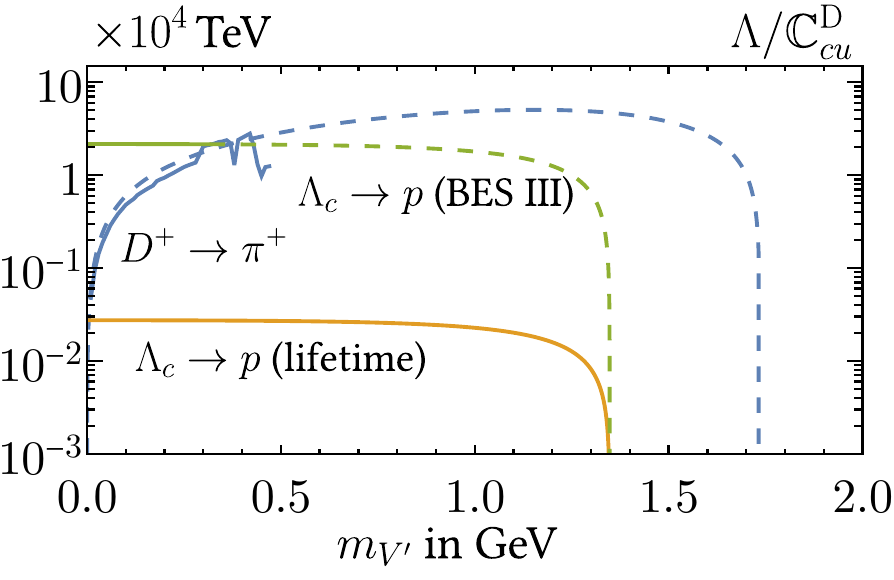}
   \quad
   \includegraphics[width=0.47\textwidth]{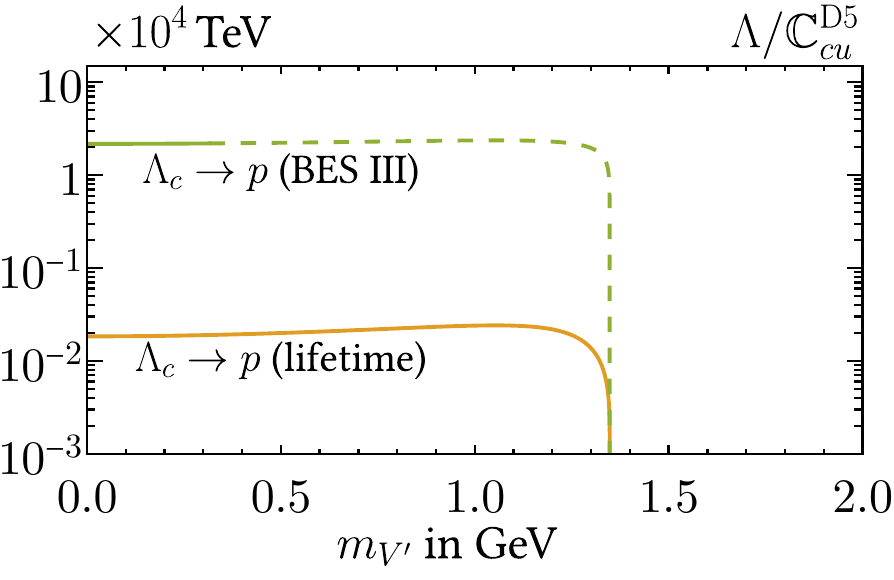}
  \end{center}
  \vspace{-1em}
  \caption{
    Lower limits on quark-flavor violating dipole couplings $\Lambda/|\CC{\rD}{ij}|$ (left  column) and $\Lambda/|\CC{\rD5}{ij}|$ (right  column)
   of the LDV for  $s \to d, b \to s, b \to d, c \to u$ transitions @$95\%$ CL$_{(\text{s})}$. See text for details.
    \label{fig:quarkdipoleVA} }
\end{figure}

\paragraph{$\boldsymbol {s \to d}$ Transitions} The bounds on the dipole couplings
$\{\CC{\rD }{sd}$, $\CC{\rD5}{sd}\}$ are set by $K\to \pi +{\rm invis.} $ and
hyperon decays, cf.~Table~\ref{tab:experimentalinput} and
Figure~\ref{fig:quarkdipoleVA}. For the two-body decay $K\to \pi
+{\rm invis.} $ we use the bound provided by the NA62
collaboration~\cite{NA62:2021zjw}.
For baryon decays there is an upper limit from BES~III~\cite{BESIII:2023utd}
  on the decay $\Sigma^+\to p + {\rm invis.}$
  with a massless invisible. We estimate the potential reach for this search by
  extending it to larger invisible masses by assuming that the same experimental limit is
  valid for the whole kinematic range. This is indicated by a dashed orange line.
  For all other baryon searches, we set upper limits on branching ratios indirectly as in
Ref.~\cite{MartinCamalich:2020dfe} by subtracting the measured branching
fractions for all relevant hyperon decay channels from unity. Due to this
rather weak limit, $K \to\pi$ sets a much more stringent constraint than
hyperon decays, limiting the UV scale $\Lambda/\CC{\rD }{ij}$ to be at least 
of the order $10^{11} \GeV$.
Note however  that the search for $\Sigma^+\to p + {\rm invis.}$
  strenghtens the upper limit by two orders of magnitude compared to the conservative
  limit estimated with the total lifetime, and thus, out of all baryon decays,
  it yields the strongest limit of order $10^7\GeV$ on the scale $\Lambda/\CC{\rD }{ij}$.

Nevertheless baryon decays with missing energy are important for two reasons.
The decays to pseudoscalar, such as $K \to\pi$, are only sensitive to the $\{\CC{\rD
}{ij}$, $\CC{\rV }{ij}\}$ couplings. Thus baryon decays are crucial
to constrain the axial coupling $\Lambda/\CC{\rD5}{ij}$ (of the order of a few
$\times 10^{7} \GeV$), as there are no two-body decays to vector particles in
$s \to d$ transitions. Moreover, the decay rates of pseudoscalar processes are
proportional to the LDV mass for the dipole interaction $\Lag_{\rD}$
(cf.~Eq.~\eqref{eq:ratePPprime}), and thus only baryon decays can constrain
$\CC{\rD }{ij}$ for small LDV masses. This can be see in
Figure~\ref{fig:quarkdipoleVA} (upper left panel), where the bounds on $\CC{\rD
}{sd}$ from hyperon decays dominate for LDV masses of $m_{\V}\approx 0$
yielding a limit of $\mathcal{O}(10^7\GeV)$ on the axial coupling
$\Lambda/\CC{\rD5}{ij}$.
This provides a strong motivation for explicit direct searches 
targeting baryon decays with invisible final states.

\paragraph{$\boldsymbol {b \to s}$ Transitions} The limits on the dipole couplings
$\{ \CC{\rD }{bs}, \CC{\rD5}{bs} \}$ are set by $B$-meson decays $B\to K/ K^*
+{\rm invis.} $ and baryon decays $\Lambda_b\to \Lambda +{\rm invis} $. 
The limits from the $B$-meson decays are obtained from our own recast of BaBar data 
(cf.~Appendix~\ref{sec:recast}), except for $B^+\rightarrow K^+ + {\rm invis.}$ for LDV
masses $m_{\V} < 3\GeV$ where we use  the recast in
Ref.~\cite{Altmannshofer:2023hkn} of the recent Belle~II measurement 
of $B^+ \to K^+ \overline{\nu} \nu$ ~\cite{Belle-II:2023esi}.
We also use the recast in Ref.~\cite{Altmannshofer:2023hkn} of the BaBar measurement of $B \to K^* \overline{\nu} \nu$ ~\cite{BaBar:2013npw} below LDV masses of 3 GeV.
The limit on unobserved
$\Lambda_b$ decays such as $\Lambda_b\to \Lambda + {\rm invis.}$ is obtained by
comparing the SM prediction for the total lifetime with the experimental one
inferred from all observed channels, ascribing the difference to the allowed
value for the two-body invisible decay~\cite{MartinCamalich:2020dfe}. 
As for $s \to d$ transitions, decays to pseudoscalar mesons such as $B^+ \to K^+ $ can
neither constrain the axial coupling $\CC{\rD5}{bs}$, nor $\CC{\rD }{bs}$ for
very small LDV masses. 
Otherwise, however, they do dominate over the constraint from $\Lambda_b\to \Lambda$.

In contrast to $s \to d$ transitions, there is also a
decay with vector mesons in the final-state, $B\to K^*$, which constrains 
both the $\CC{\rD}{bs}$ and the $\CC{\rD5}{bs}$ couplings in the entire LDV mass range, 
if kinematically allowed. 
Hence, $B\to K^*$ decays are complementary to $B\to K$
decays in constraining $\Lambda/\CC{\rD}{bs}$, setting limits on the UV scale
of the order $10^{8}\GeV$, and also dominate the bounds on
$\Lambda/\CC{\rD5}{bs}$ of similar size, up to a small region where this
channel is kinematically closed and $\Lambda_b\to \Lambda$ decays set the
strongest limit, of the order $10^{7} \GeV$. Note that there is an {\it upper}
limit of order $10^{8}\GeV$ on $\Lambda/\CC{\rD }{bs}$ at around $m_{\V}\approx
2 \GeV$ coming from $B\to K + \V$ decays~\cite{Altmannshofer:2023hkn}, due to a
$2.8\,\sigma$ excess in the latest Belle~II measurement of $B^+ \to K^+ \nu
\overline{\nu}$~\cite{Belle-II:2023esi}.

\paragraph{$\boldsymbol {b \to d}$ Transitions} The bounds on the dipole couplings
$\{ \CC{\rD }{bd}, \CC{\rD5}{bd} \}$ are obtained from $B$-meson decays $B\to
\pi/\rho +{\rm invis.} $ and baryon decays $\Lambda_b\to n +{\rm invis.}$ The
limit on $B\to \pi$ decays is obtained from our recast of BaBar data 
(cf.~Appendix~\ref{sec:recast}), while a limit on $B\to \rho$ decays from LEP data~\cite{ALEPH:2000vvi}
has been derived in Ref.~\cite{Alonso-Alvarez:2023mgc}.
Analogously to $b \to s$ transitions, the pseudoscalar decay $B\to \pi$ does 
neither constrain the axial coupling $\CC{\rD5}{bd}$ nor $\CC{\rD }{bd}$ for small 
LDV masses, while the decay to vector mesons $B\to \rho$ does.
Thus the two meson decays are complementary in setting limits on 
$\Lambda/\CC{\rD }{bd}$, of the order of $10^{8}\GeV$, while
$B\to \rho$ dominates the bounds on the limits on $\Lambda/\CC{\rD5}{bd}$ of similar
size, except for LDV masses above the kinematic threshold where $\Lambda_b\to
n$ decays take over, constraining UV scales up to $10^{7}\GeV$.

\paragraph{$\boldsymbol {c \to u}$ Transitions} Finally, the constraints on the
dipole couplings $\{ \CC{\rD }{cu}, \CC{\rD5}{cu} \}$ are set by $D\to\pi +{\rm
invis.}$ and the baryonic process $\Lambda_c\to p +{\rm invis}$. For $D \to
\pi$ and LDV masses $m_{\V}\lesssim 0.5\GeV$, we performed a recast of the CLEO
data set (analogous to the $B$-decay recasts in Appendix~\ref{sec:recast}).
The result is shown as a solid, blue line in the bottom panel of
Fig.~\ref{fig:quarkdipoleVA}.  CLEO has only collected data up to masses of
$m_{\V}\approx 0.5\GeV$, but we also show the potential bound that could be
obtained above this mass by extrapolating the bound for massless invisible
particles~\cite{MartinCamalich:2020dfe} to the whole kinematic range, which we
indicate by a dashed blue line.

For $\Lambda_c \to p$ we show two limits in the bottom panel of
Fig.~\ref{fig:quarkdipoleVA}: solid, orange lines denote the bound obtained from
simply saturating the total $\Lambda_c$ lifetime, i.e., ${\rm BR}(\Lambda_c \to
p + \V) < 1$, while the green line indicates the 95\% CL bound obtained from the
BES~III~\cite{BESIII:2022vrr} result for ``massless'' invisible particles, ${\rm
BR}(\Lambda_c \to p + V) < 8.0 \times 10^{-5}$ at 90\% CL, which in fact covers
invisible masses up to $316\MeV$ and are multiplied by a factor 1/2, see the discussion in the beginning of this section.
We estimate the potential reach for a search
extending to larger invisible masses by assuming that the same experimental
limit below $316\MeV$ is also valid above, and indicate this extrapolation by a
dashed, green line. We observe that the strongest limits on
$\Lambda/\CC{\rD}{cu}$ are set by the BES~III search for a ``massless'' LDV in
$\Lambda_c\to p$ decays, which are valid for $m_{\V} \lesssim 316\MeV$ and are
of the order of $10^7\GeV$. Between $316\MeV \lesssim m_{\V} \lesssim 500 \MeV$
a limit of similar size is obtained from $D\to\pi$ decays, recasting CLEO data
on $D \to (\tau \to \pi \nu) \nu$. The only available limit on LDV masses above
$0.5 \GeV$ arises from the total $\Lambda_c$ lifetime, which sets limits of
order $10^5\GeV$. Naively extrapolating the limits from CLEO on $D\to\pi$ and
BES~III on $\Lambda_c\to p$ decays to higher LDV masses instead suggests that
present bounds could be strengthened by two orders of magnitude, if BES~III
would either analyze the available searches for $\Lambda_c\to p$ decays with
extended signal regions, or use available data on $D\to\pi \nu \overline{\nu}$
to set a limit on the two-body decay.

Currently only $\Lambda_c\to p$ decays are capable to set constraints on the
axial coupling $\Lambda/\CC{\rD5}{cu}$, of the order of $10^{5} \GeV$ and $10^7 \GeV$ for LDV masses above and below $316\MeV$, respectively.
Besides extending the search for
$\Lambda_c \to p + \V$ to higher LDV masses, this also motivates dedicated
searches for other processes such as $D \to \rho +{\rm invis.}$ or $D_s \to K^*
+ {\rm invis.}$ at current or future experiments.

\subsection{Dark Vector Interactions\label{sec:vectorquark}}

\begin{figure}[H]
  \begin{center}
   \includegraphics[width=0.47\textwidth]{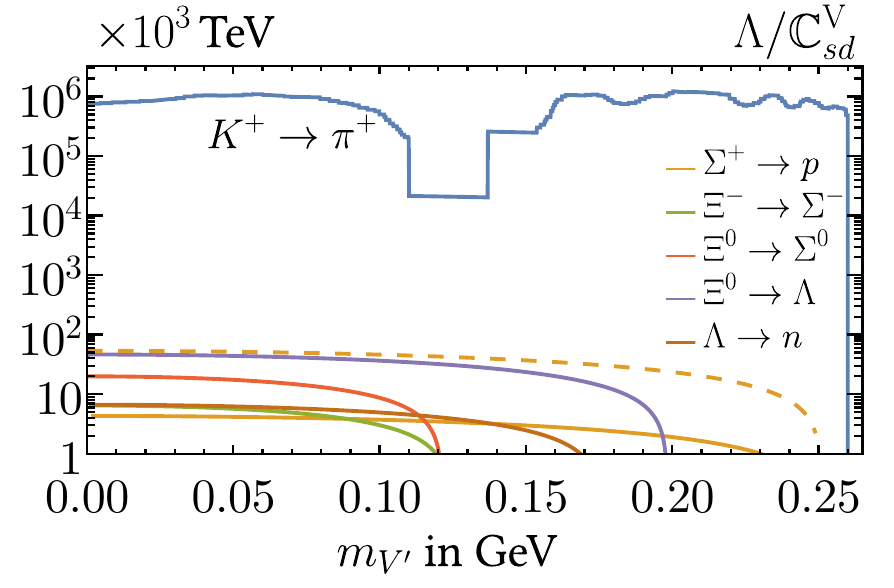}
   \quad
   \includegraphics[width=0.47\textwidth]{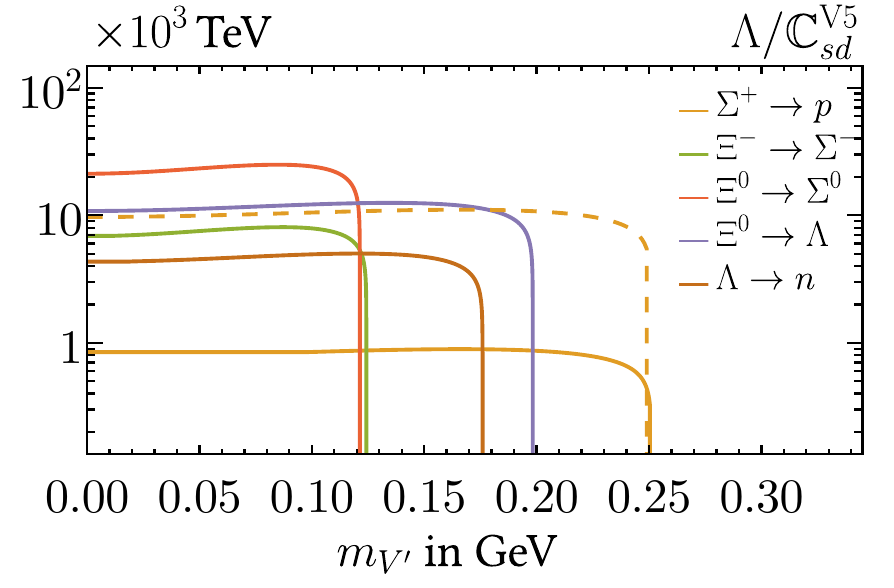}
   \\[0.5em]
   \includegraphics[width=0.47\textwidth]{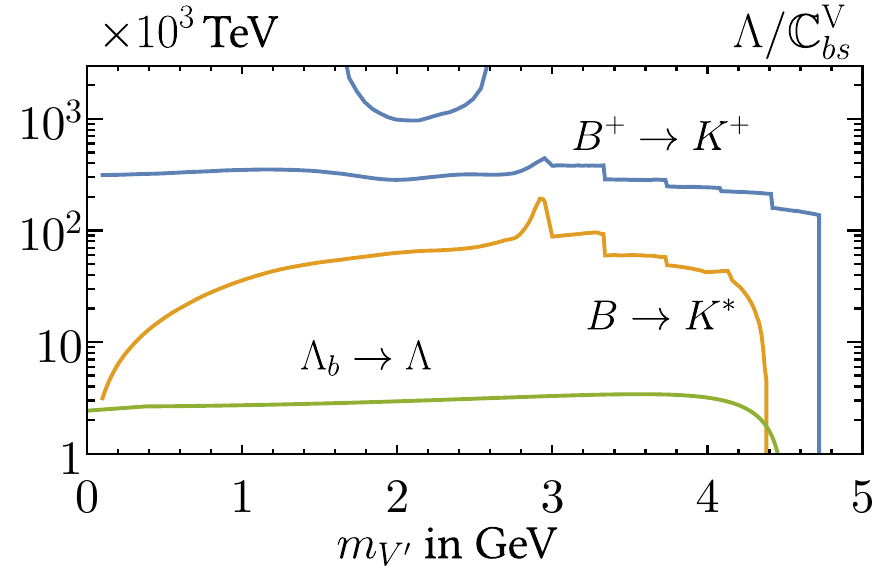}
   \quad
   \includegraphics[width=0.47\textwidth]{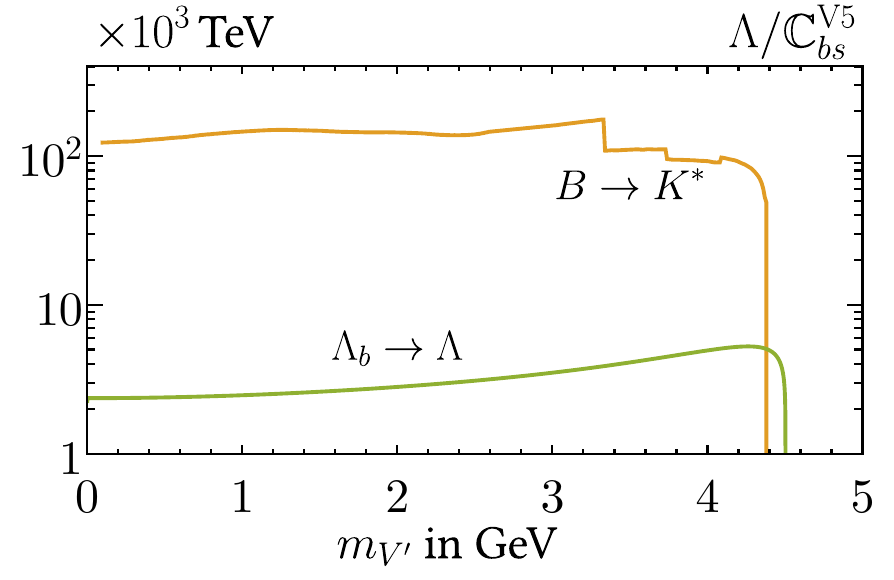}
   \\[0.5em]
   \includegraphics[width=0.47\textwidth]{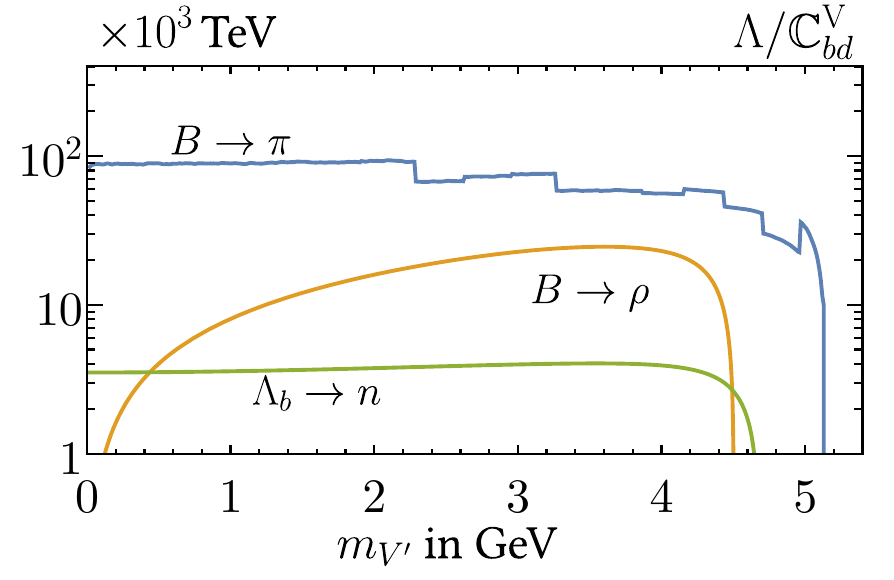}
   \quad
   \includegraphics[width=0.47\textwidth]{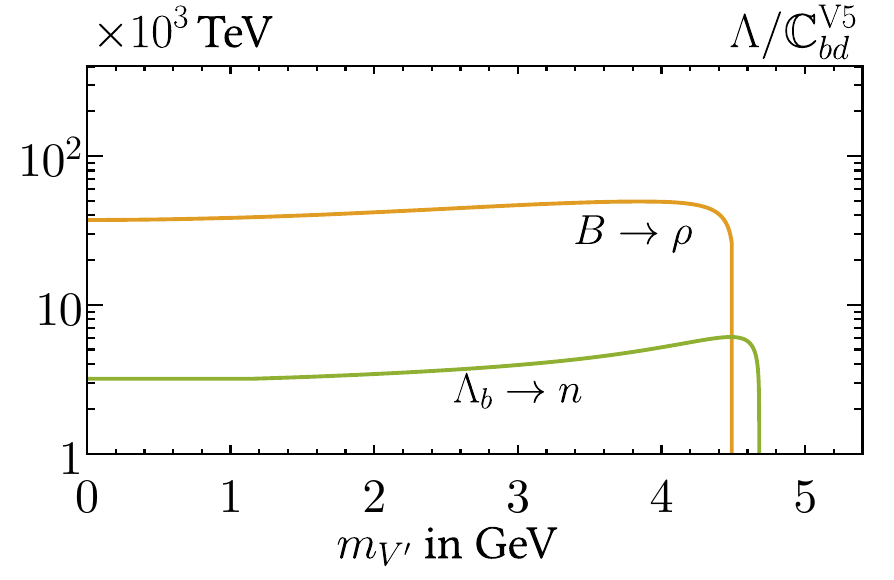}
   \\[0.5em]
   \includegraphics[width=0.47\textwidth]{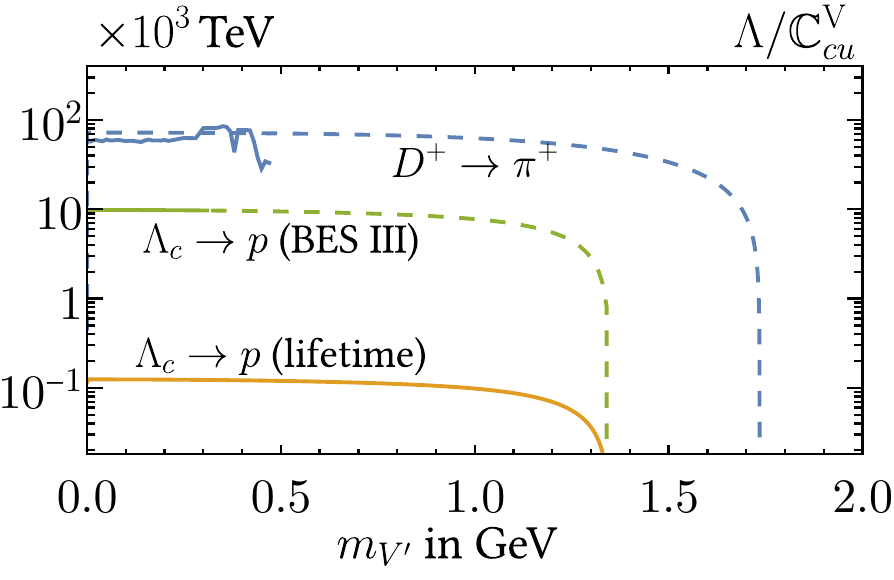}
   \quad
   \includegraphics[width=0.47\textwidth]{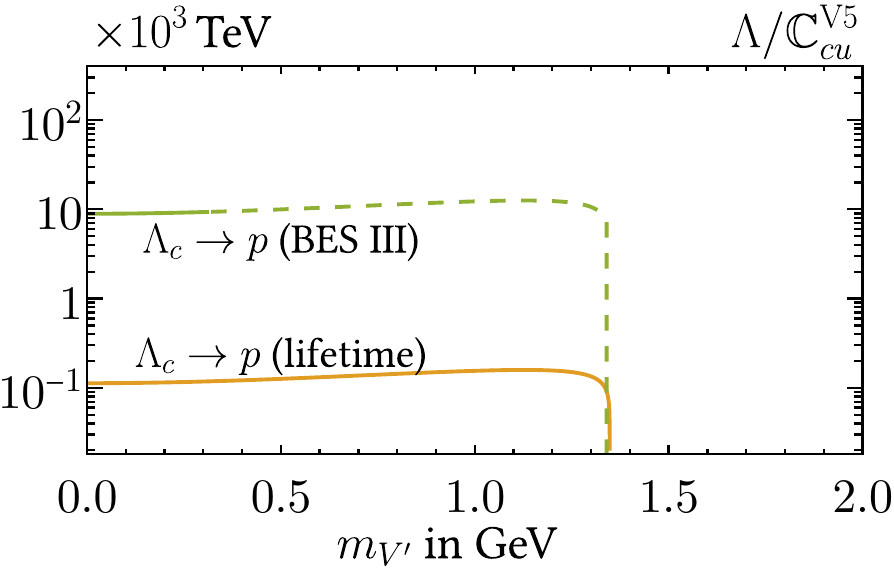}
  \end{center}
  \vspace{-1em}
  \caption{
    Lower limits on quark-flavor violating vector couplings $\Lambda/|\CC{\rV}{ij}|$ (left  column) and $\Lambda/|\CC{\rV5}{ij}|$ (right  column)
   of the LDV for  $s \to d, b \to s, b \to d, c \to u$ transitions @$95\%$ CL$_{(\text{s})}$. See text for details.
	\label{fig:quarkvectorVA}}
\end{figure}

\paragraph{$\boldsymbol {s \to d}$ Transitions} The limits on the vector couplings
$\{ \CC{\rV }{sd}$, $\CC{\rV5}{sd} \}$ are shown in
Figure~\ref{fig:quarkvectorVA}. As for dipole couplings, the relevant constraints
arise from $K\to \pi$ and hyperon decays, see
Table~\ref{tab:experimentalinput}.
Analogous to the dipole case, the limit from BES~III on the decay
  $\Sigma^+ \to p + {\rm invis.}$ for a massless invisible is tentatively assumed to be valid
  for the whole kinematic range. The limit on the scale is indicated by a dashed orange line.
$K \to\pi$ decays dominate the limits on
$\Lambda/\CC{\rV }{sd}$, restricting UV scales up to $10^{12} \GeV$, but cannot
constrain the axial coupling $\Lambda/\CC{\rV5}{sd}$, where hyperon decays set
the only available bounds of the order of $10^{7} \GeV$. All limits are
non-vanishing when the LDV mass is taken to zero, which is due to the choice of
the prefactor in $\Lag_{\rV}$ linear in the LDV mass, see Eq.~\eqref{eq:vector}.
This corresponds to the gauge-less limit where the longitudinal polarization of the LDV
is essentially a Goldstone boson. With this scaling the flavor-violating decay is
similar to the SM decay $t \to W b$, which also remains finite
in the gauge-less $g \to 0$ limit, since the top quark dominantly decays to
the charged Goldstone Higgs, which couples only via Yukawas to the quarks.
Different choices for the prefactor, corresponding to specific UV completions,
would result in bounds that would vanish in the limit of massless LDVs, with a
LDV mass dependence that can obtained by rescaling the limits presented here.

\paragraph{$\boldsymbol {b \to s}$ Transitions} The constraints on the vector
couplings $\CC{\rV }{bs}$, $\CC{\rV5}{bs}$ are obtained from $B$-meson decays
$B\to K/K^* +{\rm invis.}$ and the baryonic decays $\Lambda_b\to \Lambda +{\rm
invis.}$ $B^{+}\to K^{+}$ sets the strongest constraint on $\Lambda/\CC{\rV
}{bs}$ of the order of $10^{8} \GeV$, but cannot constrain the axial coupling
$\Lambda/\CC{\rV5}{bs}$. Here the dominant constraints are set by $B\to K^*$
decays, also of the order of $10^{8} \GeV$, apart from the region where this
channel is kinematically closed and $\Lambda_b\to \Lambda$ takes over
and sets limits on the UV scales up to $10^{6} \GeV$. 
Again there is an {\it upper} limit of order $10^{12}\GeV$ on 
$\Lambda/\CC{\rV }{bs}$ at around $m_{\V}\approx 2 \GeV$
coming from $B\to K + \V$ decays~\cite{Altmannshofer:2023hkn}, due to a
$2.8\,\sigma$ excess from the latest Belle~II measurement of $B^+ \to K^+ \nu
\overline{\nu}$~\cite{Belle-II:2023esi}.

\paragraph{$\boldsymbol {b \to d}$ Transitions} The bounds on the vector couplings
$\CC{\rV }{bd}$, $\CC{\rV5}{bd}$ arise from $B$-meson decays $B\to \pi/\rho +{\rm
invis.}$ and the baryonic decays $\Lambda_b\to n + {\rm invis.} $. Analogously
to $b \to s$ transitions $B\to \pi$ decay sets the strongest constraint on
$\Lambda/\CC{\rV }{bd}$ of the order of $10^{8} \GeV$, while
$\Lambda/\CC{\rV5}{bd}$ is limited to about the same values by $B\to \rho$
decays, up to LDV masses at the kinematic threshold where $\Lambda_b\to n$
decays dominate the bound of order $10^{6} \GeV$.

\paragraph{$\boldsymbol {c \to u}$ Transitions} Finally, the bounds on the vector
couplings $\CC{\rV }{cu}$, $\CC{\rV5}{cu}$ are set by the decays $D\to\pi +{\rm
invis.}$ and $\Lambda_c\to p + {\rm invis.}$ Meson decays $D\to\pi$ dominate
the bound on $\Lambda/\CC{\rV }{cu}$ of order $10^{8} \GeV$, while only baryon
decays $\Lambda_c\to p$ can constrain the axial coupling
$\Lambda/\CC{\rV5}{cu}$ at order $10^{5}$ and $10^7\GeV$, using the total
lifetime and the extrapolation of the BES~III measurement, respectively,
analogous to the dipole case.  Again, it would be interesting if BES~III could
extend their search for $\Lambda_c\to p + \V$ to higher invisible masses, as
this is expected to strengthen the present bound on the UV scale by two orders
of magnitude.

\section{Lepton Phenomenology of Light Dark Vectors\label{sec:leptonflavor}}
\begin{table}[t]
		\centering
	\begin{tabular}{>{\centering\arraybackslash}p{3cm}p{4.5cm}}
		\toprule
        \multicolumn{1}{c}{\bfseries LFV Transition} & {\bfseries Experimental Limit}\\
        \midrule
		 $\mu\to e$ &TWIST~\cite{TWIST:2014ymv}, Jodidio$_r$~\cite{Jodidio:1986mz, Calibbi:2020jvd}  \\
			 $\tau\to e$&Belle~II~\cite{Belle-II:2022heu} \\
			 $\tau\to \mu$&Belle~II~\cite{Belle-II:2022heu} \\
		\bottomrule
	\end{tabular}
	\caption{
      The LFV transitions relevant for the two-body decays
      $\ell\to\ell^\prime +\V$ and the corresponding relevant experimental measurements.
      The subindex ``$r$'' indicates that a recast of experimental data was needed.\label{LFVrefs}
	}
\end{table}

In this section we present the bounds on the flavor-violating couplings in
Eq.~\eqref{eq:dipole} and \eqref{eq:vector} from LFV decays $\ell \to
\ell^{\prime} + \V$ for lepton-flavor transitions $\mu \to e$, $\tau \to e$, and 
$\tau \to \mu$. 
There are three main differences to the quark-sector analysis: 
{\itshape i)} there is no hadronic input required,
{\itshape ii)} the total decay rates only depend on the combination 
$|\CC{\rD }{ij}|^2 + |\CC{\rD5}{ij}|^2$ and $|\CC{\rV }{ij}|^2 + |\CC{\rV5}{ij}|^2$, 
and  
{\itshape iii)} for the case of $\mu \to e$ transitions one can profit from
polarization in order to suppress SM background from Michel decays. This
allows us to distinguish between $\CC{\rV }{ij}$ and $\CC{\rV5}{ij}$ using the
angular distribution of the outgoing electron. 

Concretely, for $\mu\to e$ we restrict the discussion to three benchmark scenarios, depending on the angular 
dependence of the differential two-body LFV decay rate in the limit of $m_e = m_{\V} = 0$
\begin{align}
\frac{d \Gamma (\mu \to e + \V)}{d \cos \theta}  \propto (1 + A \cos \theta) \, ,
\label{Adef}
\end{align}
where $\theta$ is the angle between the outgoing electron momentum and the muon
polarization. We distinguish three benchmark cases: 
isotropic decays ($A=0$), ``$\rV-\rA$'' structure $A = -1$, and ``$\rV+\rA$'' structure $A = +1$.
Clearly polarization does not help to distinguish
an LFV signal from the SM background for the SM case $A=-1$. Thus one can only rely 
on the monochromatic electron as the signal, which leads to weaker bounds than in the other cases
$A=0, +1$~\cite{Calibbi:2020jvd}.
Interestingly, many proposals have been put forward
to look for this decay at present and future high-luminosity muon
facilities~\cite{Calibbi:2020jvd, Jho:2022snj, Knapen:2023zgi, Hill:2023dym},
which are sensitive also to invisible LDVs. We take present constraints on LFV
transitions from the references indicated in Table~\ref{LFVrefs}, and compare
them to the predictions for (polarized) lepton decay rates
calculated in Appendix~\ref{polarizeddecay}.

\begin{figure}[t]
	\centering
	\includegraphics[width=0.47\textwidth]{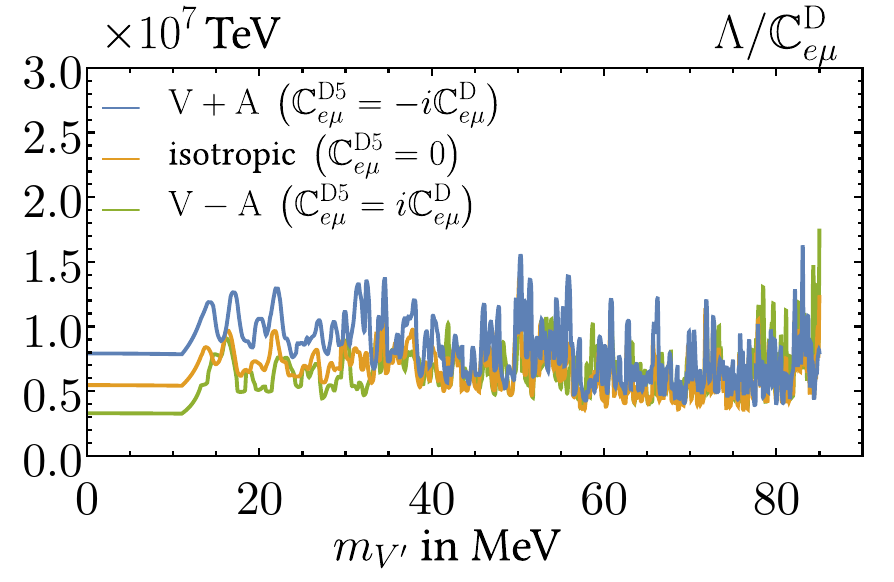}
	\quad
	\includegraphics[width=0.47\textwidth]{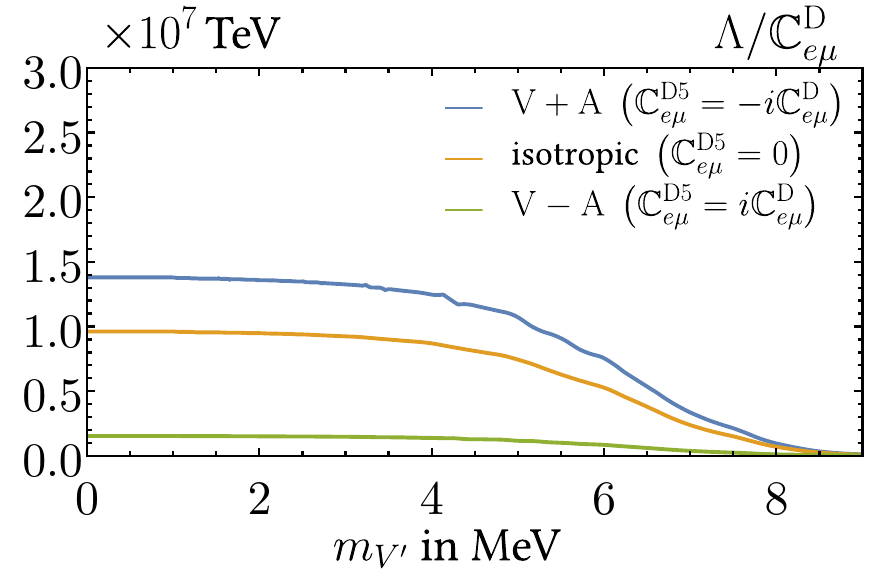}
	\includegraphics[width=0.47\textwidth]{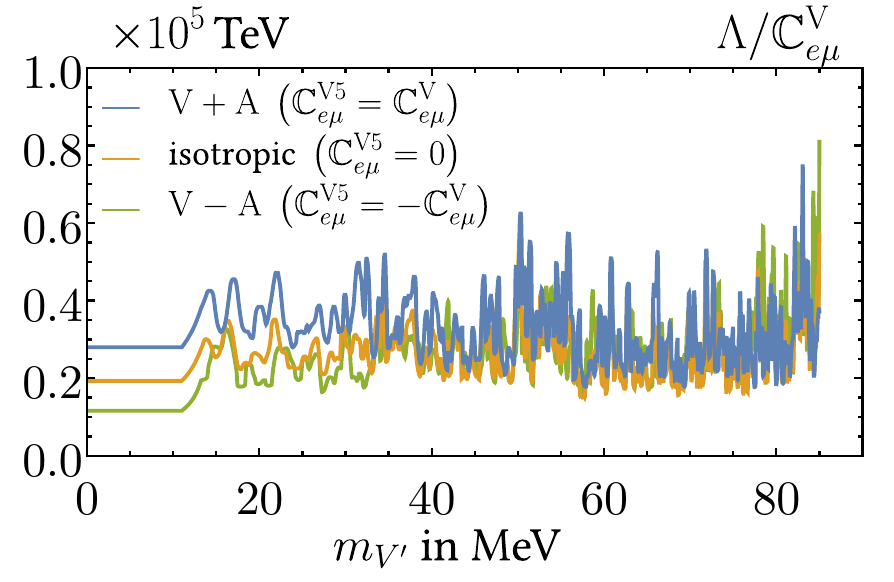}
	\quad
	\includegraphics[width=0.47\textwidth]{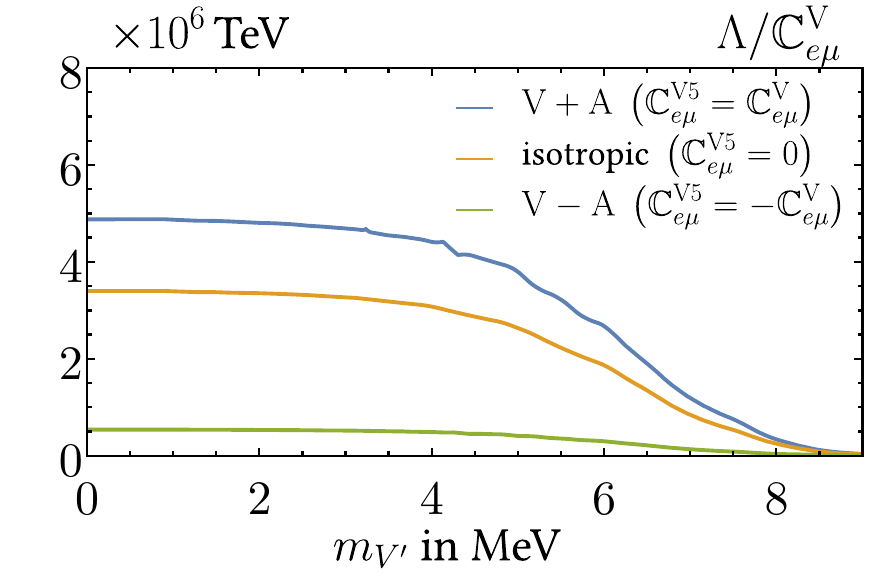}
	\caption{%
		{\it Upper panel:} Lower limits on the dipole coupling for $\mu \to e$ transitions $\Lambda/|\CC{\rD }{e\mu }|$ from
        TWIST~\cite{TWIST:2014ymv} (left panel) and Jodidio et al.~\cite{Jodidio:1986mz,Calibbi:2020jvd} (right panel).
        The bounds	are shown for three different choices for $\CC{\rD5 }{e\mu }$, corresponding to different angular distributions of the electron momentum, cf. Eq.~\eqref{Adef}:
        isotropic decay ($A=0$), alignment to SM decay ``$V\!-\!A$'' ($A=-1$) and ``$V\!+\!A$'' ($A=+1$).
	{\it Lower panel:} same for the vector coupling $\Lambda/|\CC{\rV }{e\mu }|$. See text for details. 
  \label{fig:mue}}
\end{figure}

\paragraph{$\boldsymbol {\mu \to e}$ Transitions} The bounds from $\mu \to e +
{\rm invis.}$ decays on dipole and vector couplings are shown in
Fig.~\ref{fig:mue}. We derive them employing constraints from experiments
conducted at TRIUMF, both by the TWIST collaboration~\cite{TWIST:2014ymv} in 2015 (left panel) 
and Jodidio et al.~\cite{Jodidio:1986mz} in 1986 (right panel).
For the latter, we use the
recast of Ref.~\cite{Calibbi:2020jvd}.  The three curves in Fig.~\ref{fig:mue} show
the bounds for the three benchmark scenarios for chiral structures,
corresponding to $\CC{\rD }{e\mu } = 0$ or $\CC{\rD5}{e \mu } =0$ for $A = 0$,
and $\CC{\rD }{e\mu }= \pm i \CC{\rD5}{e\mu }$ for $A \approx \pm1$ in the
upper panel, while in the lower panel they correspond to $\CC{\rV }{e\mu } = 0$
or $\CC{\rV5}{e\mu } =0$ for $A = 0$, and $\CC{\rV }{e\mu }= \pm \CC{\rV5}{e\mu
}$ for $A \approx \pm1$. For couplings that are not aligned to the SM, i.e.,
not ``$\rV-\rA$'', the dominant constraints on LDVs lighter than about 5 MeV are set by the
Jodidio experiment, which limits UV scales of the order of $10^{10} \GeV$.
Heavier LDVs are constrained only by TWIST, setting limits of the order of few
$\times 10^9 \GeV$. LDVs with ``$\rV-\rA$'' couplings are constrained by TWIST with
bounds of the same order, exceeding the corresponding Jodidio limits also 
in the light-mass regime.

\begin{figure}[t]
	\centering
	\includegraphics[width=0.47\textwidth]{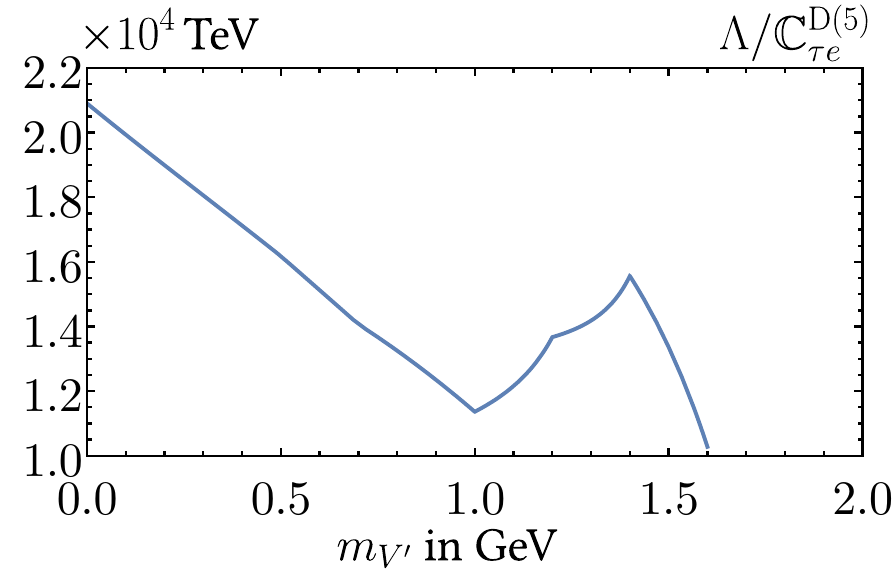}
	\quad
	\includegraphics[width=0.47\textwidth]{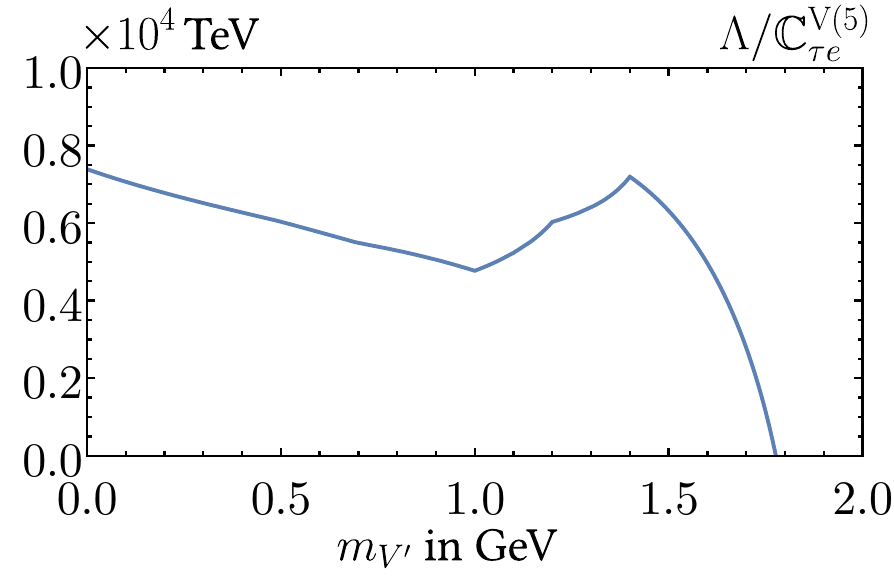}
	\includegraphics[width=0.47\textwidth]{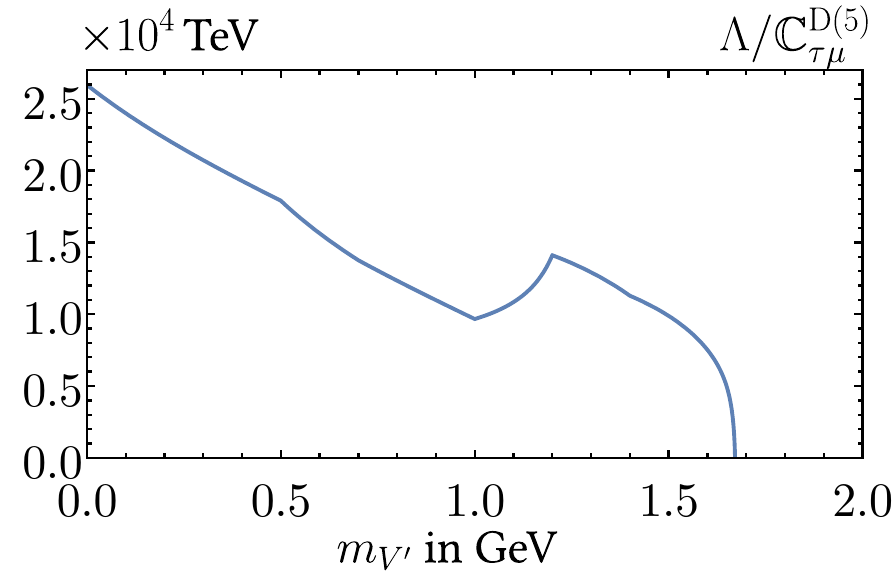}
	\quad
	\includegraphics[width=0.47\textwidth]{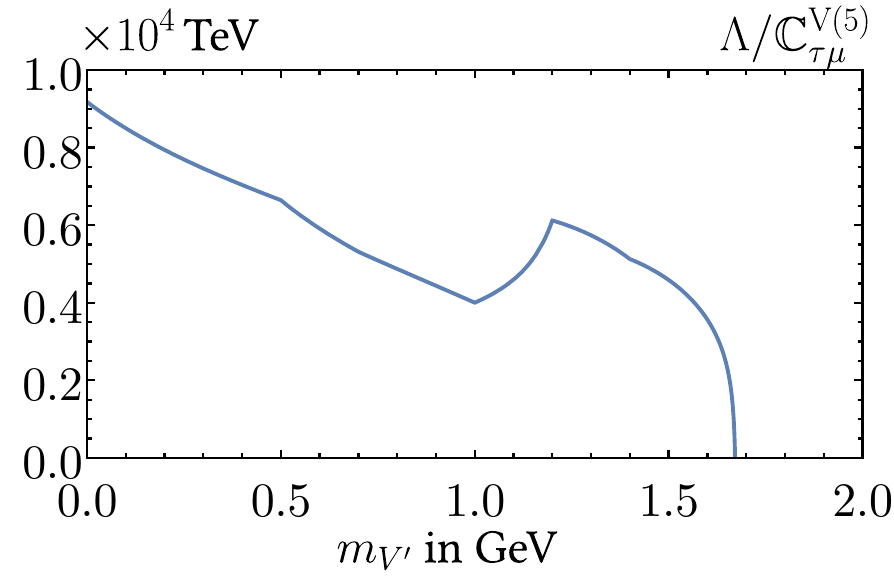}
	 \caption{%
	 {\it Upper panel:} Lower limits on the dipole (left panel) and vector (right panel) couplings for $\tau \to e$ transitions $\Lambda/|\CC{\rD }{\tau e}|$,
     $\Lambda/|\CC{\rV }{\tau e}|$ from Belle~II~\cite{Belle-II:2022heu}.
	 {\it Lower panel:} same for $\tau \to \mu$ transitions  $\Lambda/|\CC{\rD }{\tau \mu}|$,
     $\Lambda/|\CC{\rV }{\tau \mu}|$.
     Constraints on the axial couplings $\Lambda/|\CC{\rD5}{\tau \ell}|$ and 
     $\Lambda/|\CC{\rV5}{\tau\ell}|$ are essentially of the same size, as the difference
     is suppressed by $m_\ell/m_{\tau}$, cf.~Appendix~\ref{polarizeddecay}.
     \label{fig:taul}}
\end{figure}

\paragraph{$\boldsymbol {\tau \to \mu/e}$ Transitions} The limits from
Belle~II on $\tau \to \mu/e + {\rm invis.}$ decays constrain $\tau
\to e$ and $\tau \to \mu$ transitions according to Fig.~\ref{fig:taul}, where
we shows the bounds on the dipole $\Lambda/\CC{\rD }{\tau \ell }$ (left panel) 
and vector couplings $\Lambda/\CC{\rV }{\tau \ell}$ (right panel). 
Constraints on the axial couplings $\Lambda/\CC{\rD5}{\tau \ell}$ and $\Lambda/\CC{\rV5}{\tau
\ell}$ are at the same level, as the difference is suppressed by
$m_\ell/m_{\tau}$, cf. Appendix~\ref{polarizeddecay}. Bounds for $\tau \to e$
and $\tau \to \mu$ transitions are comparable, limiting UV scales of the order
of few $\times 10^7 \GeV$ for dipole couplings, and few $\times 10^6 \GeV$ for
vector couplings.

\section{Flavor-violating LDVs from the Renormalization Group\label{sec:MFV}}

In this section we study the phenomenologically interesting scenario 
in which LDV interactions with the SM are flavor-universal in the UV theory, 
so that flavor-violating couplings are generated only from the SM flavor violation
via the renomalization-group evolution.
We start right below the UV scale $\Lambda$---taken to be much above the electroweak scale---and
consider $SU(2)_\rL \times U(1)_Y$ invariant vector and dipole interactions of the $\V$ to the 
SM.
For vector couplings see the trivial $SU(2)_\rL \times U(1)_Y$ generalization of 
Eq.~\eqref{eq:LRsetup} and for dipole couplings see Eq.~\eqref{eq:EW}.
We align possible new sources of flavor-violation of the $\V$ with the flavor violation 
in the SM by taking the vector couplings to be flavor-universal, i.e., 
proportional to the identity matrix in flavor space,
and by taking the dipole couplings to be proportional to the SM
Yukawas. In both cases they are flavor diagonal in the mass basis, such that
flavor-changing interactions with the $\V$ are only induced by the renormalization-group
evolution to the EW scale and always proportional to the CKM matrix. 
Flavor-violating couplings in the IR  thus follow the paradigm of minimal flavor violation (MFV)~\cite{DAmbrosio:2002vsn}.

We do not explicitly consider kinetic mixing between the
$U(1)'$ LDV and the $U(1)_Y$ boson, as it leads only to a shift in the
flavor-universal LDV couplings after diagonalising the photon kinetic terms.
By working in this basis, our results also apply to models with kinetic
mixing, upon re-defining the flavor-universal couplings.

We discuss separately the case of dipole and vector couplings 
in Section~\ref{sec:RGEdipole} and \ref{sec:RGEvector}, respectively.

\subsection{Dipole interactions\label{sec:RGEdipole}}

\begin{table}[t]
\centering
\begin{tabular}{cclclclclclcl}
	\toprule
	$i-j$	& $2-1$  &  $1-2$ &  $2-3$&  $3-2$&  $1-3$&  $3-1$\\ \midrule
	$(\CC{\rD \rR }{u})_{i \ne j} /\Lambda$	 & $\lambda^5 y_b^2y_c$  &$\lambda^5 y_b^2y_c$&  $\lambda^2y_b^2y_t$&  $\lambda^2y_b^2y_t$ &  $\lambda^3y_b^2y_t$&  $\lambda^3y_b^2y_t$\\
	$(\CC{\rD \rR }{d})_{i \ne j}  /\Lambda$	 & $\lambda^5 y_t^2y_s$  &$\lambda^5 y_t^2y_s$& $\lambda^2y_t^2y_b$ & 	 $\lambda^2y_t^2y_b$ & $\lambda^3y_t^2y_b$ & $\lambda^3y_t^2y_b$\\
	\bottomrule
\end{tabular}
\caption{
Parametric size of leading flavor-violating contributions
at low-energy in the UV universal scenario for dipole couplings, cf.~Eq.~\eqref{couplingsDMB}.
Here $\lambda \approx 0.23$ denotes the Wolfenstein parameter and $y_f =m_f/v$ are SM Yukawas couplings.
Up-quark transitions (first line) are proportional to the high-scale coupling $c^{\rD}_{d}$, 
down-quark transitions (second line) are proportional to the high-scale coupling $c^{\rD}_{u}$, 
and all entries are multiplied by $v /\Lambda_6^2 \log (\Lambda_6/\mu) /(16 \pi^2)$.
\label{leadingFVCi}}
\end{table}

In the interaction basis, the dipole interactions of the LDV with SM fermions are
given by (cf.~Eq.~\eqref{eq:EW})
\begin{align}
\label{intdipole}
\Lag_\mathrm{int}
\supset -\left(
\overline{Q}\, Y_u \widetilde{H} u_{\rR}
+ \overline{Q}\, Y_d H d_{\rR} + \mathrm{h.c.}
\right)+ \frac{1}{\Lambda^2_6}\V_{\mu\nu}\left(
\overline{Q} C^{\rD}_{u} \sigma^{\mu\nu} \widetilde{H} u_{\rR}
+ \overline{Q} C^{\rD}_{d} \sigma^{\mu\nu} H d_{\rR}
+ \mathrm{h.c.}
\right) \,,
\end{align}
with the SM Yukawa matrices $Y_{f}$, $f=u,d$, and arbitrary $3\times 3$ matrices
$C_f^{\rD}$. 
The one-loop RG equations for the couplings $C_f^{\rD}$ and the Yukawa matrices $Y_f$ 
are listed in Appendix~\ref{sec:rges}. 
For the UV universal setup that we consider,
the initial conditions at the UV scale $\Lambda_6$ are
\begin{equation}
\label{icdipole}
C^{\rD}_{d}\big\vert_{\mu=\Lambda_6} = c^{\rD}_{d} ~Y_d\big\vert_{\mu=\Lambda_6}\,,\qquad 
C^{\rD}_{u}\big\vert_{\mu=\Lambda_6} = c^{\rD}_{u} ~Y_u\big\vert_{\mu=\Lambda_6}\,,
\end{equation}
with $c^{\rD}_{f}\in \mathbb{C}$.
By solving the RGE at leading-logarithmic accuracy and subsequently 
rotating to the mass basis for the quarks we find
the low-energy dipole couplings in the $\rL/\rR$ notation of
Eq.~\eqref{eq:LRsetup} with $f=u,d$
to be\footnote{ Since the couplings at the UV scale $\Lambda_6$ are aligned to the SM Yukawa matrices, a correction from the Yukawa RGE, given in Appendix \ref{sec:rges}, must be included.}
\begin{equation}
\label{couplingsDMB}
\begin{split}
\frac{1}{\Lambda}\CC{\rD \rR }{u}(\mu)&=\frac{v}{\Lambda_6^2}\left(c^{\rD}_{u}\hat{Y}_u-\frac{1}{16\pi^2}\left(3c^{\rD}_{u}\hat{Y}_u\hat{Y}_u^{\dagger}\hat{Y}_u -c^{\rD}_{d}V_{\text{CKM}}\hat{Y}_d \hat{Y}_d^{\dagger}V_{\text{CKM}}^\dagger\hat{Y}_u\right)\log(\Lambda_6/\mu)\right)\, ,\\
\frac{1}{\Lambda}\CC{\rD \rR }{d}(\mu)&=\frac{v}{\Lambda_6^2}\left(c^{\rD}_{d}\hat{Y}_d-\frac{1}{16\pi^2}\left(3c^{\rD}_{d}\hat{Y}_d\hat{Y}_d^{\dagger}\hat{Y}_d -c^{\rD}_{u}V_{\text{CKM}}^\dagger\hat{Y}_u \hat{Y}_u^{\dagger}V_{\text{CKM}}\hat{Y}_d\right)\log(\Lambda_6/\mu)\right)\, ,
\end{split}
\end{equation}
where $V_{\text{CKM}}$ is the CKM matrix, and $\hat{Y}_f=m_f/v$ are the diagonal SM
Yukawas. 
The left-handed couplings $\CC{\rD \rL }{f}$ are related to the ones
in Eq.~\eqref{couplingsDMB} by hermitian conjugation, $\CC{\rD \rL
}{f}=(\CC{\rD \rR }{f})^{\dagger}$. Note that indeed flavor off-diagonal entries are
generated in both the up- and the down-quark sector at one-loop. 
They are proportional to the CKM matrix and 
the UV coupling of the other sector, i.e.,
$\CC{\rD \rR}{u}\propto c_d^{D}$
and 
$\CC{\rD \rR}{d}\propto c_u^{D}$.
Carrying out the matrix multiplications, one can
identify the numerically leading contribution to a given flavor transition.
We show these leading contributions in Table~\ref{leadingFVCi} for both sectors.

Using these results, we  determine the experimental limits on the high-scale 
couplings $c^{\rD}_{d}$ and $c^{\rD}_{u}$ in Eq.~\eqref{icdipole} from
the limits on two-body meson decays discussed in Section~\ref{sec:quarkflavor}.
Note that the renormalization scale $\mu$ is set to the EW scale since
below there is no Yukawa running.

\begin{figure}[t]
\centering
\includegraphics[width=0.49\textwidth]{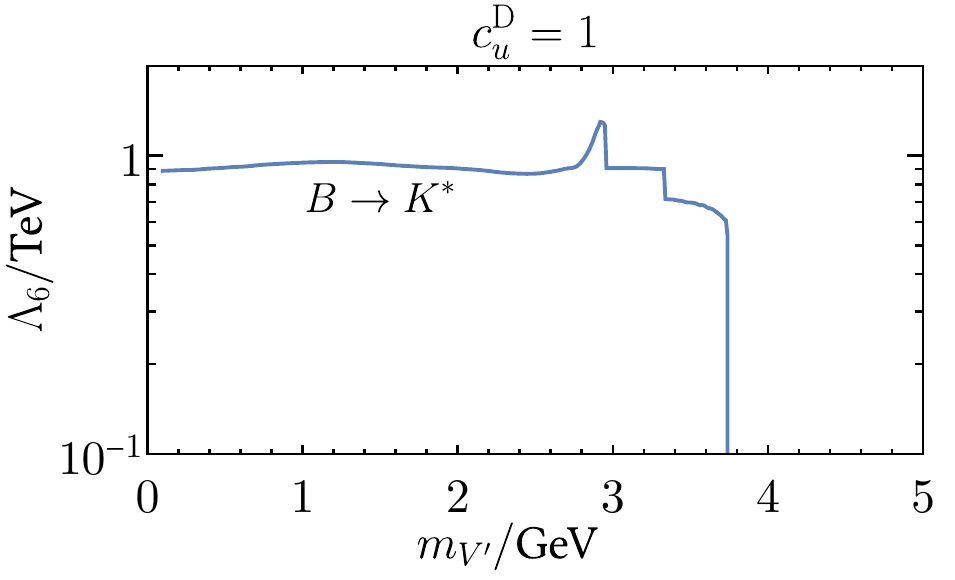}
\includegraphics[width=0.49\textwidth]{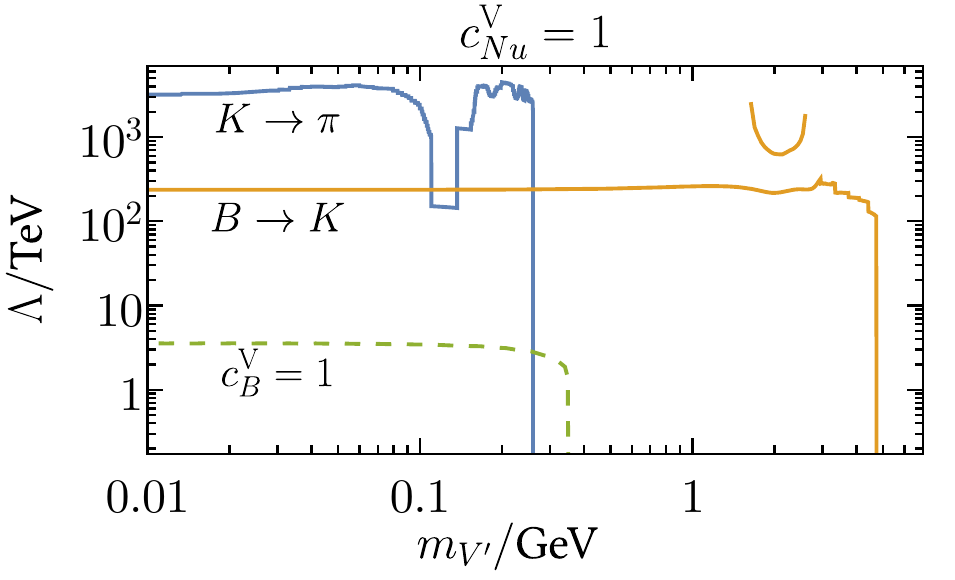}
\caption{Lower limits  on the UV scale in the UV universal scenario for dipole (left panel) 
and vector couplings (right panel), only showing the strongest constraints.
See text for details.\label{RGplots}}
\label{fig:rgebounds}
\end{figure}

As expected from the high-level of flavor suppression inherent to the setup,
the resulting bounds are very mild and
often weaker than the constraints from perturbative unitarity. For this
reason we only display in Fig.~\ref{RGplots} (left panel) the strongest
bounds, which come from $B \to K^*$ and require $\Lambda_6 \sim \TeV$ for
$c^{\rD}_{u} =1$ (for $c^{\rD}_{d} =1$ the limit on $\Lambda_6$ is far below the electroweak scale and is therefore not shown).

\subsection{Vector interaction\label{sec:RGEvector}}

In the interaction basis, the vector interactions of the LDV with the SM fermions are
given by (cf. Eq.~\eqref{eq:LRsetup})
\begin{align}
\label{intvector}
\Lag_\mathrm{int}
\supset -\left(
\overline{Q}\, Y_u \widetilde{H} u_{\rR}
+ \overline{Q}\, Y_d H d_{\rR} + \mathrm{h.c.}
\right)
+ \V_\mu\left(
\overline{Q} C_{Q}^\rV \gamma^\mu Q
+ \overline{u}_{\rR} C_{u}^\rV \gamma^\mu u_{\rR}
+ \overline{d}_{\rR} C_{d}^\rV \gamma^\mu d_{\rR}
\right) \, ,
\end{align}
with SM Yukawa matrices $Y_{f}$, $f=u,d$, and arbitrary hermitian
$3\times 3$ matrices $C_X^{\rV}$ with $X=Q,u,d$. The one-loop 
RG equations for the couplings $C_X^{\rV}$ and the Yukawa matrices $Y_f$ are listed in Appendix~\ref{sec:rges}. 
For the UV universal setup that we consider in this section, the boundary conditions at the UV scale $\Lambda$ are
\begin{align}
\label{icvector}
C^{\rV}_{Q}(\Lambda) & = c^{\rV}_{Q}~\mathbbold{1}_3\,, & 	
C^{\rV}_{u}(\Lambda) & = c^{\rV}_{u}~\mathbbold{1}_3\,, & 
C^{\rV}_{d}(\Lambda) & = c^{\rV}_{d}~\mathbbold{1}_3\, .
\end{align}
with $c^{\rV}_X$ real numbers.

By solving the RGE at leading-logarithmic accuracy and subsequently 
rotating to the mass basis for the quarks, we find
the low-energy vector couplings in the $\rL/\rR$ notation of
Eq.~\eqref{eq:LRsetup} to be
\begin{align}
\label{couplingsVMB}
\left(\frac{m_{\V}}{\Lambda}\right)\CC{\rV \rL }{u}(\mu)
&=c^{\rV}_{Q}\mathbbold{1}_3-\frac{1}{16\pi^2}\left(\left(c^{\rV}_{Q}-c^{\rV}_{u}\right)\hat{Y}_u\hat{Y}_u^{\dagger}+\left(c^{\rV}_{Q}-c^{\rV}_{d}\right)V_\text{CKM}\hat{Y}_d\hat{Y}_d^{\dagger}V_{\text{CKM}}^{\dagger}\right)\log(\Lambda/\mu) \, , \nonumber \\
\left(\frac{m_{\V}}{\Lambda}\right)\CC{\rV \rL }{d}(\mu)
&=c^{\rV}_{Q}\mathbbold{1}_3-\frac{1}{16\pi^2}\left(\left(c^{\rV}_{Q}-c^{\rV}_{u}\right)V_{\text{CKM}}^{\dagger}\hat{Y}_u\hat{Y}_u^{\dagger}V_\text{CKM}+\left(c^{\rV}_{Q}-c^{\rV}_{d}\right)\hat{Y}_d\hat{Y}_d^{\dagger}\right)\log(\Lambda/\mu) \, , \nonumber \\
\left(\frac{m_{\V}}{\Lambda}\right)\CC{\rV \rR }{u}(\mu)
	&=c^{\rV}_{u}\mathbbold{1}_3-\frac{1}{8\pi^2}\left(c^{\rV}_{u}-c^{\rV}_{Q}\right)\hat{Y}_u^{\dagger}\hat{Y}_u \log(\Lambda/\mu) \, ,  \nonumber \\
\left(\frac{m_{\V}}{\Lambda}\right)\CC{\rV \rR }{d}(\mu)
	&=c^{\rV}_{d}\mathbbold{1}_3-\frac{1}{8\pi^2}\left(c^{\rV}_{d}-c^{\rV}_{Q}\right)\hat{Y}_d^{\dagger}\hat{Y}_d \log(\Lambda/\mu)  \, , 
\end{align}
where $V_\text{CKM}$ is the CKM matrix and $\hat{Y}_f=m_f/v, f= u,d$ are the diagonal SM
Yukawas. Note that the couplings of right-handed interactions $\CC{\rV \rR }{u},
\CC{\rV \rR }{d}$ are always flavor diagonal, while flavor-violating terms in the IR are
induced in the left-handed interactions $\CC{\rV \rL }{u}$, $\CC{\rV \rL }{d}$
proportionally to $c^{\rV}_{Q}-c^{\rV}_{u}$ and
$c^{\rV}_{Q}-c^{\rV}_{d}$.  
Therefore, if the UV couplings are also universal among
the different sectors, i.e., $c^{\rV}_{Q}=c^{\rV}_{u}=c^{\rV}_{d}$, there is no
flavor violation in the IR at one-loop, as in this case the LDV actually couples to the baryon-number 
current, which is conserved at tree-level inducing flavor violation only at two-loop~\cite{Dror:2017nsg}.

We now discuss this fact in more detail, before turning to the limits. One can rewrite the interactions in Eq.~\eqref{intvector} for the case of flavor-universal UV boundary conditions in Eq.~\eqref{icvector}  in terms of the tree-level conserved (but anomalous) $U(1)_B$ current 
$J_B^{\mu} = \sum_i \left(\bar{Q}_{i} \gamma^{\mu} Q_{i} + \bar{u}_{\rR i}\gamma^{\mu}u_{\rR i} + \bar{d}_{\rR i}\gamma^{\mu}d_{\rR i}\right)$,
and the two non-conserved currents 
$J_{Nd}^{\mu}= \sum_i \bar{d}_{\rR i}\gamma^{\mu}d_{\rR i}$, and 
$J_{Nu}^{\mu}=\sum_i \bar{u}_{\rR i}\gamma^{\mu}u_{\rR i}$. As all currents are not conserved beyond tree-level, we 
take their coefficients to be proportional to the LDV mass
\begin{equation}
\Lag_\mathrm{int}
\supset -\left(
\overline{Q}\, Y_u \widetilde{H} u_{\rR}
+ \overline{Q}\, Y_d H d_{\rR} + \mathrm{h.c.}
\right) + \frac{m_{\V}}{\Lambda}  \V_\mu \Big[
  C_B^\rV J_B^{\mu}+ C_{Nd}^\rV J_{Nd}^{\mu}+C_{Nu}^\rV J_{Nu}^{\mu} \Big]\,,
\end{equation}
Matching to Eqs.~\eqref{intvector} and~\eqref{icvector} gives 
\begin{align}
\label{matching}
\frac{m_{\V}}{\Lambda}c_{B}^\rV      & = c^{\rV}_{Q}\, , & 
\frac{m_{\V}}{\Lambda}c_{Nd}^\rV  & = c^{\rV}_{d}-c^{\rV}_{Q} \, , &  
\frac{m_{\V}}{\Lambda}c_{Nu}^\rV  & = c^{\rV}_{u}-c^{\rV}_{Q}\, .
\end{align}
At the one-loop level there is no flavor violation proportional to 
$C_{B}^\rV$. However, flavor violation does arises due to the  non-conserved
currents and is thus proportional to 
the difference of couplings
$c^{\rV}_{Q}-c^{\rV}_{u}$ and $ c^{\rV}_{Q}-c^{\rV}_{d}$. Rewriting Eq.~\eqref{couplingsVMB} in terms of 
the UV coefficients $c_{B}^\rV , c_{Nd}^\rV, c_{Nu}^\rV$ with the
proper  LDV mass scaling gives finally
\begin{align}
\label{couplingsVMBfin}
\left(\frac{m_{\V}}{\Lambda}\right)\CC{\rV \rL }{u}(\mu)
&=\frac{m_{\V}}{\Lambda} \left[ c^{\rV}_{B}\mathbbold{1}_3+\frac{1}{16\pi^2}\left(c_{Nu}^{\rV}\hat{Y}_u\hat{Y}_u^{\dagger}+c_{Nd}^{\rV}V_{\text{CKM}}\hat{Y}_d\hat{Y}_d^{\dagger}V_{\text{CKM}}^{\dagger}\right) \log(\Lambda/\mu) \right] \, ,\nonumber \\
\left(\frac{m_{\V}}{\Lambda}\right)\CC{\rV \rL }{d}(\mu)
&=\frac{m_{\V}}{\Lambda} \left[ c^{\rV}_{B}\mathbbold{1}_3 + \frac{1}{16\pi^2}\left(c_{Nu}^{\rV} V_{\text{CKM}}^{\dagger}\hat{Y}_u\hat{Y}_u^{\dagger}V_\text{CKM}+c^{\rV}_{Nd}\hat{Y}_d\hat{Y}_d^{\dagger}\right)  \log(\Lambda/\mu)  \right]\, , \nonumber \\
\left(\frac{m_{\V}}{\Lambda}\right)\CC{\rV \rR }{u}(\mu)
	& = \frac{m_{\V}}{\Lambda} \left[ \left(c^{\rV}_{B}+c_{Nu}^{\rV} \right)\mathbbold{1}_3 - \frac{1}{8\pi^2}c_{Nu}^{\rV}\hat{Y}_u^{\dagger}\hat{Y}_u \log(\Lambda/\mu) \right] \, , \nonumber \\
\left(\frac{m_{\V}}{\Lambda}\right)\CC{\rV \rR }{d}(\mu)
	&= \frac{m_{\V}}{\Lambda} \left[ \left(c^{\rV}_{B}+c_{Nd}^{\rV} \right)\mathbbold{1}_3-\frac{1}{8\pi^2}c_{Nd}^{\rV}\hat{Y}_d^{\dagger}\hat{Y}_d \log(\Lambda/\mu) \right] \, ,
\end{align}

The numerically leading contributions to a given (hermitian) flavor transition in
left-handed interactions are shown in Table~\ref{leadingFVvector} for both
sectors. We display the resulting bounds on $\Lambda$
in the right panel of
Fig.~\ref{RGplots} for $c^\rV_{Nu} =1 $ (there is no constraint from $c^\rV_{Nd}$ at
one-loop), which are of order $\Lambda \ge 10^3 \TeV$ for $K \to \pi$
transitions. 
These limits are weakened by about an order of magnitude for LDV
masses above $m_K - m_\pi$, where the dominant constraint comes from $B \to K$
transitions.
In dashed green, we also show the limits coming from the flavor-violating
contribution that is induced at the two-loop level by the coupling of the LDV to
the anomalous baryon current $J_B^\mu$.  The corresponding limit on the scale
$\Lambda$ has been obtained by rescaling the result for $K\to\pi$ of Fig.~(1)
from Ref.~\cite{Dror:2017nsg}, giving $\Lambda\ge 3.5\TeV$ for $c^{\rV}_B = 1$
and $c^\rV_{Nu} = c^\rV_{Nd} = 0$. This is about three orders of magnitude weaker
than the limit one obtains if the LDV also couples to currents that are not
conserved at tree-level, i.e., taking $c^V_{Nu} = c^V_{Nd} = 1$.

\begin{table}[t]
	\centering
	\begin{tabular}{cccc}
		\toprule
		$i-j$	& $2-1$  &  $3-2$ &  $3-1$  \\ \midrule
		$(\CC{\rV \rL }{u})_{i\ne j}$	 & $\lambda^5 y_t^2$  &$\lambda^2y_t^2$&  $\lambda^3y_t^2$ \\
		$(\CC{\rV \rL }{d})_{i \ne j}$	&  $\lambda^5 y_b^2$ &   $\lambda^2 y_b^2$ & $\lambda^3 y_b^2$ \\
		\bottomrule
	\end{tabular}
\caption{
Parametric size of leading flavor-violating contributions
to the low-energy vector couplings of $\V$
in the UV universal scenario, cf. Eq.~\eqref{couplingsVMB}.
Here $\lambda \approx 0.23$ denotes the Wolfenstein parameter. 
Up-quark transitions (first line) are proportional to the high-scale 
coupling $c^{\rV}_{u}-c^{\rV}_{Q} = c_{Nu}^\rV$, 
down-quark transitions (second line) are proportional to the high-scale coupling 
$c^{\rV}_{d}-c^{\rV}_{Q}=   c_{Nd}^\rV$, 
and all entries are multiplied by $\log (\Lambda/\mu) /(16 \pi^2)$. \label{leadingFVvector}}
\end{table}

\section{Summary and Conclusions\label{sec:conclusions}}
In this work we have systematically studied the flavor phenomenology of
light dark vectors (LDVs).
We have restricted our analysis to scenarios where the LDV is directly 
linked to dark matter, and is either itself invisible or promptly decays to invisible
particles, such that the LDV appears as missing energy. Working in the context of
a general EFT, we have considered both flavor-violating dipole (see Eq.~\eqref{eq:dipole}) and vector couplings 
(see Eq.~\eqref{eq:vector}) of the LDV to SM fermions.
We have calculated the resulting predictions for the
decay rates of mesons, baryons, and polarized leptons as a function of the LDV
mass, see Section~\ref{sec:2bodydecays}. These predictions were compared to the
experimental limits on various hadronic processes
(Table~\ref{tab:experimentalinput}) and LFV transitions (Table~\ref{LFVrefs}).
For $B \to \pi/K/K^*$ decays experimental limits from B-factories are only available for
three-body decays with two invisible neutrinos, so we have recasted available
data to obtain bounds on the two-body decay with missing energy as a
function of the LDV mass, see Fig.~\ref{fig:2-body recast}. The resulting
limits on general vector and dipole interactions of the LDV are summarized in
Figs.~\ref{fig:quarkdipoleVA} and \ref{fig:quarkvectorVA} for the quark sector,
and in Fig.~\ref{fig:mue} and \ref{fig:taul} for the lepton sector. 
Vector couplings are at least dimension-five operators, which results in very 
stringent limits on the UV scale, reaching up to $10^{12} \GeV$ 
in $K \to \pi$ decays, $10^{8} \GeV$ in
$B$- and $D$-meson decays, $10^9 \GeV$ in $\mu \to e$ decays, and $10^7 \GeV$ in
$\tau \to \mu/e$ decays. 
Bounds on dipole couplings are weaker, if viewed as dimension-six operators 
above the EW scale, but they still probe UV scales of order
$10^6 \GeV$ in $K \to \pi$ and $\mu \to e$ decays. Importantly, all channels
will be improved by present or near-future experiments, such as NA62, Belle~II,
BES~III, MEG-II or Mu3e. We have also discussed a scenario where couplings in the UV are flavor-universal, so that quark-flavor violation is only induced radiatively through the CKM matrix. For this
analysis we derived the relevant renormalization-group equations (RGEs) for
both dipole and minimal couplings in Appendix~\ref{sec:rges}, and used our
previous results to convert limits on flavor-changing interactions into limits
on flavor-diagonal couplings, see Figure.~\ref{fig:rgebounds}.

To summarize, the aim of this work is to stress the importance of
flavor-violating transitions for light, dark-matter searches, which is copiously
produced in the lab as missing energy in decays of SM particles. Here we
focused on the LDV as part of a dark sector, and showed that present
constraints from precision flavor experiments already probe UV scales as large
as $10^{12} \GeV$. This underlines the important role of present and next-generation 
flavor factories in hunting down dark matter in the laboratory.

\section*{Acknowledgments}
We would like to thank Adam Falkowski, Jorge Martin Camalich and Zhijun Li for useful
discussions and Jure Zupan for  comments on the manuscript. This work has received support from the European Union's Horizon
2020 research and innovation programme under the Marie Sk{\l}odowska-Curie
grant agreement No 860881-HIDDeN and is partially supported by project B3a and C3b of
the DFG-funded Collaborative Research Center TRR257 ``Particle Physics
Phenomenology after the Higgs Discovery''.

\appendix

\section{UV Motivation of Vector Couplings\label{sec:UV}}

In this section we motivate the scaling behavior of the flavor-violating
vector coupling in the Lagrangian of Eq.~\eqref{eq:vector}, both by EFT
considerations and explicit UV-complete models. In perturbative UV
completions, the scaling is at least linear in the dark $U(1)^\prime$
breaking scale, and we will provide two example scenarios: one that gives
linear and one that gives quadratic scaling.  We begin with the EFT
discussion of the latter.

\subsection{EFT Discussion for Quadratic scaling}
For the EFT approach it is convenient to consider the coupling to the Goldstone
boson $G$ in the gauge-less limit, rather than the coupling of the 
dark vector $\V_\mu$ itself. 
They are related by the Goldstone-boson Equivalence Theorem,
which states that at sufficiently 
high energies, or equivalently sufficiently small dark vector masses $m_\V$, the vector
boson coupling is dominated by its longitudinal polarization, which in the
small $m_\V$ limit becomes the Goldstone boson. Thus one can work out the couplings of the Goldstone boson and 
recover the relevant vector-boson couplings by
replacing $\partial_\mu G \to - m_{\V}\V_\mu$ in the interaction Lagrangian. 

We, therefore, consider the case where the dark $U(1)^\prime$ gauge group is
spontaneously broken by some (SM singlet) scalar field $S$ with charge $+1$
under the $U(1)^\prime$. We take the gauge-less limit, so that $G$ is a true
Goldstone boson, contained in $S$ according to
\begin{align}
\label{Gobo}
S = \frac{v^\prime}{\sqrt{2}} \exp (i G/ v^\prime) \, ,
\end{align}
where $v^\prime/\sqrt{2}$ is the (real) VEV that breaks $U(1)^\prime$,
connected to the dark vector mass by $m_{\V}= g^\prime v^\prime$, and we have
ignored the radial mode that obtains its mass around $v^\prime$. This mode, together with all UV fields are taken to be much heavier than the electroweak scale, so that in the 
IR there is only the SM and the Goldstone boson $G \supset S$, which 
is formally invariant under global $U(1)^\prime$ transformations 
treating $S$ as a spurion with charge $+1$. 
Note that one can always realize such a  scenario by making $g'$ sufficiently small.
Writing down the general EFT for this setup, it is clear that if SM fields are
not charged under $U(1)^\prime$, the possible couplings of the Goldstone to SM
fields must involve the same powers of $S^\dagger$ and $S$.
The first such bilinear that gives a non-trivial combination containing 
the Goldstone is then 
$S^\dagger \overset{\leftrightarrow}{\partial}_\mu S \supset i v^\prime  \partial_\mu G$.
This implies that, e.g., right-handed down quarks can only couple to the Goldstone at the
level of dimension-six operators only
\begin{align}
  \Lag^{\text{EFT}}_\text{quadratic} \supset \frac{c_{ij}}{\Lambda^2 } ( i S^\dagger \overset{\leftrightarrow}{\partial}_\mu S ) \left( \overline{d}_{\rR i}  \gamma^\mu d_{\rR j} \right) & = - \frac{c_{ij} }{\Lambda^2} v^\prime \partial_\mu G  \left( \overline{d}_{\rR i}  \gamma^\mu d_{\rR j} \right) \, ,
\end{align}
where $\Lambda$ is the UV scale and in general there is flavor violation in
the (hermitian) EFT coefficients, $c_{i \ne j} \ne 0$. The coupling of the dark vector in this setup is then recovered by 
$\partial_\mu G \to - m_{\V}\V_\mu$, so is given by
\begin{align}
\label{L2EFT}
\Lag^{\text{EFT}}_\text{quadratic} \supset c_{ij} \frac{ v^\prime m_\V}{\Lambda^2} \V_\mu   \left( \overline{d}_{\rR i}  \gamma^\mu d_{\rR j} \right) \, .
 \end{align}
This analysis demonstrates that the interactions of dark vectors with
SM fields that are neutral under the $U(1)^\prime$
scale at least as $m_\V/\Lambda\times v^\prime/\Lambda$.
In particular they involve an additional factor of the $U(1)^\prime$ 
breaking scale as compared to Eq.~\eqref{eq:vector}.
Below we will confirm this expectation in an explicit UV
model, see Section~\ref{quadratic}.

\subsection{EFT Discussion for Linear scaling}
In order to have dark-vector couplings with a linear scaling in the $U(1)^\prime$
breaking scale, one necessarily has to charge SM fields under $U(1)^\prime$. In
this case the vector boson couples directly to the charged fields via
the dimension-four operator, e.g., for right-handed down quarks
\begin{align}
\label{Vcoupling}
\Lag_{\text{linear}} \supset g^\prime \V_\mu   \left( \overline{d}_{\rR}  \gamma^\mu X_d d_{\rR} \right) \, .
\end{align}
where $X_d$ is the diagonal $U(1)^\prime$ charge matrix.
To see how off-diagonal entries are generated, one has to rotate to the mass basis,
which is governed by the Yukawa couplings. It is clear that there is no flavor
violation if $X_d$ is universal, i.e., proportional to the identity matrix. 
If instead charges are non-universal, the mass matrix cannot be generic
at the renormalizable level,  i.e., it does not yield realistic
fermion masses without breaking $U(1)^\prime$.
Therefore, insertions of $S$ or $S^\dagger$ have to be
considered to obtain realistic fermion masses.

Restricting for simplicity to two generations, and charging only $d_{\rR1}$
with charge $+1$, i.e., $X_d = {\rm Diag}(1,0)$, $X_Q = X_H= 0$, 
the full Yukawa matrix requires higher-dimensional operators to have full rank
 \begin{align}
 \label{EFTlin}
 {\cal L}^{\text{EFT}}_\text{linear} \supset 
 - y_i \overline{Q}_i H d_{\rR2} 
 - z_i \frac{S^\dagger}{\Lambda} \overline{Q}_i H d_{\rR1}
 + {\rm h.c.}
 \end{align}
Thus the down-quark Yukawa matrix is given by
\begin{align}
 \label{Yuk}
 Y_d = \begin{pmatrix} z_1 \eps  & y_1 \\ z_2 \eps & y_2 \end{pmatrix}
 \quad\text{with}\quad
 \eps = \frac{v^\prime}{\sqrt{2} \Lambda}
\end{align}
We can ignore here the Goldstone in $S$, since
we already have the coupling of the gauge field in Eq.~\eqref{Vcoupling},
which leads to flavor-violating couplings with $\V$ after
rotating to the mass basis.
Nevertheless we can also reproduce this coupling with the same arguments as above:
in the gauge-less limit, we rescale $d_{\rR1} \to d_{\rR1} e^{i G/v^\prime}$, which removes $G$ from
the Yukawa sectors.
Ignoring chiral anomalies, this rescaling only affects the
kinetic terms, as it is a {\it local} $U(1)^\prime$ transformation
\begin{align}
\Lag_\text{linear}^\text{EFT} \supset i \overline{d}_{\rR1} \slashed{\partial} d_{\rR1} \to - \frac{\partial_\mu G}{v^\prime}  \overline{d}_{\rR1} \gamma^\mu d_{\rR1}   \, ,
\end{align}
which reproduces Eq.~\eqref{Vcoupling} upon $\partial_\mu G \to - m_{\V}V_\mu =
- g^\prime v^\prime V_\mu$.

We are left to diagonalize the Yukawa matrix $Y_d$ in Eq.~\eqref{Yuk}, or
rather $Y_d^\dagger Y_d$, in order to find the mixing matrix $V_d$ of
right-handed down quarks, defined as $V_Q^\dagger Y_d V_d = Y_d^{\rm diag}$.
In the limit when $\eps \ll 1$, one has
\begin{align}
 V_d \approx  \begin{pmatrix} 1  & z_2/y_2 \eps  \\ - z_2/y_2 \eps & 1  \end{pmatrix} \, ,
\end{align}
where we have set $y_1=0$ without loss of generality. Rotating the dark-vector
couplings in Eq.~\eqref{Vcoupling} to the mass basis defined by $d_{\rR} \to V_d
d_{\rR}$ gives finally
\begin{align}
\Lag_\text{linear} \supset g^\prime \V_\mu   \left( \overline{d}_{\rR}  \gamma^\mu V_d^\dagger X_d V_d d_{\rR} \right)  = g^\prime \V_\mu (V_d^*)_{1i}    (V_d)_{1j}    \left( \overline{d}_{\rR i}  \gamma^\mu d_{\rR j} \right) \, ,
\end{align}
so that indeed off-diagonal couplings are generated proportional to $g^\prime (V_d^*)_{11} (V_d)_{12} \sim g^\prime \eps \sim m_\V/\Lambda$.

To summarize, we have demonstrated that vector interactions of dark 
vectors can indeed be proportional to a single power of the $U(1)^\prime$ breaking, and thus
scale with the dark-vector mass as in Eq.~\eqref{Vcoupling}, if SM fermions have
non-universal $U(1)^\prime$ charges. This situation is quite generic
in models where SM Yukawa hierarchies are explained by non-anomalous abelian
flavor symmetries, for example simple $U(1)_F$ Froggatt-Nielsen
models~\cite{Froggatt:1978nt}, see e.g. Refs.~\cite{Allanach:2018vjg} for
examples of such models without extra heavy fermions to cancel anomalies. It is
well-known how to build UV completions for such models~\cite{Leurer:1992wg,
Calibbi:2012yj}, and below in Section~\ref{linear} we will present an
illustrative example.

\subsection{Explicit UV Model for Quadratic scaling\label{quadratic}}

We first construct an explicit renormalizable model for the scaling of vector
interactions in Eq.~\eqref{eq:vector} quadratic in the dark $U(1)^\prime$
breaking scale. We restrict the discussion for simplicity to the down-quark sector with two
generations. The field content is summarized in Table~\ref{quadtable}, and is
clearly anomaly-free.

\begin{table}[H]
\centering
\aboverulesep = 0mm
\belowrulesep = 0mm
\addtolength{\tabcolsep}{1pt}
\setlength{\extrarowheight}{1pt}
\begin{tabular}{l|ccc|ccc}
\toprule
& $\boldsymbol{Q_i}$ & $\boldsymbol{d_{\rR i}}$   
& $\boldsymbol{H}$   & $\boldsymbol{S}$   
& $\boldsymbol{\psi_{\rL i}}$    
& $\boldsymbol{\psi_{\rR i}}$\\
\midrule
$SU(2)_\rL$ & $2$   & $1$           & $2$   & $1$   & $1$               & $1$\\
$U(1)_Y$	& $1/6$ & $-1/3$        & $1/2$ & $0$   & $-1/3$            & $-1/3$\\
$U(1)'$     & $0$   & $0$           & $0$   & $1$   & $-1$              &$-1$\\
\bottomrule
\end{tabular}
\caption{
Field content of a renormalizable model featuring  quadratic scaling.
We restrict the discussion to the down-quark sector with two generations for SM quarks 
and heavy vector-like fermions $\psi_{\rL i}$, $\psi_{\rR i}$, with $i=1,2$ 
carrying $U(1)^\prime$ charges in addition to the scalar $S$.
\label{quadtable}}
\end{table}

The Lagrangian reads
\begin{align}
\Lag=\Lag_{\text{kinetic}} +\Lag_{\text{Yukawa}}+\Lag_{\text{scalar}}+\Lag_{\text{int-}\V}\, ,
\end{align}
with standard kinetic terms for all fields and
\begin{align}
\Lag_{\text{Yukawa}}&=- Y^d_{ij} \overline{Q}_i H d_{\rR j} - m_{\psi}\overline{\psi}_{\rL i} \psi_{\rR i}- \alpha_{ij} \overline{\psi}_{\rL i} d_{\rR j} S^{\dagger} +{\rm h.c.}\\
\Lag_{\text{int-}\V}&= - g'\V_{\mu} \left( \overline{\psi}_{\rL i} \gamma^{\mu}\psi_{\rL i} + \overline{\psi}_{\rR i} \gamma^{\mu}\psi_{\rR i}  \right) \, ,\\
\Lag_{\text{scalar}}&= m_H^2\vert H\vert^2+m_S^2\vert S\vert^2-\lambda_H\vert H\vert^4-\lambda_S\vert S\vert^4 - \lambda_{HS} \vert H\vert^2 \vert S \vert^2 \, .
\end{align}

For a suitable choice of parameters,
the last part in $\Lag_{\text{scalar}}$ gives a vacuum expectation value
to $S$, $\langle S \rangle = v^\prime/\sqrt{2}$, which sets the mass of the
dark vector boson to
\begin{equation}
  m_{\V}=v^\prime g^\prime \,,
\end{equation}
and induces a mixing between chiral quarks, $d_{\rR}$,  
and vector-like fermions, $\psi$, from the mixing term in $\Lag_{\text{Yukawa}}$.
In the limit of $m_\psi \gg v^\prime \gg v$ 
we can integrate out the vector-like fermions
using their equations of motion neglecting their kinetic terms
\begin{align}
\psi_{\rR i}  & = - \frac{\alpha_{ij}}{m_\psi}  d_{\rR j} S^{\dagger}\,,&\psi_{\rL i}&=0\, .
\end{align}
Plugging this back into kinetic terms and $\Lag_{\text{int}}$ lead to the EFT
\begin{align}
\label{res}
\Lag_{\text{int}}& \supset - g' \V_{\mu} \frac{S^\dagger S}{m_\psi^2} C_{ij} \left( \overline{d}_{\rR i}  \gamma^\mu d_{\rR j} \right) + \frac{S^\dagger S}{m_\psi^2} C_{ij} \left(i \overline{d}_{\rR i}  \slashed{\partial} d_{\rR j} \right) 
+  \frac{S i\partial_\mu S^\dagger}{m_\psi^2}C_{ij} \left( \overline{d}_{\rR i}  \gamma^\mu d_{\rR j} \right)  \,,
\end{align}
where $C_{ij} = ( \alpha^\dagger \alpha)_{ij} $. 
Next we integrate out the radial mode by substituting $S$ with 
the Goldstone parametrization in Eq.~\eqref{Gobo} and use the definition 
of the dark-vector mass to find
\begin{align}
\label{res2}
\Lag_{\text{int}}& \supset - \V_{\mu} \frac{m_{\V}v^\prime}{2 m_\psi^2} C_{ij} \left( \overline{d}_{\rR i}  \gamma^\mu d_{\rR j} \right) + \frac{(v^\prime)^2}{2 m_\psi^2} C_{ij} \left(i \overline{d}_{\rR i}  \slashed{\partial} d_{\rR j} \right) + \partial_\mu G \frac{v^\prime }{2 m_\psi^2} C_{ij}  \left( \overline{d}_{\rR i}  \gamma^\mu d_{\rR j} \right)  \, ,
\end{align}
recovering the gauge-invariant\footnote{
In our conventions $\V_\mu \to \V_\mu + \partial_\mu \beta/g^\prime$, 
$S \to \exp (i \beta ) S$, $G \to G + \beta v^\prime$, 
$\psi \to \exp (-i \beta ) \psi $.} combination $\V_\mu -
\partial_\mu G/m_\V$. 
Without loss of generality we can assume that $Y^d_{ij}$
is diagonal, so that we are already in the mass basis. Nevertheless, we do
need to re-diagonalize the kinetic terms due to the second term in Eq.~\eqref{res2} induced in the EFT.
In the limit of $v^\prime \ll m_\psi$ this is readily 
achieved by the rescaling $d_{\rR i} \to d_{\rR i} - (v^\prime)^2/(4 m_\psi^2)
C_{ij} d_{\rR j}$.
This leads to additional small corrections of 
  ${\cal O}(1/m_\psi^4)$ to the final dark-vector couplings, which can be
neglected, such that the leading couplings from the first term in Eq.~\eqref{res2}
remain
\begin{align}
  \Lag_{\text{quadratic}} =  - \frac{m_{\V}v^\prime}{2 m_\psi^2} C_{ij} \V_{\mu} \left( \overline{d}_{\rR i}  \gamma^\mu d_{\rR j} \right)  \, .
\end{align}
These couplings are indeed quadratic in $v^\prime$ and are in general 
flavor violating, $C_{i \ne j} \ne 0$. This matches to the EFT term in Eq.~\eqref{L2EFT} upon
identifying $C_{ij}/m_\psi^2 = - 2 c_{ij}/\Lambda^2$.

\subsection{Explicit UV Model for Linear scaling\label{linear}}
We now construct an explicit renormalizable model for the minimal scaling of
vector interactions in Eq.~\eqref{eq:vector} proportional to a single power of
the dark-vector mass. These types of models are  motivated by scenarios 
addressing the SM flavor puzzle with non-anomalous
abelian horizontal symmetries, see e.g. Ref.~\cite{Allanach:2018vjg}.
We restrict the discussion for simplicity to the down-quark sector with two generations. 
The field content is summarized in Table~\ref{lineartable}, and is 
{\itshape not} anomaly-free.
However, we can always introduce further suitably charged chiral fermions 
in the right-handed up- and charged-lepton sector in order 
to cancel color and electromagnetic anomalies, respectively.
Note that $\psi_{\rR}$ and $d_{\rR2}$ carry the same quantum
numbers.

\begin{table}[H]
\centering
\aboverulesep = 0mm
\belowrulesep = 0mm
\addtolength{\tabcolsep}{1pt}
\setlength{\extrarowheight}{1pt}
\begin{tabular}{l|cccc|ccc}
\toprule
& $\boldsymbol{Q_i}$ & $\boldsymbol{d_{\rR 1}}$   & $\boldsymbol{d_{\rR 2}}$   
& $\boldsymbol{H}$   & $\boldsymbol{S}$   
& $\boldsymbol{\psi_{\rL}}$    
& $\boldsymbol{\psi_{\rR}}$\\
\midrule
$SU(2)_\rL$	& $2$   & $1$       & $1$       & $2$   &$1$  & $1$     & $1$ \\
$U(1)_Y$	& $1/6$ & $-1/3$    & $-1/3$    & $1/2$ &$0$  & $-1/3$  & $-1/3$ \\
$U(1)'$     & $0$   & $1$       & $0$       & $0$   &$1$  & $0$     & $0$\\
\bottomrule
\end{tabular}
\caption{
Field content of a renormalizable model featuring  linear scaling.
We restrict the discussion to the down-quark sector with two generations for SM quarks and 
one family of heavy vector-like fermions $\psi_{L}$, $\psi_{R}$ 
uncharged under $U(1)^\prime$. \label{lineartable}}
\end{table}

The Lagrangian reads
\begin{align}
\Lag&=\Lag_{\text{kinetic}} +\Lag_{\text{Yukawa}}+\Lag_{\text{scalar}}+\Lag_{\text{int-}\V}\, ,
\end{align}
with standard kinetic terms for all fields and
\begin{align}
\Lag_{\text{Yukawa}}&=- y_{i} \overline{Q}_i H d_{\rR2}  - z_{i} \overline{Q}_i H \psi_{\rR} 
- m_{\psi}\overline{\psi}_{L} \psi_{R}- \alpha  \overline{\psi}_{L} d_{\rR1} S^{\dagger} 
+{\rm h.c.}\\
\Lag_{\text{int-}\V}&= g'\V_{\mu}   \overline{d}_{\rR1}  \gamma^\mu d_{\rR1} \, ,\\
\Lag_{\text{scalar}}&= m_H^2\vert H\vert^2+m_S^2\vert S\vert^2-\lambda_H\vert H\vert^4-\lambda_S\vert S\vert^4 - \lambda_{HS} \vert H\vert^2 \vert S \vert^2  \, ,
\end{align}
where we have simply defined $\psi_{\rR}$ to be that field having a mass term
with $\psi_{\rL}$. 
This already gives Eq.~\eqref{Vcoupling} and the first term
in Eq.~\eqref{EFTlin} from the EFT discussion, so it only remains to show 
that integrating out $\psi_{\rL}, \psi_{\rR}$ induces the second term 
in Eq.~\eqref{EFTlin}. 
The equations of motion  for the heavy fermions read, neglecting kinetic terms
\begin{align}
\overline{\psi}_{\rL} & = - \frac{z_i}{m_\psi} \overline{Q}_i H \, , & \psi_{\rR} &  = - \frac{\alpha}{m_\psi} d_{\rR1} S^\dagger \,,
\end{align}
and, therefore, the resulting EFT Lagrangian term is
\begin{align}
  \Lag_{\text{linear}} \supset \frac{z_i \alpha}{m_\psi} \overline{Q}_i H d_{\rR1} S^\dagger \,.
\end{align}
This indeed reproduces Eq.~\eqref{EFTlin}  with the identification of the 
UV scale as $\Lambda = - m_\psi/\alpha$. The remaining calculation follows the EFT
discussion, which shows that in these type of  UV models 
the flavor violating couplings to $\V$ scale indeed linearly with $m_{\V}/\Lambda$.

\section{Renormalization Group Equations\label{sec:rges}}

In this appendix we collect our results for the renormalization group equations
relevant for the $SU(2)_L\times U(1)_Y$ interactions of the LDV with SM quarks 
as discussed in Section~\ref{sec:MFV}. Since in the current work we focused on 
the case of the UV universal scenario, in which flavor-violation originates only from the SM
CKM matrix, we present here only the one-loop RGEs proportional to Yukawa couplings.
However, in what follows the matrices $C^{\rD}_{u}$, $C^{\rD}_{d}$ $C^{\rV}_{Q}$, $C^{\rV}_{u}$, and $C^{\rD}_{d}$ 
are generic matrices in flavor space, i.e., we have not assumed any alignment with the 
SM Yukawas. The relevant terms in the Lagrangian are the SM Yukawa interaction, the dipole, and the
vector interactions with the LDV. They respectively read:
\begin{align}
\Lag_{\text{Yukawa}} &= -\overline{Q}\, Y_u \widetilde{H} u_{\rR} - \overline{Q}\, Y_d H d_{\rR} + \mathrm{h.c.}\,,\\
\Lag_{\text{Dipole}} &=  \frac{1}{\Lambda^2}\V_{\mu\nu}\left(\overline{Q} C^{\rD}_{u} \sigma^{\mu\nu} \widetilde{H} u_{\rR}
                                                       + \overline{Q} C^{\rD}_{d} \sigma^{\mu\nu} H d_{\rR}
                                                       + \mathrm{h.c.}\right) \,,\\
\Lag_{\text{Vector}} &= \V_\mu\left(\overline{Q}       C_{Q}^V \gamma^\mu Q
                                  + \overline{u}_{\rR} C_{u}^V \gamma^\mu u_{\rR}
                                  + \overline{d}_{\rR} C_{d}^V \gamma^\mu d_{\rR}\right) \, .
\end{align}

The one-loop RGEs for Yukawa running read~\cite{Jenkins:2013wua}
\begin{equation}
\begin{split}
16\pi^2\frac{\mathrm{d}Y_{u}}{\mathrm{d} \ln\mu}&=\frac{3}{2}\left(Y_uY_u^{\dagger}Y_u-Y_dY_d^{\dagger}Y_u\right)+n_c\text{Tr}\left[Y_uY_u^{\dagger}+Y_dY_d^{\dagger}\right]Y_u\, ,\\
16\pi^2\frac{\mathrm{d}Y_{d}}{\mathrm{d} \ln\mu}&=\frac{3}{2}\left(Y_dY_d^{\dagger}Y_d-Y_uY_u^{\dagger}Y_d\right)+n_c\text{Tr}\left[Y_uY_u^{\dagger}+Y_dY_d^{\dagger}\right]Y_d\,,
\end{split}
\end{equation}
with $n_c=3$ denoting the number of colors.
The one-loop running of the Yukawas is relevant for the dipole analysis because the RG-evolved Yukawas 
contribute to the flavor-violating couplings upon rotation  to the quark mass-eigenstates 
at the EW scale~\cite{Aebischer:2020lsx}.

For the one-loop RGE of the dipole couplings proportional to the SM Yukawas
we find
\begin{equation}
	\label{rgedipole}
	\begin{split}
		16\pi^2\frac{\mathrm{d} C^{\rD}_{u}}{\mathrm{d} \ln\mu}
		&= \frac{5}{2} Y_u Y_u^\dagger C^{\rD}_{u}-\frac{3}{2} Y_d Y_d^\dagger  C^{\rD}_{u}
		- C^{\rD}_{d} Y_d^\dagger Y_u
		+ 2C^{\rD}_{u} Y_u^\dagger Y_u \\
		&\quad
		+n_c \mathrm{Tr}\left[ Y_u Y^\dagger_u + Y_d Y^\dagger_d  \right]C^{\rD}_{u}\,, \\
		16\pi^2\frac{\mathrm{d} C^{\rD}_{d}}{\mathrm{d} \ln\mu}
		&= \frac{5}{2} Y_d Y_d^\dagger C^{\rD}_{d}-\frac{3}{2} Y_u Y_u^\dagger  C^{\rD}_{d} 
		- C^{\rD}_{u} Y_u^\dagger Y_d
		+ 2C^{\rD}_{d} Y_d^\dagger Y_d \\
		&\quad
		+n_c\mathrm{Tr}\left[Y_u Y^\dagger_u + Y_d Y^\dagger_d  \right]C^{\rD}_{d}\, .
	\end{split}
\end{equation}

For the one-loop RGE of the vector couplings proportional to the SM Yukawas
we find
\begin{equation}
\label{rge}
\begin{split}
16\pi^2\frac{\mathrm{d} C^{\rV}_{Q}}{\mathrm{d} \ln\mu}
&= -Y_u C^{\rV}_{u} Y_u^\dagger - Y_d C^{\rV}_{d} Y_d^\dagger
+ \frac{1}{2}\left(Y_u Y_u^\dagger + Y_d Y_d^\dagger\right) C^{\rV}_{Q}
+ \frac{1}{2} C^{\rV}_{Q} \left(Y_u Y_u^\dagger + Y_d Y_d^\dagger\right) \,, \\
16\pi^2\frac{\mathrm{d} C^{\rV}_{u}}{\mathrm{d} \ln\mu}
&= -2 Y_u^\dagger C^{\rV}_{Q} Y_u 
   + Y_u^\dagger Y_u C^{\rV}_{u}
   + C^{\rV}_{u} Y_u^\dagger Y_u \,, \\
16\pi^2\frac{\mathrm{d} C^{\rV}_{d}}{\mathrm{d} \ln\mu}
&= -2 Y_d^\dagger C^{\rV}_{Q} Y_d
   + Y_d^\dagger Y_d C^{\rV}_{d}
   + C^{\rV}_{d} Y_d^\dagger Y_d \,.
\end{split}
\end{equation}

\section{Recast of Experimental Limits\label{sec:recast}}

\begin{figure}[H]
	\centering
	\includegraphics[width=0.47\textwidth]{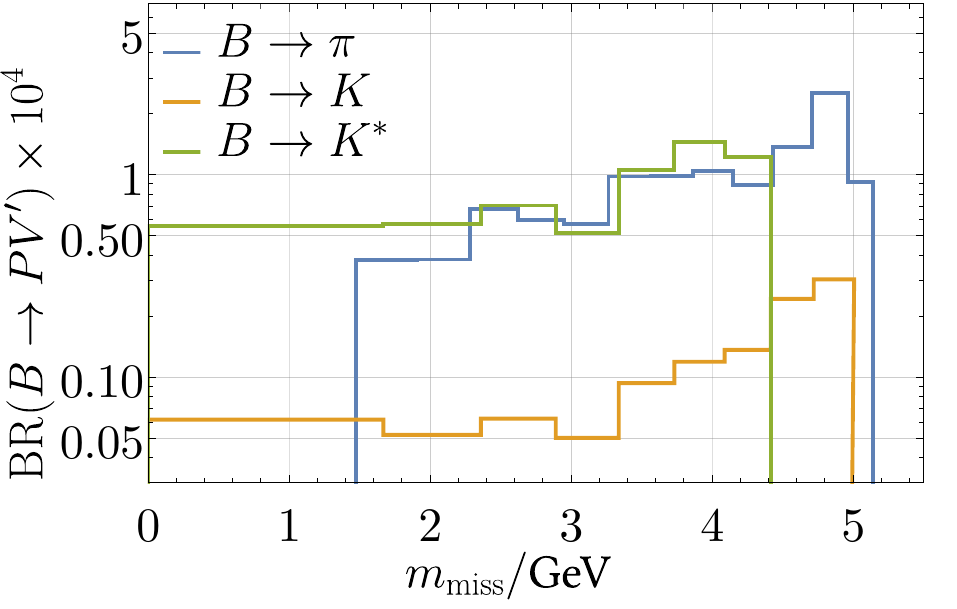}
	\caption{Upper 
      $95\,\%$ CL$_\text{s}$ limits on the two-body branching ratios
      $B\rightarrow K/K^*/\pi + \V$, as a function of the missing
      mass $m_{\rm miss} =  \sqrt{q^2}$, obtained by recasting the experimental
      three-body searches at BaBar~\cite{BaBar:2004xlo,BaBar:2013npw}, see text
      for details.\label{fig:2-body recast}}
\end{figure}
Experimental collaborations often provide only limits on the $P\rightarrow
P'+\mathrm{invisible}$ branching ratios in terms of the three body decay
$P\rightarrow P'\nu\overline{\nu}$, as a function of the squared invariant mass
of the di-neutrino system $q^2$. In order to get the experimental limits on the
two body decays $P\rightarrow P'V$, we use the event count $n_B$ per $q^2$-bin
information, if provided by the experimental collaborations. Only the BaBar
experiment~\cite{BaBar:2004xlo,BaBar:2013npw} provides all information needed
to perform a recast for two-body decays $B\rightarrow K/K^*/\pi + V$. For $B\to
K^{(*)}$ and dark-vector masses $m_{\V}<3\GeV$, we use the sophisticated
recast of Ref.~\cite{Altmannshofer:2023hkn}, otherwise we estimate upper limits
on the Wilson coefficients in terms of the CL$_\text{s}$ method as explained below.

For a given Wilson coefficient $C$, the number of signal events $s$ in a $q^2$-bin
$i$ is given as
\begin{align}
	s = \mathrm{BR}^i_{P\rightarrow P'}(C) \times N_\mathrm{tot} \times \epsilon_i \,,
\end{align}
where $N_\mathrm{tot}$ is the total number of $P$ mesons and $\epsilon_i$ the
efficiency associated to bin $i$. Further, $\mathrm{BR}^i_{P\rightarrow P'}(C)$
denotes the branching ratio of $P\rightarrow P'$ within the $q^2$-bin $i$. The
$s+b$ likelihood is then given as a Poisson distribution in the number of
signal plus background events. The efficiency $\epsilon_i$ and total number of
$P$ mesons $N_\mathrm{tot}$ are included as global observables associated to
auxiliary measurements. The uncertainty on the signal, assumed to be Gaussian,
is given by the NP theoretical prediction and is dominated by the form-factor
uncertainty. The systematic uncertainty on the background is implemented as a
Gaussian distribution. With this in mind, we denote the likelihood as $\mathcal
L(x | C, \nu)$ with $x$ being the outcome, i.e., the observed data, $C$ the
parameter of interest, i.e., the Wilson coefficient, and $\nu$ the nuisance
parameters. As a test statistics $t_C$, we choose a one-sided profile
likelihood.  Note that the parameter of interest is actually $|C|^2$ since the
branching ratio only depends on $|C|^2$ as we only consider one coupling at a
time. The $p$-value $p_C$ of the $s+b$ hypothesis for a given value of the
Wilson coefficient $C$ is then given by
\begin{align}\label{eq:pvalue}
  p_C = \int_{t_C^\mathrm{obs}}^\infty f(t_C | C, \hat{\hat{\nu}}(C) ) \,\mathrm{d}t_C \,,
\end{align}
where $t_C^\mathrm{obs}$ denotes the value of the test statistics for the
observed data, $f$ denotes the pdf of the test statistics $t_C$, and
$\hat{\hat{\nu}}(C)$ are the values of the nuisance parameter that maximise the
likelihood for a given $C$. The $\alpha\,\%$ CL$_\text{s}$ limit on the Wilson
coefficient is then given by the value of $C$ such that
\begin{align}
	\frac{p_C}{p_0} = 1-\frac{\alpha}{100} \,,
\end{align}
where $p_0$ denotes the $p$-value of the background only hypothesis. In order
to evaluate Eq.~\eqref{eq:pvalue}, one needs the pdf $f$ of the test
statistics $t_C$ for which we use the \texttt{ROOT} toolkit RooStats in order
to sample the distribution by means of a Monte Carlo method.

Taking the $\mathrm{BR}(P\rightarrow P'V)$ as a parameter of interest instead
of the Wilson coefficient $C$, we can determine a model independent limit
$\mathrm{BR}_\mathrm{exp}(P\rightarrow P'V)$ on the two body branching ratios,
see Figure~\ref{fig:2-body recast}.

\section{Limits in the $\rL/\rR$ Basis\label{sec:LRbounds}}
In this appendix we present bounds on the couplings in the $\rL/\rR$ basis
$\{\CC{\rD\rR}{ij}$, $\CC{\rD\rL}{ij}$, $\CC{\rV\rR}{ij}$, $\CC{\rV\rL}{ij}\}$, which are
obtained from the limits in the $\rV/\rA$ basis (discussed in Section \ref{sec:quarkflavor} 
and \ref{sec:leptonflavor})) using Eq.~\eqref{eq:LRbasis}.
As the decay rates are symmetric with respect to
$\CC{\rL}{} \leftrightarrow \CC{\rR}{}$ the bounds on both couplings  are the same.

\subsection{Quark Dipole Interactions}
\begin{figure}[H]
	\begin{center}
		\includegraphics[width=0.47\textwidth]{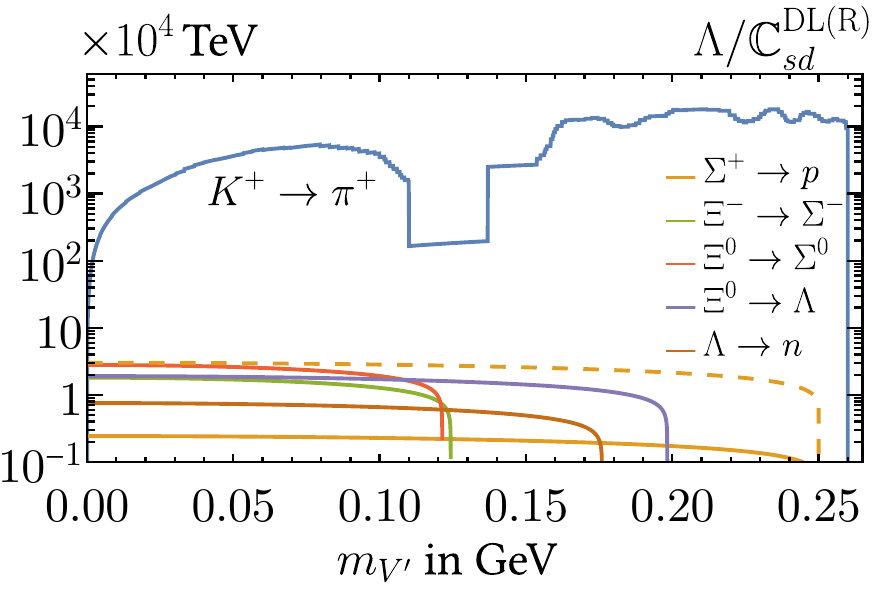}
		\qquad
		\includegraphics[width=0.47\textwidth]{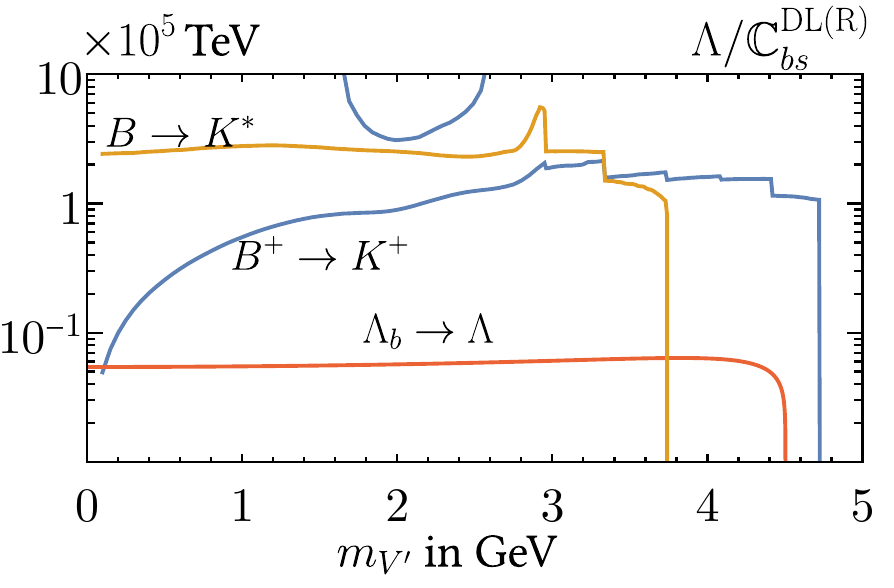}
		\\[1.5em]
		\includegraphics[width=0.47\textwidth]{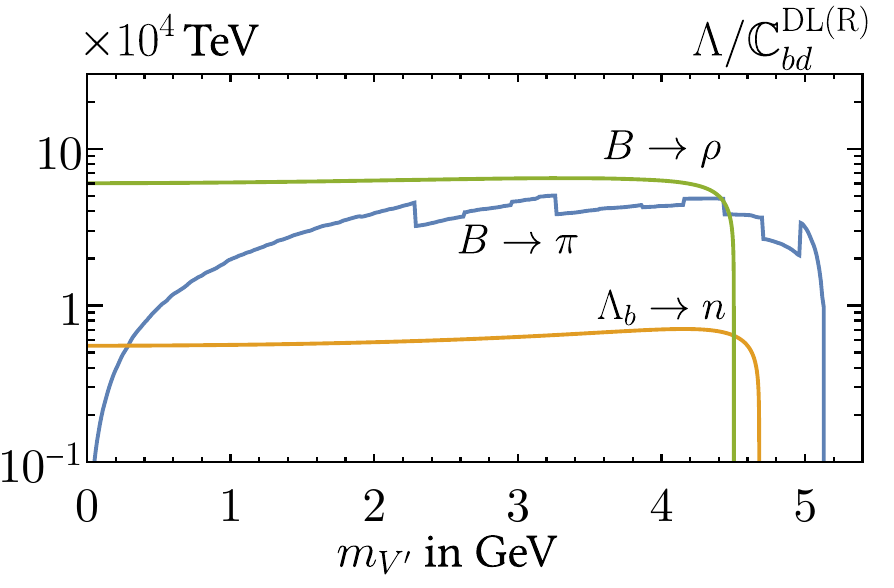}
		\qquad
		\includegraphics[width=0.47\textwidth]{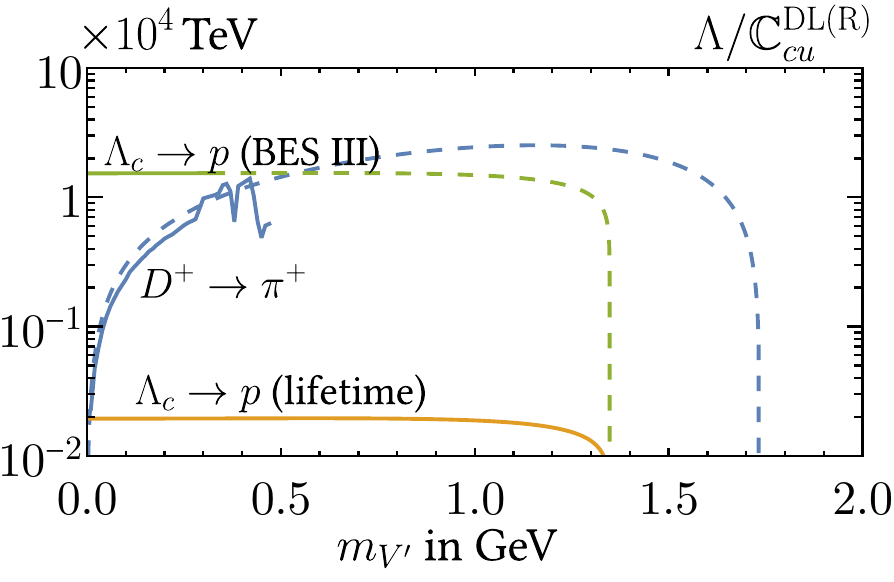}
		\\[1.5em]
	\end{center}
	\vspace{-1em}
	\caption{
		Upper limits on quark-flavor violating dipole couplings $\Lambda/|\CC{\rD \rL }{ij}|$, for $s \to d, b\to s, b \to d$ and $c \to u$ transitions. Bounds on $\Lambda/|\CC{\rD \rR }{ij}|$ are identical.}
\end{figure}
\subsection{Quark Vector Interactions}
\begin{figure}[H]
	\begin{center}
		\includegraphics[width=0.47\textwidth]{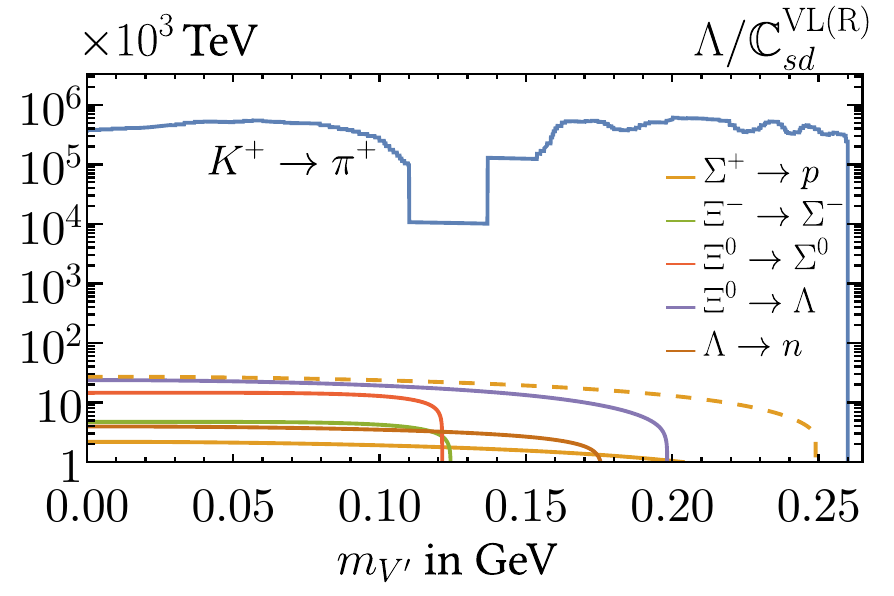}
		\qquad
		\includegraphics[width=0.47\textwidth]{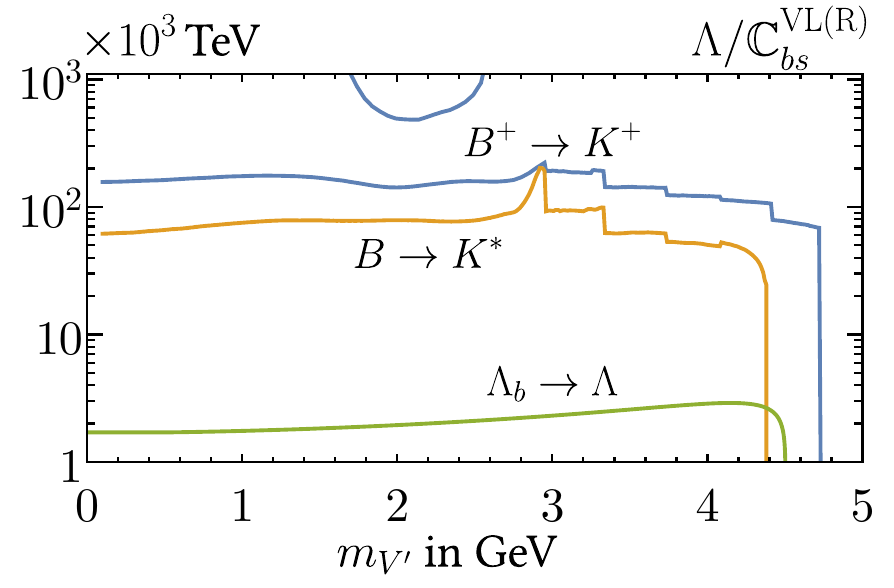}
		\\[1.5em]
		\includegraphics[width=0.47\textwidth]{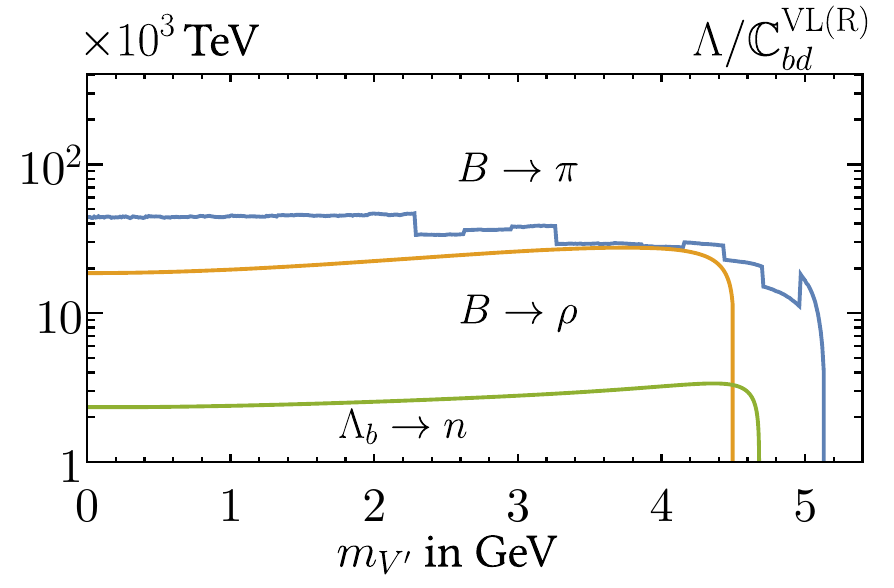}
		\qquad
		\includegraphics[width=0.47\textwidth]{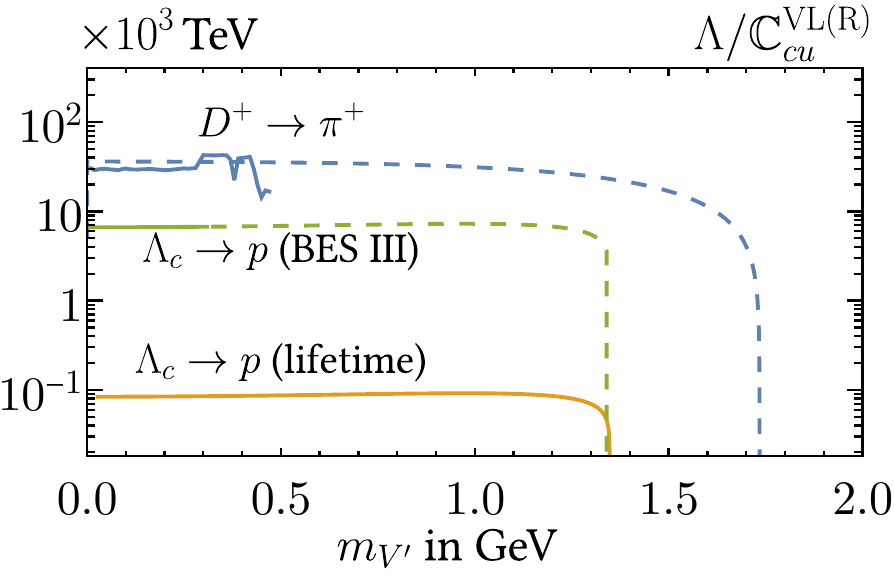}
		\\[1.5em]
	\end{center}
	\vspace{-1em}
	\caption{
		Upper limits on quark-flavor violating vector couplings $\Lambda/|\CC{\rV \rL }{ij}|$, for $s \to d, b\to s, b \to d$ and $c \to u$ transitions. Bounds on $\Lambda/|\CC{\rV \rR }{ij}|$ are identical.}

\end{figure}

\begin{figure}[H]
	\centering
	\includegraphics[width=0.47\textwidth]{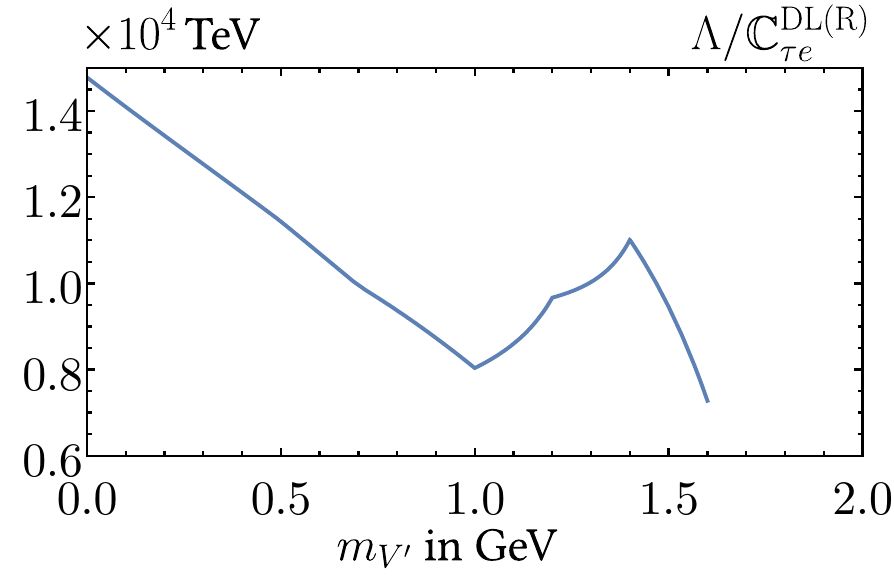}
	\includegraphics[width=0.47\textwidth]{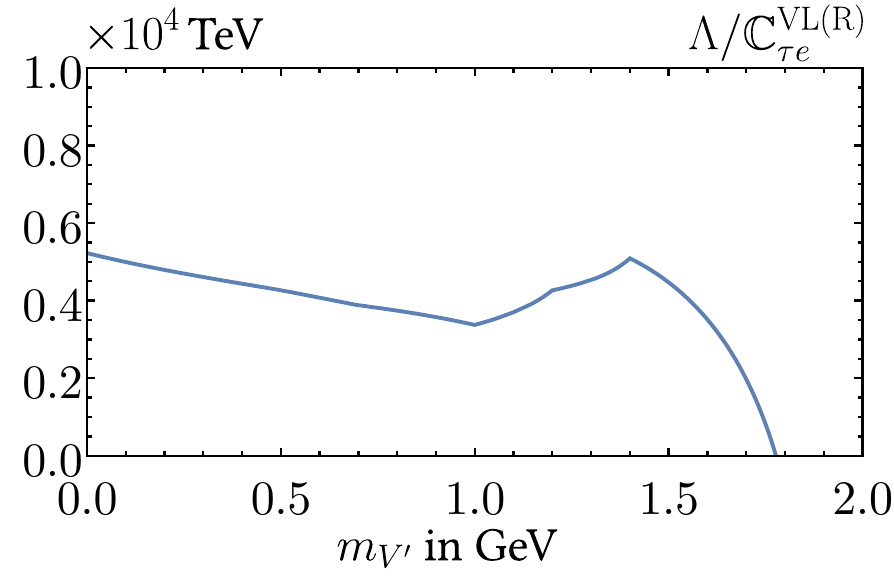}
	\includegraphics[width=0.47\textwidth]{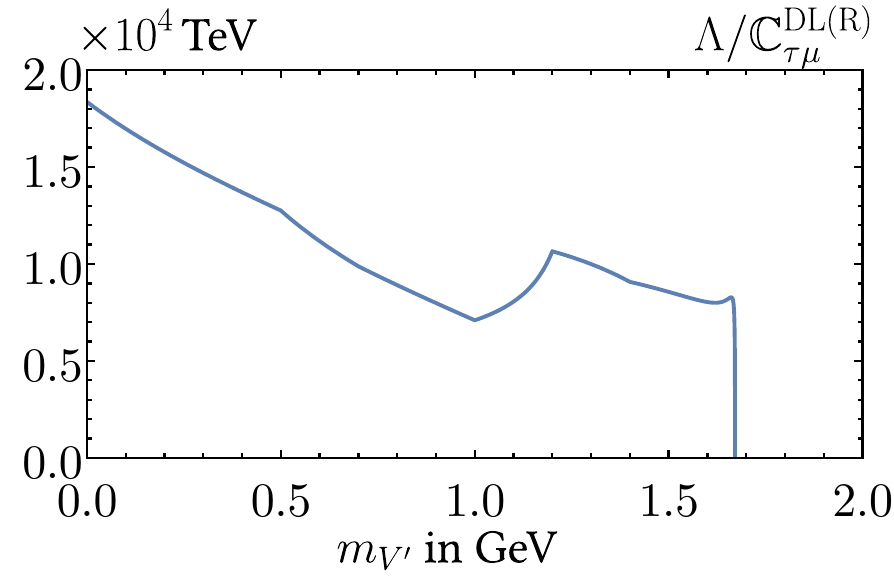}
	\includegraphics[width=0.47\textwidth]{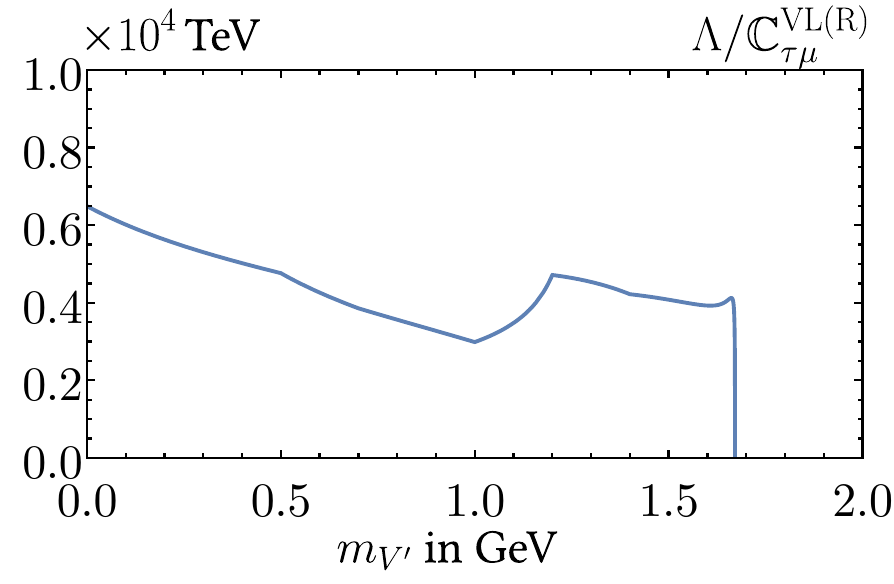}
	\caption{%
	 {\it Upper panel:} Lower limits on the dipole (left panel) and vector (right panel) couplings for $\tau \to e$ transitions $\Lambda/|\CC{\rD \rL(\rR) }{\tau e}|$, $\Lambda/|\CC{\rV \rL(\rR) }{\tau e}|$ from Belle~II~\cite{Belle-II:2022heu}.
	 {\it Lower panel:} same for $\tau \to \mu$ transitions  $\Lambda/|\CC{\rD \rL(\rR) }{\tau
\mu}|$, $\Lambda/|\CC{\rV \rL(\rR) }{\tau \mu}|$.}
\end{figure}

\section{Two-body decays to Light Dark Vectors\label{sec:2bodydecays}}

\begin{figure}[H]
	\centering
    \includegraphics[]{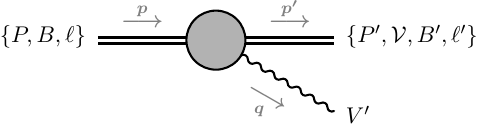}
	\caption{Two-body decays $\{P,B,\ell\}\to\{P^{\prime},\mathcal{V},B^{\prime},\ell^{\prime}\}+\V$. 
    The blob represents the  non-perturbative QCD effects for the hadronic decays.
  \label{fig:twobodykinematics}}
\end{figure}

In this appendix we present the full expressions for the two-body decays to a LDV that
enter our analysis, namely
\begin{itemize}
  \item $P\to P^{\prime}+\V$: pseudoscalar meson to pseudoscalar meson and LDV,
  \item $P\to \mathcal{V}+\V$: pseudoscalar meson to vector meson and LDV,
  \item $B\to B^{\prime}+\V$: baryon to baryon and LDV,
  \item $\ell\to \ell^{\prime}\V$: lepton to lepton and LDV.
\end{itemize}
For the hadronic processes illustrative Feynman diagrams are shown in 
Figure~\ref{fig:twobodydecays}, while throughout this appendix we define the two-body
kinematics for all decays as in Figure~\ref{fig:twobodykinematics}, namely as
\begin{equation}
  \text{SM} (p)\to \text{SM}^\prime(p') + \V(q)
\end{equation}
with $q=p-p^\prime$ and $q^2=(p-p^\prime)^2=m_{\V}^2$.
In the next subsection we collect the parametrization of all the relevant form factors
for the hadronic processes considered, and in the subsequent subsections 
we present the expressions for the rates. The numerical values for the form-factors are always taken from the most recent
work referenced.

\subsection{Form Factors\label{sec:formfactors}}
\subsubsection*{$\boldsymbol{P \to P^\prime + \V}$}
For these decays the hadronic matrix elements for the 
vector and axial-vector currents read~\cite{Gubernari:2018wyi}
\begin{equation}
\begin{split}
\langle P^\prime (p^\prime) | \overline{q}^\prime \gamma^\mu q | P(p) \rangle &= (p + p^\prime)^\mu f_+^{P P^\prime} (q^2) +  (p - p^\prime)^\mu f_-^{P P^\prime} (q^2)\, , \\
\langle P^\prime (p^\prime) | \overline{q}^\prime \gamma_5\gamma^\mu q | P(p) \rangle &= 0\, .
\end{split}
\end{equation}
The corresponding matrix elements for tensor and pseudo-tensor currents read~\cite{Gubernari:2018wyi}
\begin{equation}
\label{FFsigma}
\begin{split}
\langle P^\prime (p^\prime) | \overline{q}^\prime \sigma_{\mu \nu} q | P(p) \rangle & = \frac{2}{m_P + m_{P^\prime}} \left( p^\prime_\mu p_\nu - p^\prime_\nu p_\mu \right) f_\rT^{P P^\prime} (q^2) \, , \\
\langle P^\prime (p^\prime) | \overline{q}^\prime \sigma^{\mu \nu} \gamma_5 q | P(p) \rangle & = \frac{2i}{m_P + m_{P^\prime}} \eps^{\mu \nu \rho \sigma} p^\prime_\rho p_\sigma f_\rT^{P P^\prime} (q^2) \,,
\end{split}
\end{equation}
where here and throughout we use the $\epsilon_{0123}= - \epsilon^{0123} = +1$ convention for the Levi-Civita 
tensor.

\subsubsection*{$\boldsymbol{P \to \mathcal{V} +\V}$}
For the pseudoscalar decays to two vectors with $\mathcal{V}$ 
denoting the vector-meson, the hadronic matrix element for the vector and 
axial-vector currents are parametrized as~\cite{Gubernari:2018wyi}
\begin{align}
\langle \mathcal{V} (p^\prime,\lambda)  | \overline{q}^\prime \gamma^\mu\left(1\mp\gamma_5\right)  q | P(p) \rangle = 
P_1^{\mu}\mathcal{V}_1(q^2)\pm P_2^{\mu}\mathcal{V}_2(q^2)\pm P_3^{\mu}\mathcal{V}_3(q^2)\pm P_P^{\mu}\mathcal{V}_P(q^2)\, ,
\end{align}
where $\lambda$ denotes the polarization of $\mathcal V$. The kinematic functions read
\begin{align}
P_P^{\mu}&=i(\epsilon^{*}\!\cdot\!q) q^{\mu}\,,& 
P_1^{\mu}&=2\epsilon^{\mu}_{~~\alpha\beta\gamma}\epsilon^{*\alpha}p^{\prime\beta}q^{\gamma} \, ,\nonumber\\
P_2^{\mu}&=i\left[\left(m_P^2-m_{{\mathcal{V}}}^2\right)\epsilon^{*\mu}-(\epsilon^{*}\!\cdot\!q)\left(p^{\prime}+p\right)^{\mu}\right]\,,&
P_3^{\mu}&=i(\epsilon^{*}\!\cdot\!q) \big[q^{\mu}-\frac{q^2}{m_P^2-m_{\mathcal{V}}^2}(p^{\prime}+p)^{\mu}\big]\,,
\end{align}
where $\epsilon^*_\mu=\epsilon_\mu^*(p',\lambda)$ denotes the polarization vector of the 
outgoing $\mathcal V$. The scalar form factors can be further parametrized as
\begin{align}
\begin{split}
\mathcal{V}_P(q^2)&=\frac{-2m_{\mathcal{V}}}{q^2}A_0(m_{\V}^2)\,,\qquad
\mathcal{V}_1(q^2) =\frac{-V(q^2)}{m_P+m_{\mathcal{V}}}\,,\qquad
\mathcal{V}_2(q^2) =\frac{-A_1(q^2)}{m_P-m_{\mathcal{V}}}\,,\\
\mathcal{V}_3(q^2)&=\frac{m_P+m_{\mathcal{V}}}{q^2}A_1(q^2)-\frac{m_P-m_{\mathcal V}}{q^2}A_2(q^2)
                    \equiv\frac{2m_{\mathcal{V}}}{q^2}A_3(q^2)\,,
\end{split}
\end{align}
with $A_3(0)=A_0(0)$, which ensures finite matrix elements at $q^2=0$, i.e., for massless LDV.

The corresponding matrix elements for tensor and pseudo-tensor currents read~\cite{Gubernari:2018wyi}
\begin{equation}
\begin{split}
\langle \mathcal{V} (p^\prime, \lambda) | \overline{q}^\prime \sigma^{\mu \nu} q | P(p) \rangle & = - i \eps^*_\alpha  T^{\alpha \mu \nu } (q^2) \, , \\
\langle \mathcal{V} (p^\prime, \lambda) | \overline{q}^\prime \sigma_{\mu \nu} \gamma_5 q | P(p) \rangle & = \frac{1}{2} \eps^*_\alpha \eps_{\mu \nu \rho \sigma} T^{\alpha \rho \sigma }(q^2)  \, ,
\end{split}
\end{equation}
where
\begin{equation}
\begin{split}
T^{\alpha \mu \nu }(q^2) &= 
\eps^{\alpha \mu \nu \beta}  \left[ \left( p_\beta + p^\prime_\beta - q_\beta \frac{m_P^2 - m_{\mathcal{V}}^2}{q^2} \right) T_1^{P\mathcal{V}} (q^2)  
                                  + q_\beta \frac{m_P^2 - m_{\mathcal{V}}^2}{q^2} T_2^{P\mathcal{V}}(q^2) \right]\\
& + \frac{2 p^\alpha}{q^2}  \eps^{\mu\nu\beta\gamma} p_\beta p^\prime_\gamma \left( T_2^{P\mathcal{V}}  (q^2) - T_1^{P\mathcal{V}}(q^2) 
  + \frac{q^2}{m_P^2 - m_{\mathcal{V}}^2} T_3^{P\mathcal{V}}(q^2) \right)  \, .
\end{split}
\end{equation}
For vanishing momentum transfer $q^2 = 0$, i.e., massless LDV, the scalar form-factors satisfy
\begin{align}
T_1^{P\mathcal{V}}(0) & = T_2^{P\mathcal{V}}(0)  \equiv T \, ,
\end{align}
while the contribution proportional to  $T_3(0)$ vanishes.

\subsubsection*{$\boldsymbol{B \to B^\prime+\V}$}
For the baryon decays the matrix elements for vector and axial-vector currents
are parametrized by~\cite{Detmold:2016pkz, Detmold:2015aaa, Meinel:2017ggx}
\begin{equation}
\label{BFF}
\begin{split}
\langle B^\prime (p^\prime) | \overline{q}^\prime \gamma_{\mu} q | B(p) \rangle &= 
\overline{u}_{B^\prime}(p^\prime)\left(f_1(q^2)\gamma_{\mu}-i\frac{f_2(q^2)}{m_B}\sigma_{\mu\nu}q^{\nu}+\frac{f_3(q^2)}{m_B}q_{\mu}\right) {u}_{B}(p)\, ,\\
\langle B^\prime (p^\prime) | \overline{q}^\prime \gamma_{\mu}\gamma_5 q | B(p) \rangle &= 
\overline{u}_{B^\prime}(p^\prime)\left(g_1(q^2)\gamma_{\mu}-i\frac{g_2(q^2)}{m_B}\sigma_{\mu\nu}q^{\nu}+\frac{g_3(q^2)}{m_B}q_{\mu}\right) \gamma_5{u}_{B}(p)\,,
\end{split}
\end{equation}
with $u_B(p)$ and $u_{B'}(p')$ the spinor functions for $B$ and $B'$ respectively. For $\Lambda$ decays the values of the form factors are taken from~\cite{Detmold:2016pkz, Detmold:2015aaa, Meinel:2017ggx}, while for hyperon decays they are taken from~\cite{Ledwig:2014rfa,
	Cabibbo:2003cu, Gaillard:1984ny}.

The corresponding matrix elements for tensor and pseudo-tensor currents have the form
\cite{Camalich:2020wac, Gupta:2018lvp}
\begin{equation}\label{eq:ffsdipolehyperon}
\begin{split}
\langle B^\prime (p^\prime) | \overline{q}^\prime \sigma^{\mu \nu} q | B(p) \rangle &= 
g_\rT^{B B^\prime} \overline{u}_{B^\prime} (p^\prime) \sigma^{\mu \nu}  u_B(p) \, , \\
\langle B^\prime (p^\prime) | \overline{q}^\prime \sigma_{\mu \nu} \gamma_5 q | B(p) \rangle &= 
\frac{i}{2} g_\rT^{B B^\prime} \eps_{\mu \nu \alpha\beta} \overline{u}_{B^\prime} (p^\prime) \sigma^{\alpha\beta}  u_B(p) \,,
\end{split}
\end{equation}
which is an approximation  valid for $m_{\V}^2=0$, which we use for the hyperon decays.
For the baryon $\Lambda_b\to \Lambda$, $\Lambda_b\to n$, and
$\Lambda_c\to p$ decays we use the available full parametrization, given by
\cite{Detmold:2016pkz, Meinel:2017ggx}
\begin{equation}
  \label{eq:FFBBdipoleq}
  \begin{split}
\langle B^\prime (p^\prime) | \overline{q}^\prime i\sigma^{\mu\nu}q_{\nu} q | B(p) \rangle &= 
\overline{u}_{B^\prime}(p^\prime)\left(\frac{f_1^{\rTV}(q^2)}{m_B}\left(\gamma^{\mu}q^2-q^{\mu}\slashed{q}\right)-f_2^{\rTV}(q^2)i\sigma^{\mu\nu}q_{\nu}\right) {u}_{B}(p)\, ,\\
\langle B^\prime (p^\prime) | \overline{q}^\prime i\sigma^{\mu\nu}q_{\nu}\gamma_5 q | B(p) \rangle &= 
\overline{u}_{B^\prime}(p^\prime)\left(\frac{f_1^{\rTA}(q^2)}{m_B}\left(\gamma^{\mu}q^2-q^{\mu}\slashed{q}\right)-f_2^{\rTA}(q^2)i\sigma^{\mu\nu}q_{\nu}\right)\gamma_5 {u}_{B}(p)\, .\nonumber
\end{split}
\end{equation}
\\

Having collected all hadronic input used in the analysis we next present the full 
expressions for the two-body rates. We show separately the contributions from dipole and 
vector interactions with the LDV, c.f.~Eqs.~\eqref{eq:dipole} and \eqref{eq:vector}.
For brevity we drop the argument in all form factors since it is always $q^2=m^2_{\V}$ in
two-body decays. To shorten the expression we also introduce the notations
\begin{equation*}
  \kappa_x \equiv m^2_x / M^2 \qquad \text{and}\qquad
  \lambda_{xy}\equiv (1- \kappa_x- \kappa_y)^2 - 4 \kappa_x \kappa_y\,,
\end{equation*}
with $m_x$ indicating the mass of the final-state particle $x$ and $M$ the mass of the 
decaying particle.

\subsection{Partial width for $\boldsymbol{P\to P'+\V}$}
The partial width for the decay ${P\to P'+\V}$ with an underlying 
$q \to q^\prime$ flavor-changing transition is given respectively 
for dipole and vector interaction by
\begin{align}
  \label{eq:ratePPprime}
\Gamma (P \to P^\prime \V)\Big\vert_\rD &=
\frac{ \kag m_P^3  }{ 4\pi \Lambda^2 }
\frac{ \lambda^{3/2}_{P^\prime  \V} }{ (1 + \sqrt{\kap})^2 } |f_\rT^{PP'}|^2 |\CC{\rD}{q^\prime q}|^2\,,\\
\Gamma (P \to P^\prime \V)\Big\vert_\rV &=
\frac{m_P^3}{16\pi\Lambda^2}
			\lambda^{3/2}_ {P^\prime  \V }|f_+^{P P^\prime}|^2 | \CC{\rV }{q^\prime q } |^2\,,
\end{align}
Note that due to the parity conservation of strong interactions
the rate is independent of the axial couplings $\CC{\rV5}{ij}$ and $\CC{\rD5}{ij}$.
Therefore, $P\to P^{\prime}+\V$ decays are only sensitive to the
$\CC{\rV(\rD)}{ij}$ couplings. In the $\{\rL,\rR\}$ basis, these decays are
sensitive to both $\CC{\rD \rL(\rR)}{ij}, \CC{\rV \rL(\rR)}{ij}$ couplings. 

In the limit for massless LDV, the leading in $m_{\V}$ contributions to the decay rates read
\begin{align}
\lim_{m_{\V} \to 0} \Gamma(P \to P^\prime +\V)\Big\vert_\rD &=
\frac{m^2_{\V}  m_P  }{4\pi \Lambda^2}\left(1-\sqrt{\kap}\right)^3 |f_\rT^{PP'}|^2 | \CC{\rD }{ q^\prime q } |^2\,,\\
\lim_{m_{\V} \to 0} \Gamma(P \to P^\prime +\V)\Big\vert_\rV &=
\frac{ m_P^3}{16\pi\Lambda^2} \left(1-\kap\right)^3 |f_+^{P P^\prime}|^2| \CC{\rV }{ q^\prime q } |^2\,.
\end{align}
While the rate originating from dipole interactions vanishes in the massless limit,
the contribution of the vector interaction remains constant due to the linear scaling $m_{\V}/\Lambda$ introduced 
and discussed in Eq.~\eqref{eq:vector}.

\subsection{Partial width for $\boldsymbol{P\to {\mathcal V}+\V}$}
The partial width for the decay ${P\to {\mathcal V}+\V}$ with an underlying 
$q \to q^\prime$ flavor-changing transition is given respectively 
for dipole and vector interaction by
\begin{align}
\begin{split}
\Gamma(P \to \mathcal{V}+ \V)\Big\vert_\rD &= 
\frac{m_P^3      }{2\pi \Lambda^2} \lambda^{1/2}_{\mathcal{V} \V} \left( A_{\rD} |\CC{\rD }{ q^\prime q }|^2  + A_{\rD5}   |\CC{\rD5}{q^\prime q }|^2  \right)\,,\\
\Gamma(P \to \mathcal{V}+ \V)\Big\vert_\rV &= 
\frac{m_P^3\kag^2}{8\pi\Lambda^2}\lambda^{1/2}_{\mathcal{V} \V}\left(A_{\rV}|\CC{\rV }{q^\prime q }|^2+A_{\rV5}|\CC{\rV5}{q^\prime q }|^2\right)\, , \\
\end{split}
\end{align}
with the coefficients $A_X$ given by
\begin{align}
A_{\rD}  &= |T^{P\mathcal{V}}_1|^2 \lambda_{\mathcal{V}\V} \,,\\
A_{\rD5} &= |T^{P\mathcal{V}}_2|^2 \frac{8 \kaV \left( 1- \kaV \right)^2 + \kag \left( 1 + 3 \kaV \right)^2 - 2 \kag^2 (1 + 3 \kaV) + \kag^3 }{8 \kaV}\nonumber\\
         &+ |T^{P\mathcal{V}}_3 |^2 \lambda^{2}_{\mathcal{V}\V} \frac{\kag }{8 \kaV (1 - \kaV)^2}
          - {\rm Re}(T^{P\mathcal{V}}_2T^{P\mathcal{V *}}_3 ) \lambda_{\mathcal{V}\V} \frac{\kag (1+ 3 \kaV - \kag)}{4 \kaV (1 -\kaV)}\, , \\
A_{\rV}  &=  |V|^2 \frac{\lambda_{\mathcal{V} \V}}{\left(1+\sqrt{\kaV}\right)^2}\,, \\
A_{\rV5} &=|A_1|^2\frac{\kag^3-2\kag^2(1+3\kaV)+\kag(1+3\kaV)^2+8(1-\kaV)^2\kaV}{8\kaV\left(1-\sqrt{\kaV}\right)^2}\nonumber\\
         &+|A_3|^2\frac{\lambda_{\mathcal{V}\V}^2}{2\kag\left(1-\kaV\right)^2}
          +{\rm Re}(A_1A_3^{*}) \frac{\sqrt{1+\kaV}}{2\sqrt{\kaV}\left(1-\kaV\right)^2} \lambda_{\mathcal{V}\V}  \left(1-\kag+3\kaV\right)\,.
\end{align}

In the limit of a massless LDV, the decay rates reduce to
\begin{equation}
\begin{split}
  \lim_{m_{\V} \to 0} \Gamma (P \to \mathcal{V} \V)\Big\vert_\rD &=
        \frac{m_P^3}{2\pi \Lambda^2}\left(1-\kaV\right)^3 |T|^2
        \left( |\CC{\rD }{ q^\prime q }|^2  +  |\CC{\rD5}{q^\prime q }|^2 \right)\,,\\
  \lim_{m_{\V} \to 0} \Gamma (P \to \mathcal{V} \V)\Big\vert_\rV &=
		\frac{m_P^3}{16\pi\Lambda^2}\left(1-\kaV\right)^3\left(|A_3|^2 |\CC{\rV5}{ q^\prime q }|^2+\frac{2\kag |V|^2}{\left(\sqrt{\kaV}+1\right)^2}|\CC{\rV}{ q^\prime q }|^2\right)
        \,,
 \end{split}
\end{equation}
which illustrates that the sensitivity to $\CC{\rV}{q^\prime q }$ weakens for very light LDVs.

\subsection{Partial width for $\boldsymbol{B\to B'+\V}$}

For baryon decays $B\to B'+\V$ with an underlying  
$q \to q^\prime$ transition the contribution
to the partial width from the dipole and vector interaction read
\begin{align}
\begin{split}
\Gamma (B \to B^\prime \V)\Big\vert_\rD  &= \frac{m_B^3}{4\pi\Lambda^2}\lambda_{B^\prime \V}^{1/2}
\Big[
\left(|f_1^{\rTV}|^2 \hat{A}_{\rD 1}^{-}+|f_2^{\rTV}|^2\hat{A}_{\rD 2}^{-}+\hat{A}_{\rD 12}^{-}{\rm Re}(f_1^{\rTV}f_2^{\rTV *})\right)
|\CC{\rD }{ q^\prime q}|^2\\
&\hspace{5em}+
\left(|f_1^{\rTA}|^2 \hat{A}_{\rD 1}^{+}+|f_2^{\rTA}|^2\hat{A}_{\rD 2}^{+}+\hat{A}_{\rD 12}^{+}{\rm Re}(f_1^{\rTA}f_2^{\rTA *})\right)
|\CC{\rD5}{ q^\prime q}|^2
\Big] \, ,\\
\Gamma (B \to B^\prime \V)\Big\vert_\rV &=
\frac{m_B^3}{16\pi\Lambda^2}\lambda^{1/2}_{B^\prime \V}
\Big[\left(|f_1|^2 \hat{A}_{\rV 1}^{-}+|f_2|^2 \hat{A}_{\rV 2}^{-}+\hat{A}_{{\rV 12}}^{-}{\rm Re}(f_1f_2^{*})\right) |\CC{\rV }{ q^\prime q}|^2\\
&\hspace{6em}             +\left(|g_1|^2 \hat{A}_{\rV 1}^{+}+|g_2|^2 \hat{A}_{\rV 2}^{+}+\hat{A}_{{\rV 12}}^{+}{\rm Re}
(g_1g_2^{*})\right) |\CC{\rV5}{ q^\prime q}|^2 \Big]\, ,
\end{split}
\end{align}
with the kinematic coefficients
\begin{equation}
\begin{split}
\hat{A}_{\rD 1}^{\pm}   &=\kag\left(\kaB^2+\kaB\left(\kag-2\right)\pm 6\sqrt{\kaB}\kag-2\kag^2+\kag+1\right)\, , \nonumber\\
\hat{A}_{\rD 2}^{\pm}   &=2\kaB^2-\kaB\left(\kag+4\right)\pm 6\sqrt{\kaB}\kag-\kag^2-\kag+2\, , \nonumber\\
\hat{A}_{\rD 12}^{\pm}&=6\kag\left(\sqrt{\kappa_{B^\prime}}\mp 1\right)\left(1+\kappa_{B^\prime}\pm 2\sqrt{\kappa_{B^\prime}}-\kag\right)\, , \nonumber\\
\hat{A}_{\rV 1}^{\pm} & = \left(1+\kappa_{B^\prime}\pm 2\sqrt{\kappa_{B^\prime}}-\kag\right)\left(1+\kappa_{B^\prime}\mp 2\sqrt{\kappa_{B^\prime}}+2\kag\right)\, ,\nonumber\\
\hat{A}_{\rV 2}^{\pm} & =\kag\left(1+\kappa_{B^\prime}\pm 2\sqrt{\kappa_{B^\prime}}-\kag\right)\left(2+2\kappa_{B^\prime}\mp 4\sqrt{\kappa_{B^\prime}}+\kag\right) \, ,\nonumber\\
\hat{A}_{\rV 12}^{\pm} &=6 \kag\left(\kaB\pm 2\sqrt{\kaB}-\kag+1\right)\left(\sqrt{\kaB}\mp 1\right) \, .\\
\end{split}
\end{equation}
In the limit of a massless LDV, the rates reduce to
\begin{align}
\begin{split}
\lim_{m_{\V} \to 0}\Gamma (B \to B^\prime \V)\Big\vert_\rD & = 
\frac{m_B^3}{2\pi\Lambda^2}\left(1-\kappa_{B^\prime}\right)^3
\Big[
|f_2^{\rTV}|^2 |\CC{\rD }{ q^\prime q}|^2+|f_2^{\rTA}|^2|\CC{\rD5}{ q^\prime q}|^2
\Big] \, ,\\
\lim_{m_{\V} \to 0}\Gamma (B \to B^\prime \V)\Big\vert_\rV &= 
\frac{ m_B^3}{16\pi\Lambda^2}\left(1-\kappa_{B^\prime}\right)^3\Big(|f_1|^2|\CC{\rV }{ q^\prime q}|^2+|g_1|^2|\CC{\rV5}{q^\prime q}|^2\Big)\, .
\end{split}
\end{align}
For hyperon decays we use the form factor parametrization of Eq.~\eqref{eq:ffsdipolehyperon}, valid for $m_{\V}=0$.
Nonetheless, we consider a massive LDV for the kinematics for completeness.  The decay rate reads
\begin{align}
\begin{split}
\Gamma (B \to B^\prime \V)\Big\vert_\rD &=
\frac{m_B^3  }{4 \pi \Lambda^2} \lambda^{1/2}_{B^\prime \V}
(g_\rT^{B B^\prime})^2\left(\hat{A}^{-}_{\rD}|\CC{\rD }{ q^\prime q}|^2 + \hat{A}^{+}_{\rD} |\CC{\rD5}{q^\prime q}|^2  \right)\, ,
\end{split}
\end{align}
with the kinematic coefficients
\begin{equation}
\begin{split}
\hat{A}^{\pm}_{\rD}   & =\left(\kaB\pm 2\sqrt{\kaB}-\kag+1\right)\left(2\kaB\mp 4\sqrt{\kaB}+\kag+2\right) \, .
\end{split}
\end{equation}
In the limit of a massless LDV, the rate reduces to
\begin{align}
\begin{split}
\lim_{m_{\V} \to 0}\Gamma (B \to B^\prime \V)\Big\vert_\rD &= 
\frac{m_B^3  }{2 \pi \Lambda^2}\left(1-\kappa_{B^\prime}\right)^3|g_\rT^{B B^\prime}|^2\Big( |\CC{\rD }{ q^\prime q}|^2  +  |\CC{\rD5}{q^\prime q}|^2  \Big)\, .
\end{split}
\end{align}

For a fully polarized initial $B$, the differential width  read
\begin{align}
\begin{split}
	\frac{d\Gamma (B \to B^\prime \V)}{d\cos\theta}\bigg\vert_\rD
  &= \frac{m_B^3}{8\pi\Lambda^2}\lambda_{B^\prime \V}^{1/2}\Big[ \\
	&\quad
    \phantom{+}\ \left(|f_1^{\rTV}|^2 \hat{A}_{\rD 1}^{-}+|f_2^{\rTV}|^2\hat{A}_{\rD 2}^{-}+\hat{A}_{\rD 12}^{-}{\rm Re}(f_1^{\rTV}f_2^{\rTV *})\right)
	  |\CC{\rD }{q'q}|^2 \\
	&\quad
    + \left(|f_1^{\rTA}|^2 \hat{A}_{\rD 1}^{+}+|f_2^{\rTA}|^2\hat{A}_{\rD 2}^{+}+\hat{A}_{\rD 12}^{+}{\rm Re}(f_1^{\rTA}f_2^{\rTA *})\right)
	  |\CC{\rD5}{q'q}|^2 \\
	&\quad
    - 2\lambda^{1/2}_{B^\prime \V}\cos\theta \left(\hat{B}_{\rD 11}^{}{\rm Im}(f_1^{\rTV}f_1^{\rTA *})+\hat{B}_{\rD12}^{-}{\rm Im}(f_1^{\rTV}f_2^{\rTA *})\right.\\ 
	&\left.\hspace{7em}
      +\hat{B}_{\rD22}^{}{\rm Im}(f_2^{\rTV}f_2^{\rTA *})+\hat{B}_{\rD12}^{+}{\rm Im}(f_2^{\rTV}f_1^{\rTA *})\right){\rm Re}( {\CC{\rD }{q'q} }\CC{\rD5*}{q'q} )\\
	&\quad
    -2\lambda^{1/2}_{B^\prime \V}\cos\theta\left(\hat{B}_{\rD 11}^{}{\rm Re}(f_1^{\rTV}f_1^{\rTA *})+\hat{B}_{\rD12}^{-}{\rm Re}(f_1^{\rTV}f_2^{\rTA *})\right.\\ 
	&\left.\hspace{7em}
      +\hat{B}_{\rD22}^{}{\rm Re}(f_2^{\rTV}f_2^{\rTA *})+\hat{B}_{\rD 12}^{+}{\rm Re}(f_2^{\rTV}f_1^{\rTA *})\right) {\rm Im}( {\CC{\rD }{q'q} }\CC{\rD5*}{q'q} )
	\Big]\, ,\\
  \frac{d\Gamma (B \to B^\prime \V)}{d\cos\theta}\bigg\vert_\rV
  &= \frac{m_B^3}{32\pi\Lambda^2}\lambda^{1/2}_{B^\prime \V}\Big[ \\
  &\quad
    \phantom{+}\ \left(|f_1|^2 \hat{A}_{\rV 1}^{-}+|f_2|^2 \hat{A}_{\rV 2}^{-}+\hat{A}_{{\rV 12}}^{-}{\rm Re}(f_1f_2^{*})\right)
    |\CC{\rV }{ q^\prime q}|^2 \\
  &\quad
    + \left(|g_1|^2 \hat{A}_{\rV 1}^{+}+|g_2|^2 \hat{A}_{\rV 2}^{+}+\hat{A}_{{\rV 12}}^{+}{\rm Re}(g_1g_2^{*})\right)
    |\CC{\rV5}{ q^\prime q}|^2 \\
  &\quad
    - 2\lambda^{1/2}_{B^\prime \V}\cos\theta \left(\hat{B}_{\rV11}^{}{\rm Re}(f_1g_1^{*})+\hat{B}_{\rV12}^{+}{\rm Re}(f_2g_1^{*})\right.\\ 
  &\left.\hspace{7em}
      +\hat{B}_{\rV12}^{-}{\rm Re}(f_1g_2^{*})+\hat{B}_{\rV22}^{}{\rm Re}(f_2g_2^{*})\right){\rm Re}( {\CC{\rV }{q'q} }\CC{\rV5*}{q'q} )\\
  &\quad
    +2\lambda^{1/2}_{B^\prime \V}\cos\theta\left(\hat{B}_{\rV11}^{}{\rm Im}(f_1g_1^{*})+\hat{B}_{\rV12}^{+}{\rm Im}(f_2g_1^{*})\right.\\ 
  &\left.\hspace{7em}
      +\hat{B}_{\rV12}^{-}{\rm Im}(f_1g_2^{*})+\hat{B}_{\rV22}^{}{\rm Im}(f_2g_2^{*})\right) {\rm Im}( {\CC{\rV }{q'q} }\CC{\rV5*}{q'q} )\Big]\, ,
\end{split}
\end{align}
with the kinematic coefficients
\begin{align}
\hat{B}_{\rD11}^{}&=\kag\left(\kaB+2\kag-1\right),&
\hat{B}_{\rD22}^{}&=2\kaB+\kag-2,    &
\hat{B}_{\rD12}^{\pm}&=-\kag\left(3\sqrt{\kaB} \pm 1\right),\\
\hat{B}_{\rV11}^{}&=\kaB+2\kag-1 ,     &
\hat{B}_{\rV22}^{}&=\kag\left(2\kaB+\kag-2\right) ,&
\hat{B}_{\rV12}^{\pm}&=\kag \left(3\sqrt{\kaB}\pm 1\right). 
\end{align}

In the limit of a massless LDV, the rate reduces to
\begin{align}
\begin{split}
\lim_{m_{\V} \to 0}\frac{d\Gamma (B \to B^\prime \V)}{d\cos\theta}\bigg\vert_\rD &=\frac{m_B^3}{4\pi\Lambda^2}\left(1-\kappa_{B^\prime}\right)^3
\Big[
|f_2^{\rTV}|^2 |\CC{\rD }{q'q}|^2+|f_2^{\rTA}|^2|\CC{\rD5}{q'q}|^2\\ 
&
+2\cos\theta \left({\rm Im}(f_2^{\rTV}f_2^{\rTA *}){\rm Re}( {\CC{\rD }{q'q} }\CC{\rD5*}{q'q} )+{\rm Re}(f_2^{\rTV}f_2^{\rTA *}){\rm Im}( {\CC{\rD }{q'q} }\CC{\rD5*}{q'q} )\right) \Big] \, ,\\
\lim_{m_{\V} \to 0}\frac{d\Gamma (B \to B^\prime \V)}{d\cos\theta}\bigg\vert_\rV &=	\frac{ m_B^3}{32\pi\Lambda^2}\left(1-\kappa_{B^\prime}\right)^3\Big(|f_1|^2|\CC{\rV }{q'q}|^2+|g_1|^2|\CC{\rV5}{q'q}|^2\\ 
&
+2\cos\theta \left({\rm Re}(f_1g_1^{*}){\rm Re}( {\CC{\rV }{q'q} }\CC{\rV5*}{q'q} )-{\rm Im}(f_1g_1^{*}){\rm Im}( {\CC{\rV }{q'q} }\CC{\rV5*}{q'q} )\right)\Big)\, .
\end{split}
\end{align}

\subsection{Polarized lepton distributions and rates $\boldsymbol{\ell\to \ell'+\V}$\label{polarizeddecay}}
Next we consider the decays $\ell\to\ell^\prime+\V$ for the case
in which lepton-flavor violating dipole or vector interactions 
with the LDV are present.
In this case there is experimental sensitivity to the polarization 
of the initial lepton by the measurement of the angular distribution of 
the angle $\theta$, defined as the angle between the polarization 
vector of $\ell$ and the three-momentum of $\ell^\prime$.
For the different LDV interactions we find
for the differential width of a fully polarized initial $\ell$

\begin{equation}
\label{totaldecayrate}
\begin{split}
\frac{d\Gamma (\ell \to \ell^\prime\V )}{d\cos\theta}\bigg\vert_\rD&=
\frac{ m_{\ell}^3}{{8 \pi \Lambda^2 }}\lambda^{1/2}_{\ell^\prime\V} 
\Big[
  (\tilde{A}^{\rD}_{+}+\tilde{A}^{\rD}_{-})\left|\CC{\rD }{\ell^\prime\ell}\right|^2
+ (\tilde{A}^{\rD}_{+}-\tilde{A}^{\rD}_{-})\left|\CC{\rD5}{\ell^\prime\ell}\right|^2\\
&\hspace*{6em}
+ \tilde{A}^{\rD}_{\theta}\cos\theta\cdot {\rm Im}( \CC{\rD }{\ell^\prime\ell}\CC{\rD5*}{\ell^\prime\ell})
\Big]\,,\\
\frac{d\Gamma (\ell \to \ell^\prime\V )}{d\cos\theta}\bigg\vert_\rV&=
\frac{m_{\ell}^3}{32 \pi   \Lambda^2} \lambda^{1/2}_{\ell^\prime\V} 
\Big[
  (\tilde{A}^{\rV}_{+}+\tilde{A}^{\rV}_{-})\left|\CC{\rV }{\ell^\prime\ell} \right|^2 
+ (\tilde{A}^{\rV}_{+}-\tilde{A}^{\rV}_{-})\left|\CC{\rV5}{\ell^\prime\ell}\right|^2\\ 
&\hspace*{6em}
+ \tilde{A}^{\rV}_{\theta}\cos\theta\cdot {\rm Re}( {\CC{\rV }{\ell^\prime\ell} }\CC{\rV5*}{\ell^\prime\ell} )
\Big]\, ,
\end{split}
\end{equation}
with the kinematic coefficients
\begin{align}
  \tilde{A}^{\rD}_{+}      &=2\left(1-\kal \right)^2-\kag  \left(1+\kal \right)- \kag^2\,,&
  \tilde{A}^{\rV}_{+}      &=\left(1-\kal\right)^2+\kag\left(1+\kal\right)-2\kag^2  \,, \nonumber\\
  \tilde{A}^{\rD}_{-}      &=- 6 \sqrt{\kal} \kag \, , &
  \tilde{A}^{\rV}_{-}      &=- 6 \sqrt{\kal} \kag \, , \\
  \tilde{A}^{\rD}_{\theta} &= 2 \lambda^{1/2}_{\ell^\prime \V} \left(2-2 \kal - \kag\right)\,,&
  \tilde{A}^{\rV}_\theta   &= 2 \lambda^{1/2}_{\ell^\prime \V} \left(1-2 \kag -\kal\right)\,.
  \nonumber
\end{align}

In the limit of massless LDV, the polarized differential two-body rate reduces to
\begin{equation}
\begin{split}
\lim_{m_{\V}\to 0}\frac{d\Gamma (\ell \to \ell^\prime \V) }{d\cos\theta}\bigg\vert_\rD&=
\frac{m_{\ell}^3}{{4 \pi \Lambda^2 }}\left(1-\kal\right)^3\left(\left|\CC{\rD }{\ell^\prime\ell}\right|^2  + \left|\CC{\rD5}{\ell^\prime\ell}\right|^2 +2 \cos\theta\cdot {\rm Im}( \CC{\rD }{\ell^\prime\ell}\CC{\rD5*}{\ell^\prime\ell})\right)\\
\lim_{m_{\V}\to 0}\frac{d\Gamma (\ell \to \ell^\prime \V) }{d\cos\theta}\bigg\vert_\rV&=
\frac{ m_\ell^3}{32\pi\Lambda^2}\left(1-\kal\right)^3\left(\left|\CC{\rV }{\ell^\prime\ell} \right|^2  + \left|\CC{\rV5}{\ell^\prime\ell}\right|^2 +2 \cos\theta\cdot {\rm Re}( \CC{\rV }{\ell^\prime\ell} \CC{\rV5*}{\ell^\prime\ell})\right)\, .
\end{split}
\end{equation}
Finally, after integrating over $\theta$ and averaging over the 
initial- and final-state polarizations, the total decay rates read
\begin{align}
\Gamma(\ell \to \ell^\prime \V)\Big\vert_\rD &=
\frac{\lambda^{1/2}_{\ell^\prime \V}  m_{\ell}^3}{{4 \pi \Lambda^2 }}\left(
  \left|\CC{\rD }{\ell^\prime\ell}\right|^2 (\tilde{A}^{\rD}_{+}+\tilde{A}^{\rD}_{-})
+ \left|\CC{\rD5}{\ell^\prime\ell}\right|^2 (\tilde{A}^{\rD}_{+}-\tilde{A}^{\rD}_{-}) \right)\,,\\
\Gamma(\ell \to \ell^\prime \V)\Big\vert_\rV &=
\frac{\lambda^{1/2}_{\ell \V}  m_{\ell}^3}{16 \pi\Lambda^2} \left(
  \left|\CC{\rV }{\ell^\prime\ell}\right|^2 (\tilde{A}^{\rV}_{+}+\tilde{A}^{\rV}_{-})
+ \left|\CC{\rV5}{\ell^\prime\ell}\right|^2 (\tilde{A}^{\rV}_{+}-\tilde{A}^{\rV}_{-}) \right)\,,
\end{align}
which in the limit of massless LDVs reduces to
\begin{equation}
\begin{split}
\lim_{m_{\V}\to 0} \Gamma (\ell \to \ell^\prime \V)\Big\vert_\rD&=
\frac{m_{\ell}^3}{{2 \pi \Lambda^2 }}\left(1-\kal\right)^3\left(
\left|\CC{\rD }{\ell^\prime\ell}\right|^2  + \left|\CC{\rD5}{\ell^\prime\ell}\right|^2  
\right) \,,\\
\lim_{m_{\V}\to 0} \Gamma (\ell \to \ell^\prime \V)\Big\vert_\rV&=
\frac{m_{\ell}^3}{{16 \pi \Lambda^2 }}\left(1-\kal\right)^3\left(
\left|\CC{\rV }{\ell^\prime\ell} \right|^2 + \left|\CC{\rV5}{\ell^\prime\ell}\right|^2
\right) \,.
\end{split}
\end{equation}

\newpage
\addcontentsline{toc}{section}{References}
\bibliographystyle{JHEP}
\bibliography{references}

\end{document}